\documentclass[a4paper,12pt]{article}
\usepackage{amsmath}
\usepackage{amsfonts}
\usepackage{lscape}
\usepackage{fullpage}
\usepackage{bigfoot}
\usepackage{graphicx,color} 
\usepackage{graphics}
\usepackage{subcaption}
\usepackage{ae}
\usepackage[english]{babel}
\usepackage[T1]{fontenc}
\usepackage{pdflscape}
\usepackage{url}
\usepackage{setspace}
\usepackage{tabularx}
\usepackage{dcolumn}
\usepackage{rotating}
\usepackage[lmargin=2cm, rmargin=2cm,tmargin=2cm,bmargin=2.5cm]{geometry}
\usepackage{fullpage}
\usepackage{multirow}
\usepackage{longtable}
\usepackage{geometry}
\usepackage{rotating}
\usepackage{bigstrut}
\usepackage{comment}
\usepackage{lscape}
\usepackage{authblk}
\usepackage{cleveref}
\usepackage{lmodern}
\usepackage{pdflscape}
\usepackage{algorithm}
\usepackage{algpseudocode}
\usepackage{changepage}

\usepackage{caption} 

\usepackage[round]{natbib}

\usepackage{enumerate}
\usepackage{physics}
\usepackage{tikz}
\usepackage{mathdots}
\usepackage{yhmath}
\usepackage{cancel}
\usepackage{color}
\usepackage{siunitx}
\usepackage{array}
\usepackage{multirow}
\usepackage{amssymb}
\usepackage{gensymb}
\usepackage{extarrows}
\usepackage{booktabs}
\usetikzlibrary{fadings}
\usetikzlibrary{patterns}
\usetikzlibrary{shadows.blur}
\usetikzlibrary{shapes}

\newcolumntype{d}[1]{D{.}{.}{#1}}

\usepackage{amssymb}

\begin{document}
	
	
	\title{\bfseries {\Large Variable selection for minimum-variance portfolios}}
	
	\author[1]{Guilherme V. Moura}
	\author[2]{André P. Santos}
	\author[3]{Hudson S. Torrent}
		
	\affil[1]{\small Department of Economics, Universidade Federal de Santa Catarina, Florianópolis/SC, Brazil. e-mail: guilherme.moura@ufsc.br \bigskip}
	\affil[2]{\small  CUNEF Universidad, Madrid, Spain. e-mail: andre.santos@cunef.edu (corresponding author) \bigskip}
	\affil[3]{Department of Statistics, Universidade Federal do Rio Grande do Sul, Porto Alegre/RS, Brazil. e-mail: hudson.torrent@ufrgs.br}

	\date{\small \today}

	\maketitle
	
	
	\begin{abstract}
	
		Machine learning (ML) methods have been successfully employed in identifying variables that can predict the equity premium of individual stocks. In this paper, we investigate if ML can also be helpful in selecting variables relevant for optimal portfolio choice. To address this question, we parameterize minimum-variance portfolio weights as a function of a large pool of firm-level characteristics as well as their second-order and cross-product transformations, yielding a total of 4,610 predictors. We find that the gains from employing ML to select relevant predictors are substantial: minimum-variance portfolios achieve lower risk relative to sparse specifications commonly considered in the literature, especially when non-linear terms are added to the predictor space. Moreover, some of the selected predictors that help decreasing portfolio risk also increase returns, leading to minimum-variance portfolios with good performance in terms of Shape ratios in some situations.  Our evidence suggests that ad-hoc sparsity can be detrimental to the performance of minimum-variance characteristics-based portfolios.\\

		

		
		
	\end{abstract}

	
	\doublespacing
	
	\newpage
	
\section{Introduction}

Which variables are important to the portfolio construction problem, and how can they be selected? Addressing this question is difficult considering the availability of a myriad of predictors that seem to explain the cross-sectional patterns of stock returns \citep*[see, for instance,][]{cochrane2011presidential,harvey2016and,mclean2016does,hou2020replicating}. The problem becomes even more challenging when non-linear transformations, such as interactions between covariates and high-order moments, are taken into account, which results in an exponential increase in the predictor space. In this context, the number of features can easily surpass the number of data points, and the application of standard statistical methods, such as OLS, is known to perform poorly, see \cite{nagel2021machine}.  In this paper, we assemble a data set with 4,610 firm-level predictors and demonstrate that machine learning (ML) variable selection methods are capable of selecting subsets of these features that are important for the portfolio construction problem.

Much of the literature on variable selection in empirical asset pricing considers ad-hoc sparse specifications based on a few predetermined state variables in order to explain the cross-section patterns of stock returns \citep*[e.g.][]{fama1993common,fama2015five,hou2015digesting} and to determine how they affect optimal portfolio allocations \citep{ait2001variable}.  \cite*{kozak2020shrinking} refer to this approach as ``characteristics-sparse models''. For instance, \cite*{chan1999portfolio} and \cite*{brandt2009parametric} propose portfolio allocation methods based on low-dimensional factor specifications containing three characteristics. An obvious limitation of this approach is that many important variables can be omitted. Two notable and recent exceptions are the works of \cite*{demiguel2017portfolio} and \cite{kozak2020shrinking} who employ high-dimensional data sets containing a large number of characteristics and find that the number of relevant ones for the portfolio construction problem is larger  in comparison to the standard specifications based on only a few ad-hoc characteristics. \cite{nagel2021machine} argue that ``One interpretation of this expansion in the number of factors is that the literature is slowly adjusting to the fact that there are, indeed, relevant omitted factors.''

In this paper, we present a methodological framework for high-dimensional predictor selection and provide empirical evidence that enables us to establish four main conclusions that are useful for academics and market practitioners. First and foremost, the choice of variable selection method matters: there are substantial differences in terms of the \textit{preference} of each method with respect to specific features and to the \textit{number} of features each method selects. Second, the number of selected features that are relevant for the portfolio construction problem exceeds the number of variables used in popular low-dimensional factor models. Third, augmenting the predictor space with non-linear transformations of the original characteristics leads to optimal portfolios with improved performance.\footnote{Non-linear transformations have been extensively studied in the asset pricing literature. For instance, \cite{harvey2000conditional} assume that the stochastic discount factor is quadratic in the market return. \cite*{hong2000bad} show that momentum interact with firm size and analyst coverage: momentum strategies declines with firm size and work better among stocks with low analyst coverage. \cite{medhat2022short} show that previous month's return and share turnover also interact: low-turnover stocks exhibit significant short-term reversal whereas high-turnover stocks exhibit short-term momentum.}  Fourth, working with a subset of important predictor variables results in minimum-variance portfolios that outperform their benchmarks in terms of a lower standard deviation of out-of-sample portfolio returns.

We add to the existing literature by extending the works of \cite{kozak2020shrinking} and \cite{demiguel2017portfolio} along two dimensions. First, we raise the statistical challenge by assembling a set of predictor variables containing the 95 firm characteristics considered in \cite{gu2020empirical} and augment it with the second-order and cross-product (interaction) terms among all characteristics, yielding a total of 4,610 predictors.  This allows us to study not only \textit{whether} and to \textit{what extent} non-linear functions of features that are important for portfolio construction, but also \textit{what type} non-linearity (i.e. non-linearities of individual predictors or interactions between covariates) that are most important in helping the investor improve portfolio performance. Existing empirical evidence \citep*{moritz2016tree,bryzgalova2020forest,chen2020deep,freyberger2020dissecting,gu2020empirical,gu2021autoencoder} suggests that non-linear functions of firm characteristics can be an important source \textit{of predictability of security returns}. 
Thus, it is natural to argue whether non-linearities are relevant and what type they may be \textit{from a portfolio choice perspective}. 

Our second contribution is to expand the set of variable selection methods used to identify the most important predictors. Specifically, we consider three classes of variable selection methods: \textit{regularization}, \textit{ensembles}, and \textit{Bayesian} methods. While regularization methods perform variable selection by shrinking the coefficients using the $L1$ regularization (or a combination of $L1$ and $L2$ regularizations), the ensemble method considered in the paper operates via a stagewise additive modeling procedure that iteratively estimates the model by sequentially adding new components, with early stopping of the algorithm to avoid overfitting. The Bayesian variable selection, on the other hand, performs regularization by specifying a  prior over the parameters, which shrinks weak signals toward zero while allowing strong signals to remain unshrunk. Finally, as a robustness check, we implement the marginal screening method of \cite{fan2008sure}, which is computationally efficient and suitable for ultra-high dimensions of predictors. 

It is also worth highlighting three important methodological aspects of our work. First, consistent with our interest in the portfolio choice problem, we focus on \textit{directly} parameterizing the objective of interest - \textit{portfolio weights} - as a function of lagged firm-level characteristics and their cross-products and second-order transformations, instead of predicting \textit{stock returns}. This difference is important because variables that lead to better predictions of individual security returns in terms of higher out-of-sample $R$-squared do not necessarily lead to better portfolios in terms of standard metrics such as the out-of-sample standard deviation of portfolio returns or the portfolio's Sharpe ratio; see \cite{nagel2021machine} for a discussion. The reason is that portfolio performance depends on the properties of the covariance matrix of asset returns, and $R$-squared measures are uninformative about those properties.\footnote{\cite{pastor2000comparing} show that a model that is better for pricing is not necessarily better for investing, because investors are usually subject to margin requirements and model uncertainty that prevent them from implementing certain investment strategies suggested by asset pricing models. \cite*{fabozzi2016difference} builds on the approach developed by \cite{pastor2000comparing} and find that, although the 4-factor model of \cite*{hou2015digesting} is better than the 5-factor model of \cite{fama2015five} at explaining anomalies, the two models have similar performance in terms of investing.} 

A second important aspect of our methodology is the use of the parametric portfolio policy approach of \cite{brandt2009parametric} to parameterize portfolio weights as a linear function of non-linear transformations of lagged firm characteristics. We borrow analytical results obtained in \cite{demiguel2017portfolio} and reformulate the parametric portfolio approach as a penalized regression model that allows us to study the role played by alternative variable selection strategies in identifying relevant predictors for the portfolio construction problem.

Finally, the third important aspect of our methodology is the focus on minimum-variance portfolios. Several reasons motivate this choice. First, minimum-variance strategies are among the most popular smart-beta strategies in the U.S. \citep*{ang2017crowding}.\footnote{A total of 18 minimum-volatility ETFs are traded in the U.S. They have \$52.08 billion in assets under management as of July 2024 (see \url{etf.com/topics/low-volatility}).} Second, minimum-variance portfolios are less sensitive to estimation errors in expected returns and often lead to robust performance in practice \citep{demiguel2009portfolio}. Third, the  minimum-variance  parametric portfolios can be cast as an unconstrained regression problem, which allows for the deployment of a wide range of variable selection methods in a natural way. Finally, arbitrary characteristic sparsity can be particularly harmful to these portfolios. This is the case because a potentially high number of predictors can help reduce portfolio risk, even if they do not contribute to increasing portfolio mean returns. For instance, a predictor with a low covariance with the benchmark portfolio or with other predictors is likely to be important for constructing a minimum-variance portfolio, even if it does not aid in increasing average portfolio returns. We circumvent the arbitrary sparsity by considering a very large cross-section of firm characteristics and employing variable selection methods to identify predictors that are relevant for the minimum-variance construction problem. 



We perform an empirical exercise in which we use the selected predictors from each variable selection method to construct minimum-variance portfolios for the universe of NYSE, NASDAQ, and NYMEX stocks and conduct an out-of-sample evaluation. The results point to a clear superiority of the portfolios obtained with the variable selection methods in comparison to those obtained with ad hoc sparse specifications such as the 3- and 5-factor models of \cite{fama1993common,fama2015five} as well as classic benchmark strategies such as the value-weighted and equally weighted portfolios. For instance, the minimum-variance portfolios with predictors selected using the best-performing method delivered out-of-sample returns with an annualized standard deviation of 9.1\%, and this result is further improved to 8.3\% when using the augmented data set with all non-linear predictor transformations. These figures are substantially and statistically lower than those obtained with the 3-factor (15\%) and 5-factor (14.6\%) models.



We also investigate the importance of individual predictors. We find that the main effects and interactions among market beta, different measures of return volatility, return momentum, liquidity, and bid-ask spread are among the most important predictors. We provide an interpretation of the predictor importance by i) identifying  which aspects of the investor's utility the predictors contribute to the most, and ii) showing how predictor importance translates to portfolio allocations. For instance, stocks with lower liquidity receive higher weights, and this result is stronger for stocks with lower idiosyncratic volatility. Stocks with lower market betas receive higher weights, and this result is stronger for stocks with a higher standard deviation of liquidity. Our results are in line with those reported in \cite{scherer2011note}, which finds that risk-based characteristics such as market beta and idiosyncratic volatility can explain the allocations of minimum-variance portfolios. Our paper extends those results and shows that different measures of size, return momentum, and liquidity are also important, as well as their interactions with different risk-based characteristics.

To assess the economic significance of the most important predictors, we examine whether variations in their estimated parametric-portfolio coefficients are associated with fundamental state variables that reflect market conditions, such as investor sentiment, market-wide liquidity, and economic recession. Specifically, we use the investor sentiment index from \cite{baker2006investor}, the liquidity index from \cite{pastor2003liquidity}, and the NBER recession indicator. Our analysis shows that variations in the estimated coefficients of the most influential predictors are indeed related to these state variables. Furthermore, the variations in coefficients associated with the main effects of predictors and the interactions between predictors tend to be linked to different state variables, which suggests that different classes of predictors capture distinct market attributes.

Our paper is connected to a growing literature that employs machine learning methods to support portfolio choice decisions; see \cite{kelly2023financial} for a recent review of the literature. \cite*{li2020selecting}, \cite*{demiguel2023machine}, and \cite*{kaniel2023machine} employ ML methods to construct prediction-based portfolios of mutual funds, whereas \cite{gu2020empirical} focuses on prediction-based stock portfolios. ML methods have also been used in economic and financial forecasting. For instance, \cite*{rapach2013international}, \cite*{chinco2019sparse}, and \cite*{freyberger2020dissecting} use the lasso and the elastic net to select regressors for stock return prediction, whereas \cite*{dong2022anomalies} employs ML to predict the aggregate market excess return. \cite{bai2009boosting}, \cite{buchen2011forecasting},  \cite*{dopke2017predicting}, and \cite{kauppi2021boosting} use the boosting method for macroeconomic forecasts. A common denominator across these studies is that employing variable selection methods leads to improved results compared to those obtained with standard statistical methods.

Our paper is mostly connected with the works of \cite*{simon2022deep} and \cite{lamoureux2024empirical}. \cite{simon2022deep} formulates a mean-variance parametric portfolio policy using a non-linear functional form based on deep neural networks. In our paper, we preserve the linear functional form and model non-linearities via second-order and cross-product variable transformations. While the approach of \cite{simon2022deep} can capture non-linearities in a  flexible way, our method is highly interpretable and allows us to understand not only which \textit{how many} predictors survive the different variable selection processes but also how \textit{why} a selected predictor is important according to a given method. \cite{lamoureux2024empirical} uses the parametric portfolio approach to study whether an investor with power utility can exploit the predictability contained in a set of six characteristics. Our paper, on the other hand, focuses on minimum-variance portfolios and considers a much larger cross-section of predictors, which allows us to understand the extent to which ad-hoc sparsity is detrimental to portfolio construction.

The rest of the paper is organized as follows. Section \ref{section:data} describes the dataset used in the paper. Section \ref{section:methodology} details the portfolio strategies as well as the methods for variable selection and regularization. Section \ref{section:empirical_application} presents the results of an empirical application. Finally, Section \ref{section:conclusions} concludes.

\section{Data}\label{section:data}
	
We use the dataset of 95 monthly firm characteristics utilized in \cite{gu2020empirical}. The database was downloaded from Dacheng Xiu's homepage (\url{https://dachxiu.chicagobooth.edu/download/datashare.zip}).\footnote{Table A.6 of the supplementary material of \cite{gu2020empirical} brings additional details of the characteristics. The supplementary material can be download at \url{https://dachxiu.chicagobooth.edu/download/ML_supp.pdf}. Although the original data set used in \cite{gu2020empirical} contains 94 characteristics, the data set available Dacheng Xiu's homepage contains one additional characteristic, market value of equity ($mve0$). Although $mve0$ is highly correlated with another measure of firm size in the data set ($mvel1$), we opted to retain both variables.} Stock returns are sourced from CRSP. Our sample spans from January 1970 to June 2019 (a total of 594 months). 

We perform feature engineering to augment the original dataset containing the main effects of the 95 predictors. Specifically, for each firm characteristic, we obtain the square values and cross-products among all predictors; that is, we define $pred_{i,t}^2$ and 
$pred_{i,t} \times pred_{j,t}$ 
for $i\ne j$ to denote the second-order and interaction effects, respectively.\footnote{The predictors Convertible debt indicator (\textit{convind}), Dividend initiation (\textit{divi}), Dividend omission (\textit{divo}), R\&D increase (\textit{rd}), Secured debt indicator (\textit{securedind}), and Sin stocks (\textit{sin}) are defined as binary variables and therefore we do not consider their second-order effects. Additionally, the squared value of the predictor \textit{beta} is not computed since its second-order effect is represented by the predictor \textit{betasq}.} The augmented dataset contains the predictors' main effects, the second-order effects, and the interaction effects. The final set contains 4,610 predictors. The total number of firms in our sample is 15,701, with the fewest firms in January 1973 (1,443 firms) and the most firms in December 1997 (5,128 firms). 

To alleviate the impact of extreme observations, we cross-sectionally winsorize each predictor at the 1st and 99th cross-sectional percentiles, as in \cite*{green2017characteristics}. We also follow \cite{brandt2009parametric} and standardize each predictor so that its cross-sectional mean is zero and its standard deviation is one. Missing characteristic values are set to zero. The resultant standardized predictor is a long-short portfolio that goes long on stocks whose characteristics are above the cross-sectional average and short on stocks whose characteristics are below the cross-sectional average.

We report in Table \ref{table:descriptive_stats} descriptive statistics of the predictor returns, which are defined as  
$$rpred_{t+1} = \frac{pred_t^\intercal r_{t+1}}{N_t},$$
where $pred_t$ is the standardized predictor long-short portfolio at time $t$, $r_{t+1}$ is the vector of stock returns at time $t+1$, and $N_t$ is the number of stocks at time $t$. The Table reports the average monthly return, the standard deviation of monthly returns, the 25th and 75th percentiles, and the correlation of the predictor return with the equally weighted and value-weighted portfolio returns. The first four rows report aggregate values across i) all predictors, ii) predictors' main effects, iii) predictor's second-order effects, and iv) predictors' interaction effects. The remaining rows report individual statistics for the  top and bottom 10 predictors within each category, sorted in terms of average monthly return. 

The overall average monthly return of all predictors is slightly negative at -0.003\%, with a standard deviation of 0.523\% and a near-zero correlation with the equally weighted and value-weighted portfolio returns. We find similar results for the subgroups of main effects, second-order effects, and interaction effects, except that second-order effects display higher correlations with the equally weighted and value-weighted portfolio returns. The analysis of the top and bottom 10 predictors within each category, on the other hand, reveals a wide variation in performance. For instance, among the top-10 main effects, industry momentum (\textit{indmom}) shows the highest average return of 0.3\%, while book-to-market (\textit{bm}) shows the lowest at 0.1\%. The interaction between illiquidity and size ($ill \times mve0$) earns the highest average return (0.9\%), whereas the interaction between short-term reversal and financial statement score ($mom1m \times ps$) earns the lowest (-0.4\%). We also observe substantial variation (between -0.4 and 0.6) in the correlations of the top and bottom 10 predictors with the equally weighted and value-weighted portfolio returns.

\section{Methods}\label{section:methodology}

We now describe the portfolio optimization framework considered in the paper, as well as the ML methods used to identify predictors that are helpful for the portfolio minimum-variance construction problem. First, we review the parametric portfolio policy of \cite{brandt2009parametric} and its minimum-variance reformulation as an unconstrained regression problem. Second, we discuss how alternative variable selection methods can be introduced into this framework. Finally, we discuss the approach to interpret the importance of the selected predictors.

The parametric policy is a portfolio optimization task aimed at maximizing utility. This is achieved by adjusting a benchmark portfolio (such as equally weighted or value-weighted portfolios) using a series of asset-specific predictors to enhance the investor's utility. Specifically, the benchmark portfolio is modified by adding a weighted sum of long-short portfolios. These portfolios are derived from $K$ predictors that are normalized cross-sectionally to have a mean of zero and a standard deviation of one. The parametric policy strictly involves equity investments, explicitly excluding any allocation to risk-free assets. 

The composition of the parametric portfolio at a given time $t$, denoted as $w_t(\theta)\in \mathbb{R}^{N_t}$, is expressed as follows:
\begin{equation}\label{eq:PP}
	w_t(\theta) = w_{b_t} + \frac{1}{N_t}\sum_{k=1}^{K}x_{k,t}\theta_k = w_{b_t} + \frac{X_t \theta}{N_t},
\end{equation}
where $\theta^\intercal = (\theta_1, \theta_2, \dots, \theta_K) \in \mathbb{R}^{K}$ is a $1 \times K$ vector of coefficients, $X_t = [x_{1,t}, x_{2,t}, \ldots, x_{K,t}] \in \mathbb{R}^{N_t \times K}$ represents a matrix of asset characteristics at time $t$, $w_{b,t} \in \mathbb{R}^{N_t}$ is the benchmark portfolio at time $t$, $x_{k,t}\in \mathbb{R}^{N_t}$ is the standardized long-short portfolio for the $k$th predictor, $\theta_k$ is the associated coefficient for the $k$th predictor, and $N_t$ is the total number of stocks at time $t$. The standardization ensures that the average of $x_{k,t}\theta_k$ across all assets is zero, leading to an optimal portfolio where deviations from the benchmark weights sum to zero, and the total portfolio weights are always equal to one.

The return of this parametric portfolio at a subsequent time point, $t+1$, denoted as $r_{p,t+1}$, is given by
\begin{equation}\label{eq:port.ret}
	r_{p,t+1}(\theta) = r_{b,t+1} + \theta^\intercal r_{c,t+1},
\end{equation}
where $r_{t+1}\in \mathbb{R}^{N_t}$ is the return vector at time $t + 1$, $r_{b,t+1}=w_{b,t}^\intercal r_{t+1}$ represents the benchmark portfolio's return, and $r_{c,t+1} = \frac{X_t^\intercal r_{t+1}}{N_t}$ is the predictor return vector at that time. This predictor return vector encapsulates the returns from the long-short portfolios related to the $K$ predictors, adjusted by the number of assets, $N_t$. The equation reveals that the return on the parametric portfolio is a sum of the benchmark return and the yield from a linear combination of the characteristic portfolios.

\subsection{Minimum-variance parametric portfolios}\label{section:MVSPP}


A very large body of literature on portfolio optimization considers the minimum-variance policy; see, for instance, \cite*{green1992will}, \cite*{jagannathan2003risk}, \cite*{clarke2006minimum}, \cite*{demiguel2009portfolio}, and \cite*{clarke2011minimum} just to name a few. This strategy performs well as it is more robust than the mean-variance policy since it does not require estimating mean returns, which is a notoriously difficult task.

We assume a minimum-variance investor who solves the following problem:
\begin{align}\label{eq:RSPP1C}
	\min_{\theta}&\,\,\,\frac{1}{2}\text{var}\left[r_{p,t+1}(\theta)\right] \equiv
	\frac{1}{2} E\left[(\dot{r}_{p,t+1}(\theta))^2\right],
\end{align}	
where $\dot{r}_{p,t+1}(\theta)$ is the portfolio return vector centered on zero, i.e. $\dot{r}_{p,t+1}(\theta) = r_{p,t+1}(\theta) - \bar{r}_{p}(\theta)$ and $\bar{r}_{p}(\theta)$ is the mean portfolio return.
In practice, one minimizes the empirical version of eq.~(\ref{eq:RSPP1C}) using a sample with $T$ observations, defined as
\begin{equation}\label{eq:emp.obj.fun}
	\min_{\theta}\,\,\,\frac{1}{2}
	\frac{1}{(T - 1)} \sum_{t = 1}^{T - 1} 
	\left(\dot{r}_{b,t+1} + \theta^\intercal \dot{r}_{c,t+1}\right)^2,
\end{equation}
where $\dot{r}_{b,t+1}$ and $\dot{r}_{c,t+1}$ are, respectively, the mean-centered benchmark return and predictor return. It is straightforward to note that the portfolio loss function in \eqref{eq:emp.obj.fun} can be formulated as a regression problem. Specifically, the optimal parameter $\theta^*$ in eq.~\eqref{eq:emp.obj.fun} can be estimated with a time-series regression without an intercept,  
\begin{equation}\label{regression}
	\dot{r}_{b, t}=-\theta^{\intercal} \dot{r}_{c, t}+\epsilon_t,
\end{equation}
where $\epsilon_t$ is the error term. The corresponding ordinary least squares (OLS) solution of \eqref{eq:emp.obj.fun} is given by
\begin{equation}\label{eq:solution_OLS}
	\hat \theta = - \hat \Sigma_{c}^{-1} \hat \sigma_{bc},
\end{equation}
where $\hat \Sigma_c$ is the sample covariance matrix of the predictor-return vector $r_c$, and $\hat \sigma_{bc}$ is the sample vector of covariances between the benchmark portfolio return $r_b$ and the predictor-return vector $r_c$.\footnote{The regression formulation of the parametric minimum-variance portfolio problem in \eqref{regression} differs from existing portfolio-regression formulations. For instance, \cite{britten1999sampling} shows that the tangency portfolio can be obtained by a regression of a constant $\mathbf{l}$ onto a set of asset's excess returns, without an intercept term. \cite{kempf2006estimating} show that the global minimum-variance portfolio can be obtained by regressing the returns of a given asset $i$ onto the differences between the returns of asset $i$ and all other assets. The portfolio-regression formulation adopted in the paper is closer to that developed in \cite{demiguel2017portfolio} as it is based on the parametric-portfolio formulation.}

The OLS solution of the empirical minimum-variance portfolio problem in \eqref{eq:solution_OLS} suffers from a major drawback: when $K>T$, that is, when the number of predictors exceeds the number of data points, the OLS solution is poor or unfeasible since the sample covariance matrix $\hat \Sigma_c$ is no longer positive definite. This poses a serious limitation to the use of standard regression methods in estimating the parameters of the minimum-variance parametric portfolios when the dimension of the predictor space is high. 

\subsection{Variable selection methods}\label{section:ML_methods}

We will now discuss the three classes of variable selection methods that help solve the empirical minimum-variance parametric portfolio problem in eq.~\eqref{eq:emp.obj.fun} in situations where the number of predictors can exceed the number of data points. The three major classes of variable selection procedures considered are regularization methods, Bayesian methods, and ensemble methods.

\subsubsection{Regularization}

We consider a formulation of the unconstrained regression in \eqref{eq:emp.obj.fun} in which a regularization term is added to encourage sparsity,
\begin{equation}\label{eq:emp.obj.fun.reg}
	\min_{\theta}\,\,\,\frac{1}{2}
	\frac{1}{(T - 1)} \sum_{t = 1}^{T - 1} 
	\left(\dot{r}_{b,t+1} + \theta^\intercal \dot{r}_{c,t+1}\right)^2 + \Omega_{\lambda}\left(\theta\right),
\end{equation}
where $\Omega_{\lambda}\left(\theta\right)$ defines the penalization function and $\lambda$ denotes the penalization hyperparameters. Next, we describe the five penalty-based methods used in the paper.

\begin{description}
	\item[Ridge] \cite{hoerl1970ridge} proposed the ridge estimator to reduce mean squared error of the OLS at the cost of some bias. The idea is to add a penalty proportional to the \textit{squared magnitude} of the coefficients. The ridge penalization is defined as
	\begin{equation}\label{Ridge}
		\Omega_{\lambda}\left(\theta\right) =  \lambda \sum_{k=1}^{K} \theta_k^2,
	\end{equation}
where $\lambda$ is the regularization hyperparameter that controls the strength of the penalty. The decrease in mean squared error compared to the OLS occurs for some intermediate value of $\lambda$, for which the reduction in variance of $\widehat{\theta}$ that solves \eqref{Ridge} surpasses the bias induced by the regularization. Ridge regression offers the advantage of a computationally straightforward analytic solution. However, in this approach, coefficients associated with less relevant predictors are gradually shrunk toward zero but never precisely reach it. Consequently, ridge regression is not well suited for predictor selection \citep*[see, for example,][]{masini2023machine}.


	\item[Least absolute sum of squares operator (lasso)] With the objective of performing variable selection jointly with regularization, \cite{tibshirani1996} proposed the lasso, where the penalty is the \textit{absolute magnitude} of the coefficients. The formulation becomes	
	\begin{equation}
		\Omega_{\rho}\left(\theta\right)  = \rho \sum_{k=1}^{K} |\theta_k|,
	\end{equation}
	where $\rho$ is the regularization hyperparameter.	Due to the specific nature of the constraint in the lasso, reducing $\rho$ sufficiently will result in some coefficients being precisely zero. Consequently, the lasso exhibits a form of continuous subset selection, providing a \textit{sparse} estimator  of the paremeter vector $\theta$ \citep*[see, for example,][]{hastie2009elements}. 

\item[Adaptive lasso (adalasso)] Although the lasso is a consistent method for variable selection under certain conditions, there exist scenarios where the lasso is inconsistent for this purpose. \cite{zou2006adaptive} proposed the adalasso regression, which is a modified version of the lasso regression based on adaptive weights used to penalize coefficients differently, 
\begin{equation}
	\Omega_{\rho,\phi}\left(\theta\right)  =  \rho \sum_{k=1}^{K} \phi_k |\theta_k| ,
\end{equation}
where $\phi_k$ are weights typically set as the inverse of the absolute values of the estimates from an initial ridge regression, and $\rho$ is the regularization parameter. As the lasso, the adalasso also delivers sparsity and efficient estimation algorithm, but enjoys the oracle property \citep[][]{fan2001variable}, meaning that it has the same asymptotic distribution as the OLS conditional on knowing the variables that should enter the model.

\item[Elastic net] Introduced by \cite{zou2005regularization}, the elastic net strategically combines the merits of both lasso and ridge regression, and the penalization function becomes
	\begin{equation}
	\Omega_{\lambda,\rho}\left(\theta\right)= \lambda\rho\ \sum_{k=1}^K |\theta_k| + \lambda(1-\rho)\ \sum_{k=1}^K \theta_k^2.
\end{equation}
The $L1$-norm term ($\lambda\rho\ \sum_{k=1}^K |\theta_k|$) can be used to control the sparsity of the estimated parameter vector $\theta$ and the $L2$-norm term ($\lambda(1-\rho)\ \sum_{k=1}^K \theta_k^2$) to increase its stability. For the case with $\rho=0$, the objective function includes only the  $L2$-norm term, and thus, elastic net is equivalent to ridge regression. If, on the other hand, $\rho=1$, the objective function includes only the $L1$-norm term, and lasso regression is performed. The elastic net regression can offer advantages over using either lasso or ridge alone, particularly in the presence of correlated predictors, when it outperforms the lasso \citep[see, for example,][]{zou2009adaptive}.

\item[Smoothly clipped absolute deviation (SCAD)] The SCAD penalization introduced by \cite{fan2001variable} combines the strengths of both ridge and lasso by offering a penalty that varies with the coefficient's magnitude. The objective function can be expressed as
\begin{equation}\label{scad}
	\Omega_{p}\left(\theta\right) = \sum_{k=1}^{K} p_{\lambda}(\theta_k),
\end{equation}
where $p_{\lambda}(\theta)$ is the SCAD penalty function defined as
$$
p_\lambda(\theta)= \begin{cases}\lambda|\theta| & \text { if }|\theta| \leq \lambda, \\ -\frac{|\theta|^2-2 a \lambda|\theta|+\lambda^2}{2(a-1)} & \text { if } \lambda<|\theta| \leq a \lambda, \\ \frac{(a+1) \lambda^2}{2} & \text { if }|\theta|>a \lambda,\end{cases}
$$
where $a>2$ is a fixed parameter, typically set to 3.7 based on empirical evidence and $\lambda$ controls the penalty's intensity. Unlike ridge, SCAD allows some coefficients to be exactly zero and, unlike lasso, the larger coefficients are shrunk less severely. A drawback of the SCAD penalty is that it is non-convex, which makes computation more difficult. We adopt the coordinate descent algorithm for non-convex penalized regression proposed in \cite{breheny2011} who provide a fast, efficient and stable algorithm for solving \eqref{scad}. 

\end{description}

\subsubsection{Bayesian variable selection}

From a Bayesian point of view, variable selection is accomplished by specifying a prior distribution over the model parameters. A Bayesian counterpart to \eqref{regression} can be written as
\begin{equation}\label{Bayesregression}
	\dot{r}_{b, t}=-\theta^{\intercal} \dot{r}_{c, t}+\epsilon_t,\qquad \epsilon_t\sim N(0,\sigma^2_{\epsilon}),\qquad \theta\sim p(\theta|\omega),
\end{equation}
where $p(\theta|\omega)$ defines the prior distribution over $\theta$, and $\omega$ collects all hyperparameters. It is interesting to note that the maximum a posteriori probability estimator, obtained by minimizing the negative log-posterior, is equivalent to classical approaches of penalized regression, and specific choices of prior distributions can generate all classical penalized regressions presented above \citep*[see, for example,][]{bhadra2019lasso}. A successful global-local shrinkage prior is the horseshoe \citep*[][]{carvalho2010}, which can be defined as
\begin{equation}\label{horseshoe}
	\theta\sim p(\theta|\lambda_i,\tau)\sim N(0,\lambda_i^2), \quad \lambda_i|\tau\sim C^+(0,\tau),\quad \tau\sim C^+(0,1),
\end{equation}
where $C^+$ denotes the half-Cauchy density, whose support is the non-negative real line.

The main advantages of the horseshoe prior are its tail- and sparse-robustness properties. These properties are a byproduct of its spike at zero and the heavy tails produced by the scale mixture of normals defined by \eqref{horseshoe}, which shrink weak signals toward the origin while still allowing strong signals to remain unshrunk due to their heavy tails. 

\subsubsection{Ensembles}

The ensemble approach to variable selection consists of integrating  multiple models to identify and select the most significant variables in a given data set. Techniques such as \textit{bagging} \citep*{breiman1996bagging} and \textit{boosting} \citep*{freund1996schapire,zhang2005boosting} are commonly used, where models are trained on various subsets or reweighted instances of the data. In this paper, we consider the $L2$-boosting method of \cite{BuhlmannYu2003} which falls under the umbrella of boosting algorithms. The $L2$-boosting method is based on the principle of improving a model's predictive power by sequentially adding weak learners across a maximum of $m^*$ iterations, where $m^*$ is a hyperparameter that must be tuned. When the number of covariates $K$ in a data set is large (and when selecting a small number of relevant covariates is desirable), boosting is usually superior to standard estimation techniques for regression models (such as backward stepwise linear regression, which, for example, cannot be applied if $K$ is larger than the number of observations $T$).  Algorithm \ref{algo_L2} provides a pseudocode to implement the $L2$-boosting for minimum-variance parametric portfolios.\footnote{In Algorithm \ref{algo_L2}, $\hat \theta^{[m]}$ is the estimated value of $\theta$ obtained \textit{during} the $m$-th iteration of the boosting algorithm, and $\hat \theta_{m}$ is the estimated value of $\theta$ obtained \textit{after} $m$ iterations of the boosting algorithm.}

\subsubsection{Choice of hyperparameters}

In each estimation window, the regularization hyperparameters for ridge, lasso, elastic net, and SCAD are determined through five-fold cross-validation, as discussed in \citet[Chapter~7]{hastie2009elements}. This involves setting a range of potential values for the hyperparameters. The sample is segmented into five segments, referred to as ``folds.'' For each $j$ in the range from 1 to 5, the $j$th fold is excluded, and the remaining four are used to generate predictions for each hyperparameter value. The prediction error, or cross-validation error, for each hyperparameter value is then calculated on the excluded $j$th fold. This procedure is repeated across all five folds. The hyperparameter value that results in the lowest average cross-validation error is chosen. As for the $L2$-boosting method, the hyperparameter of interest is the maximum number of boosting iterations, which is analogous to an early stopping criterion. Stopping the algorithm too early will not capture important features of the data. Terminating the process too late can lead to the well-known issue of overfitting. We follow \cite{Hurvich_etal_1998} and \cite{SCHMID2008298} and employ the corrected $AIC$ criterion to select the optimal number of iterations, $m^{*}$,
\begin{equation}\label{eq:aic}
	AIC(m) = \log \left(\text{vâr}\left[r_{p,t+1}^{m}\right]\right) + \frac{1 + df_{m} / t}{1 - (df_{m} + 2) / t},
\end{equation}
where $df$ is the effective number of degrees of freedom after $m$ iterations of the boosting algorithm.\footnote{We also experimented with the five-fold cross-validation method in order to tune the maximum number of boosting iterations for the $L2$-boosting method. The results in terms of portfolio performance are slightly worse than those obtained with the corrected AIC criterion.} 
As $m$ increases, $\text{vâr}\left[r_{p,t+1}^{m}\right]$ decreases, and $df_m$ increases. Therefore, the optimal number of iterations, $m^{*},$ is the one that corresponds to the smallest AIC. Finally, we set the step size $\nu$ to a fixed value of 0.10.

Finally, we perform full hierarchical Bayes estimation of the Bayesian regression model in \eqref{Bayesregression} and \eqref{horseshoe} and learn about the hyperparameter $\tau$ in each estimation window. Estimation is performed using the efficient algorithm proposed by \cite*{bhattacharya2016fast}, and the global-local parameters $\tau$ and $\lambda_i$ are updated using the slice sampler of \cite*{polson2014bayesian}.

\subsection{Interpreting predictor importance}\label{section:importance}

We are also interested in understanding how and why a selected predictor is important from a minimum-variance portfolio construction perspective. Drawing on the approach developed in \cite{demiguel2017portfolio}, we rewrite the minimum-variance optimization problem in eq.~\eqref{eq:RSPP1C} as a quadratic optimization problem,
\begin{equation}\label{eq:RPP2}
	\min_{\theta}\,\,\,(1/2)\theta^\intercal\hat \Sigma_c\theta + \theta^\intercal \hat \sigma_{bc},
\end{equation}
where $\hat \Sigma_c$ is the sample covariance matrix of the predictor return vector $r_c$, and $\hat \sigma_{bc}$ is the sample vector of covariances between the benchmark portfolio return $r_b$ and the predictor-return vector $r_c$. Eq.~\eqref{eq:RPP2} decomposes the minimum-variance utility function into  the variance of the predictor return vector, $(1/2)\theta^\intercal\hat \Sigma_c\theta$, and the covariance of the predictor returns with the benchmark returns, $\theta^\intercal \hat \sigma_{bc}$. This decomposition reveals that the utility achieved by a given method is fundamentally determined by two complementary aspects: i) the ability to select predictors that contribute to decrease the variance-covariance of the predictor return vector, and ii) the ability to select predictors that contribute to decrease the covariance of the predictor return vector with the benchmark return vector. 

Eq.~\eqref{eq:RPP2} allows us to evaluate the importance of a \textit{individual} predictor. For that purpose, we calculate their \textit{marginal contributions} using the first-order derivative of eq.~\eqref{eq:RPP2},
\begin{align}\label{eq:FOC}
	\text{Predictor importance} &= \frac{\partial}{\partial \theta} \left( \frac{1}{2} \theta^\intercal \hat{\Sigma}_c \theta + \theta^\intercal \hat{\sigma}_{bc} \right)\\
	& = \underbrace{\text{diag}(\hat \Sigma_c)\theta}_{\text{own-var.(pred.)}} + \underbrace{\hat \Sigma_c-\text{diag}(\hat \Sigma_c)\theta}_{\text{cov.(pred.)}}+\underbrace{\hat\sigma_{bc}.}_{\text{cov.(bench.)}}\nonumber
\end{align} 
The $i$th, $i=1,\ldots,K$, component of the right-hand side in  \eqref{eq:FOC} is the marginal contribution of the $i$th term to the parametric portfolio minimum-variance utility; that is, the marginal change in minimum-variance utility associated with a unit increase in the weight that the parametric portfolio assigns to the $i$th term. The marginal contribution of an individual predictor can be decomposed into three terms: the predictor's own variance, $\text{diag}(\hat \Sigma_c)\theta$, the predictor's covariance with other predictors, $(\hat \Sigma_c-\text{diag}(\hat\Sigma_c))\theta$, and the covariance between the predictor and benchmark portfolios, $\hat\sigma_{bc}$.

\section{Empirical application}\label{section:empirical_application}

We now perform an empirical analysis of the performance of minimum-variance parametric portfolios when employing alternative regularization and variable selection methods discussed in Section \ref{section:ML_methods}. The parametric portfolios are obtained by assuming an equally weighted (EW) benchmark portfolio.  Two reasons motivate our choice of this benchmark portfolio. First, the well-documented good performance of this strategy relative to more sophisticated ones, as extensively discussed in \cite{demiguel2009optimal}. Second, \cite{lassance2020shrinking} shows that the EW strategy emerges naturally in a mean-variance context since an optimal mean-variance portfolio can be viewed as a linear combination of the EW portfolio and an arbitrage portfolio.\footnote{We obtain qualitatively similar results when assuming a value-weighted benchmark portfolio.} 

We consider two alternative configurations of the data set discussed in Section \ref{section:data}. First, we implement parametric portfolios using only the original 95 predictors mentioned in \cite{gu2020empirical,gu2021autoencoder}. Second, we utilize the augmented set of 4,610 predictors, which includes second-order and cross-product transformations. This enables us to understand the extent to which non-linear transformations of firm characteristics improve portfolio performance. 

Given our interest in the minimum-variance strategy, the most relevant performance metric is the out-of-sample ex-post (i.e. realized) standard deviation of out-of-sample portfolio returns. However, it is also common in the literature to report performance in terms of risk-adjusted returns measured by the Sharpe ratio (SR) both before and after transaction costs. The out-of-sample evaluation works as follows. First, we choose a window over which to perform the estimation. The total number of monthly observations in the dataset is $T_{tot} = 594$ and we choose an estimation window of $T = 120$. Second, using the return data over the estimation window, we compute the minimum-variance parametric portfolios using the ML methods detailed in Section \ref{section:ML_methods}. Third, we repeat this rolling-window procedure for the next month by including the data for the next month and dropping the data for the earliest month. We continue doing this until the end of the dataset is reached. At the end of this process, we have generated $T_{tot}-T=474$ portfolio weight vectors, $w^j_t$, for $t=T,\ldots,T_{tot}-1$ and for each strategy $j$. Holding the portfolio $w^j_t$ for one month gives the out-of-sample return net of transaction costs at time $t+1$:
$$
r_{t+1}^{j}=\left(w_{t}^{j}\right)^{\top} r_{t+1}-c\times |w_{t}^{j}-\left(w_{t-1}^{j}\right)^{+}|,
$$
where $|w_{t}^{j}-\left(w_{t-1}^{j}\right)^{+}|$ denotes the monthly portfolio turnover,  $c$ is the level of transaction costs, and $\left(w_{t-1}^{j}\right)^{+}$ is the portfolio for the $j$th strategy before rebalancing at time $t$, that is
$$
\left(w_{t-1}^{j}\right)^{+}=w_{t-1}^{j} \circ\left(e_{t-1}+r_{t}\right),
$$
where $e_{t-1}$ is the $N_{t-1}$ dimensional vector of ones and $x \circ y$ is the element-wise product of vectors $x$ and $y$. Then, for each portfolio we study, we compute the annualized out-of-sample standard deviation and the SR of returns net of transaction costs:
$$
\begin{aligned}
	\hat{\sigma}^{j} &=\left(\frac{12}{T_{t o t}-T} \sum_{t=T}^{T_{\mathrm{tot}}-1}\left(\left(w_{t}^{j}\right)^{\top} r_{t+1}-\hat{\mu}^{j}\right)^{2}\right)^{1 / 2}, \quad  \\
	\widehat{\mathrm{SR}}^{j} &=\frac{\hat{\mu}^{j}-Rf_t}{\hat{\sigma}^{j}},
\end{aligned}
$$
where $\hat{\mu}^{j} =\frac{12}{T_{\text {tot }}-T} \sum_{t=T}^{T_{\text {tot }}-1}\left(w_{t}^{j}\right)^{\top} r_{t+1}$ and $Rf_t$ denote the risk-free rate at time $t$.\footnote{The risk free rate was obtained from Ken French's data library web site.}

We consider two levels of transaction costs: 0 basis points (b.p.) and 10 b.p. We also test for the statistical significance of the differences in the portfolio variances and the Sharpe Ratio (SR) of the two portfolios by using the two-sided $p$-value of the prewhitened HAC$_{PW}$ test described by \cite{LW2011} and \cite{ledoit2008robust} for the portfolio variance and the SR, respectively. Specifically, we test for differences in portfolio variances and SR of the various strategies in relation to those of the minimum-variance parametric portfolio obtained when all 4,610 predictors are used in conjunction with the lasso variable selection method. 

\paragraph{Competing strategies.} We also implement alternative portfolio strategies for comparison purposes. First, we implement the classic equally weighted (EW) and value-weighted (VW) strategies. These two strategies do not require estimating model parameters and are easily scalable to large cross-sections of assets. In a parametric portfolio context, investing in these two strategies corresponds to an extreme regularization of the parameters, in which all coefficients are set to zero. Second, we implement the minimum variance parametric portfolios using the characteristics from the 3-factor (market, size, and value) and 5-factor (market, size, value, profitability, and investment) models of \cite{fama1993common} and \cite{fama2015five}, respectively. Finally, we implement the minimum variance portfolios using the 95 characteristics along with the traditional OLS method to estimate the coefficients of the parametric policy.

\subsection{Results}


We report in Table \ref{table:out_sample_std_dev} the out-of-sample standard deviation of minimum-variance portfolio returns. The Table  allows us to draw several important conclusions. First and foremost, the choice of the variable selection approach matters: the different methods lead to minimum-variance parametric portfolios with varying performance. We observe that the lasso method results in minimum-variance portfolios with lower portfolio risk compared to all other methods for both configurations of the data set. Three methods emerge as the best performers in terms of the standard deviation of portfolio returns: lasso, elastic net, and $L2$-boosting. The best overall performance in terms of annualized standard deviation of portfolio returns is achieved when using the 4,610 predictors and the lasso method (8.3\%). This figure, however, is statistically indistinguishable from that obtained with the elastic net (8.4\%) and with the $L2$-boosting (8.8\%) methods. 

Second, all variable selection methods lead to minimum-variance parametric portfolios that outperform those obtained with ad-hoc selected characteristics, as well as portfolios obtained using the OLS method. Specifically, using the characteristics from the popular 3-factor and 5-factor models leads to a portfolio standard deviation of 15\%, whereas using the OLS method leads to a portfolio standard deviation of 18\%, and both figures are substantially higher relative to those obtained with ML-based portfolios. Furthermore, the classic EW and VW strategies also exhibit higher standard deviations than the ML-based portfolios.

Third,  we observe a notable decrease in the standard deviation of portfolio returns performance when utilizing the expanded set of 4,610 predictors compared to the original 95 predictors. For the lasso method, the standard deviation decreases by approximately 9\%, dropping from 9.1\% to 8.3\% -- and the difference is statistically significant. Similarly, the adalasso method shows a decrease of about 5\%, whereas the elastic net method also experiences a reduction of approximately 9\%. The $L2$-boosting method exhibits a modest decrease in standard deviation of about 2\% with the augmented data set, while the Bayesian horseshoe method yields similar results in both configurations of the data set.

It is interesting to note that the opposite result is obtained when using the ridge method: the portfolio standard deviation \textit{increases} when using the augmented data set. The rationale for this result lies in the inherent features of the ridge approach: the dense estimation obtained with the ridge methods implies that all predictors are   
Incorporated into the model to some extent. With such a large number of predictors, the dense estimation demands very strong regularization to properly estimate the model. As a result, all coefficients are shrunk to very small values, and the ridge method with the augmented data set begins to resemble that of the classic EW strategy, which is the benchmark portfolio policy used in the parametric portfolio; see eq.~\eqref{eq:PP}. Not surprisingly, the standard deviation of the ridge-based portfolios with the augmented data set is close to that obtained with the EW and VW strategies. This outcome highlights a limitation of the ridge method in scenarios where the predictor set is vastly expanded relative to the number of data points.

Table \ref{table:out_sample_performance} reports the annualized Sharpe ratios both before and after transaction costs of 10 basis points, along with the average monthly turnover of all strategies. We observe that portfolios based on the lasso, elastic net, and SCAD methods perform poorly in terms of Sharpe ratio, both before and after transaction costs. For example, the after-fee Sharpe ratio obtained with the lasso method varies between -0.12 and -0.32 depending on the data set configuration. When transaction costs are taken into account, only the $L2$-boosting and ridge methods yield parametric portfolios with positive Sharpe ratios (1.076 and 0.814 when using the 95 characteristics).\footnote{Similar to the findings in Table \ref{table:out_sample_std_dev}, the performance of ridge-based portfolios in terms of risk-adjusted returns is comparable to that obtained with the equally weighted portfolio strategy.} The $L2$-boosting portfolio is the best performer in terms of before- and after-fee Sharpe ratios among all competing strategies. We also observe that the Sharpe ratios obtained with the augmented data set are often \textit{worse} than those obtained with the restricted data set, indicating that increasing the predictor space can negatively affect performance in terms of risk-adjusted returns due to lower average portfolio returns.

The results in Table \ref{table:out_sample_performance} reveal a much larger dispersion in risk-adjusted performance across variable selection methods relative to the dispersion observed in the standard deviation of portfolio returns, as shown in Table \ref{table:out_sample_std_dev}. This suggests that differences in the Sharpe ratio are driven mostly by the \textit{average} portfolio returns rather than the \textit{standard deviation} of portfolio returns. Moreover, Table \ref{table:out_sample_performance} indicates an apparent superiority of the $L2$-boosting method. To help understand the differences in risk-adjusted performance across methods, eq. \eqref{eq:port.ret} shows that the differences in parametric portfolio returns are fully determined by differences in $\theta^\intercal r_{c}$, since the term $r_b$ is the same for all parametric portfolio strategies. In other words, the predictors selected by each method and how each method invests in the selected predictors are the main drivers of performance in terms of returns.

\begin{landscape}

\begin{figure}[h]
	
	\caption{\textbf{Predictor returns}}\label{figure:predictor_returns_all_methods}
	
	\scriptsize{The figure plots the average values of the aggregate monthly returns for the predictors selected with each method. Average values of $\theta^\intercal r_{c}$ calculated across all rolling estimation windows. }\\
	
	\centering
	\includegraphics[width=7in]{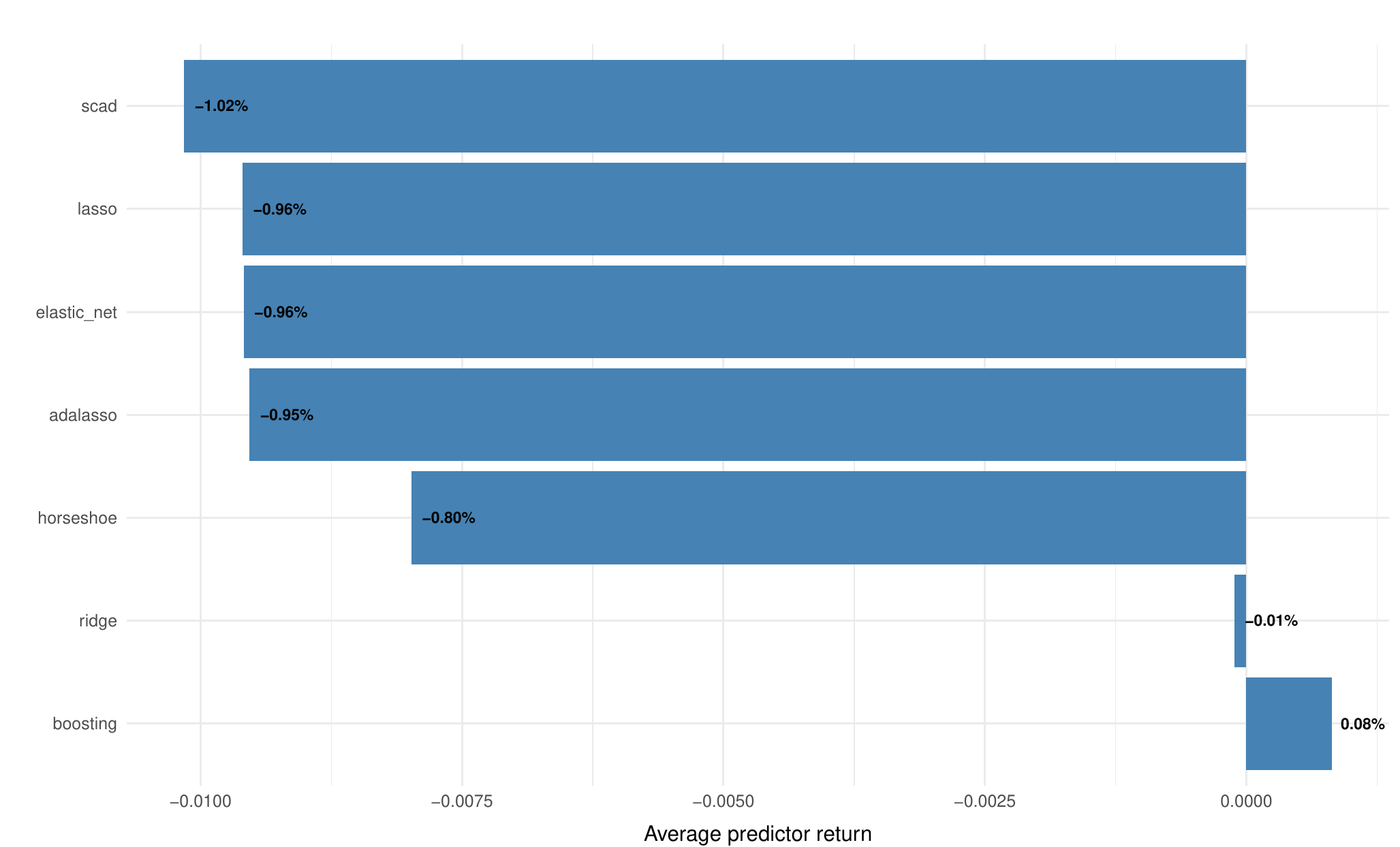}
\end{figure}

 \end{landscape}

Figure \ref{figure:predictor_returns_all_methods} plots the average values of the aggregate monthly returns for the predictors selected with each method. Specifically, the figure displays the average values of $\theta^\intercal r_{c}$ calculated across all rolling estimation windows. Consistent with the results reported in Table \ref{table:out_sample_performance}, minimum-variance portfolios obtained with the $L2$-boosting method achieve the highest value (1\% per month), whereas the SCAD method achieves the lowest (-1\% per month). As expected, the predictor return obtained with the ridge method is very close to zero, as its performance resembles that of the benchmark strategy.

\begin{figure}[h]
	
	\caption{\textbf{Individual predictor returns}}\label{figure:individual_predictor_returns}
	
	\scriptsize{The figure plots the returns on the top-10 and bottom-10 individual predictors for each method.}\\
	
	\centering
	\includegraphics[width=3.3in]{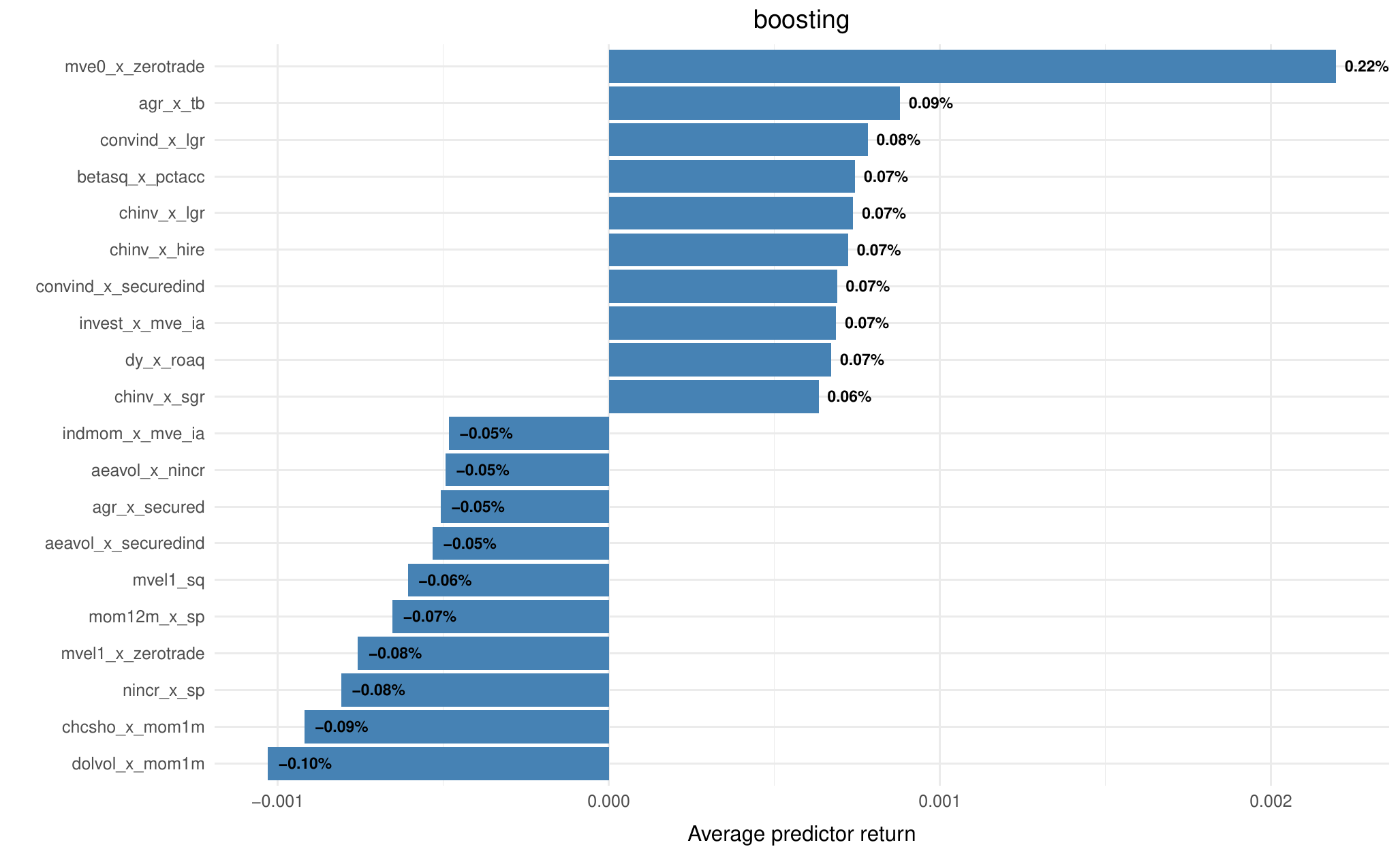}
	\includegraphics[width=3.3in]{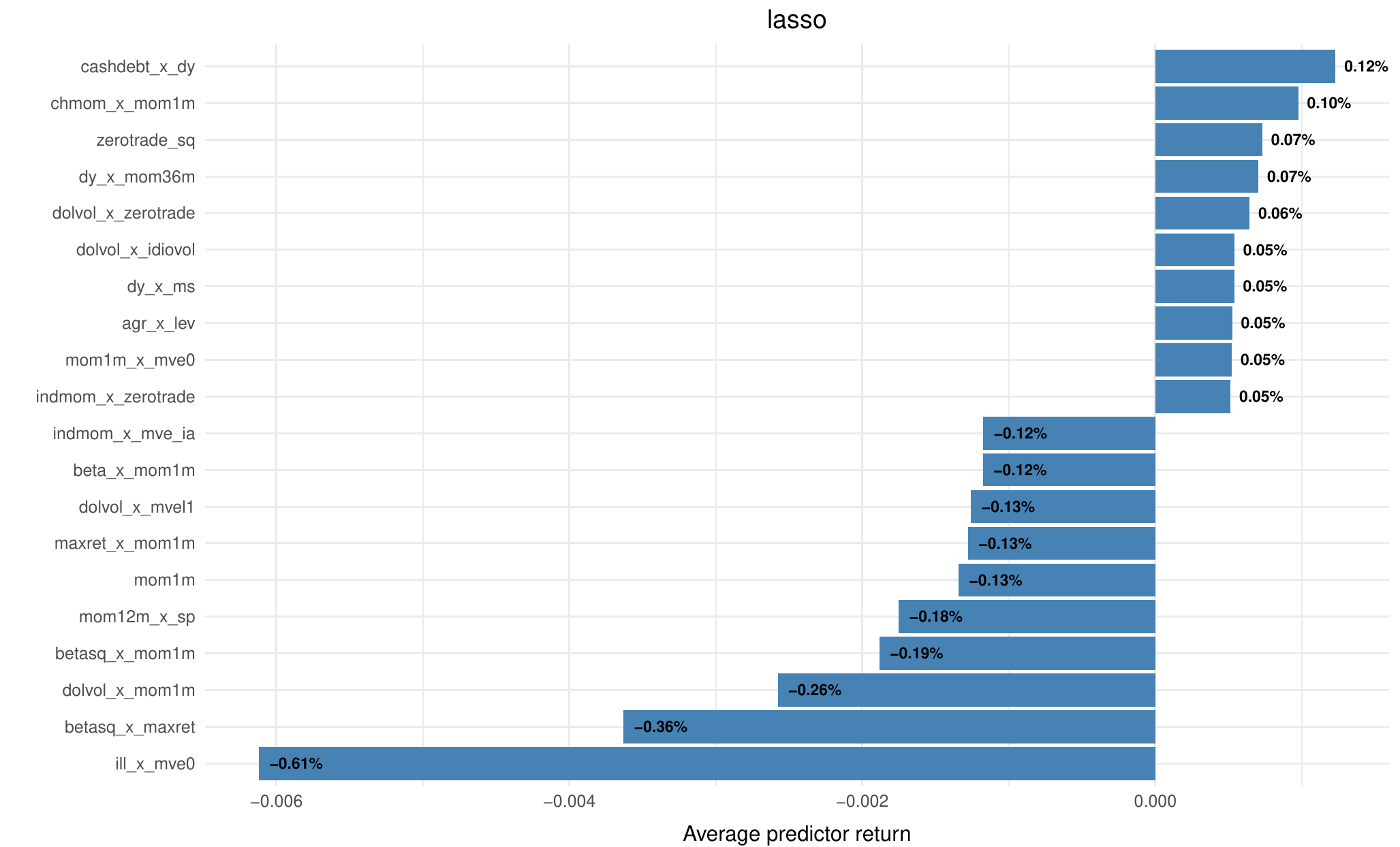}\\[3cm]
	\includegraphics[width=3.3in]{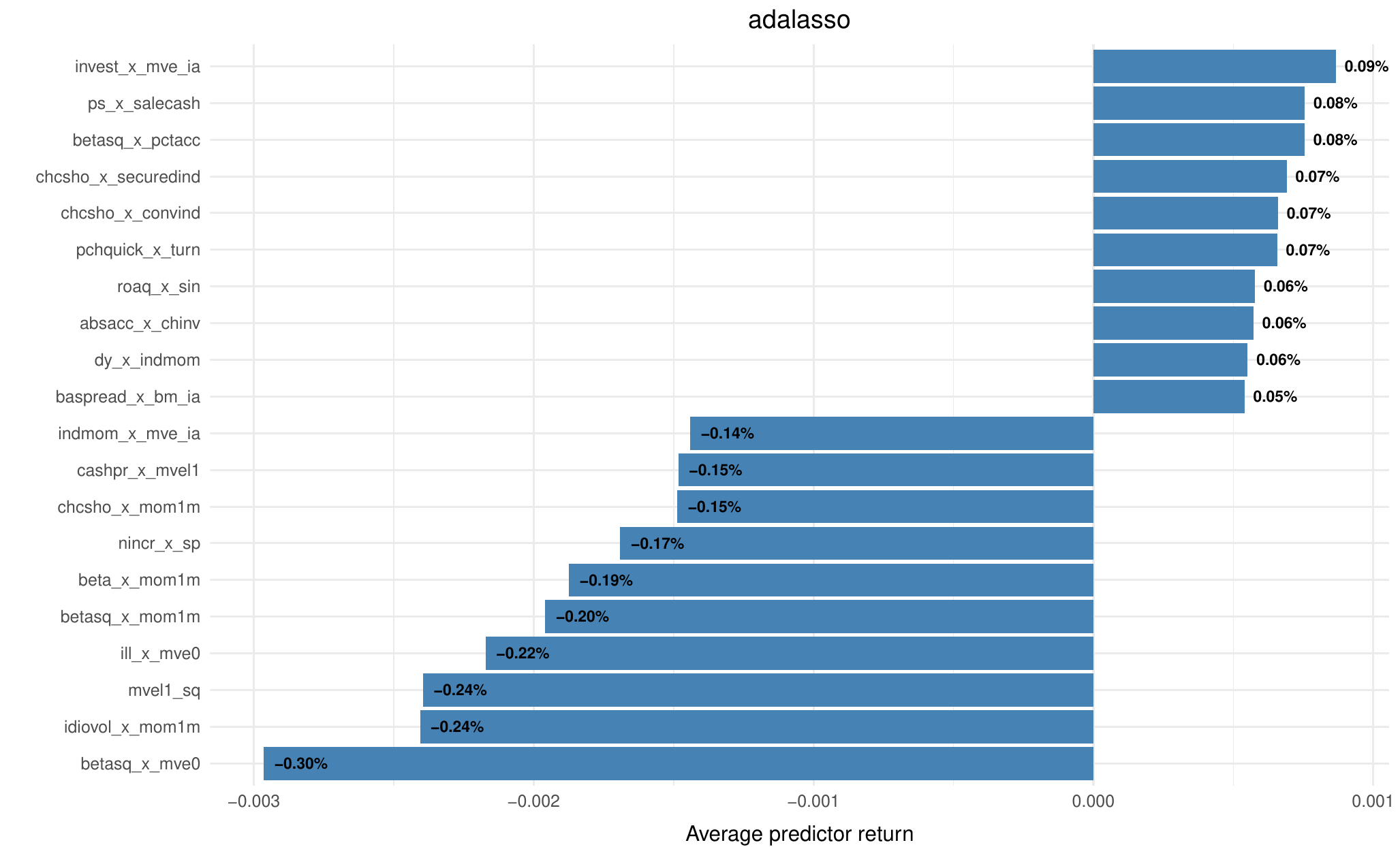}
	\includegraphics[width=3.3in]{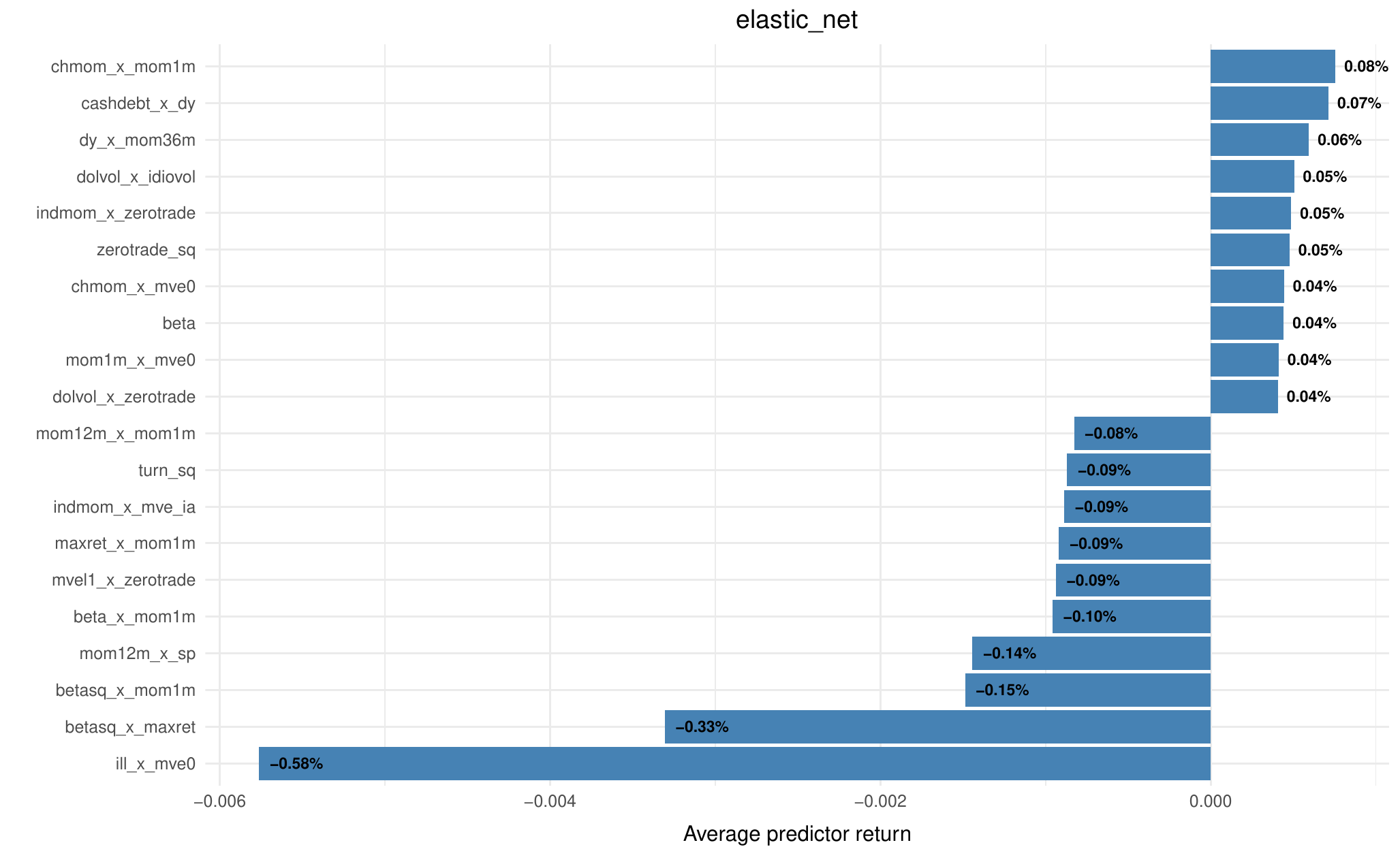}\\[3cm]
	\includegraphics[width=3.3in]{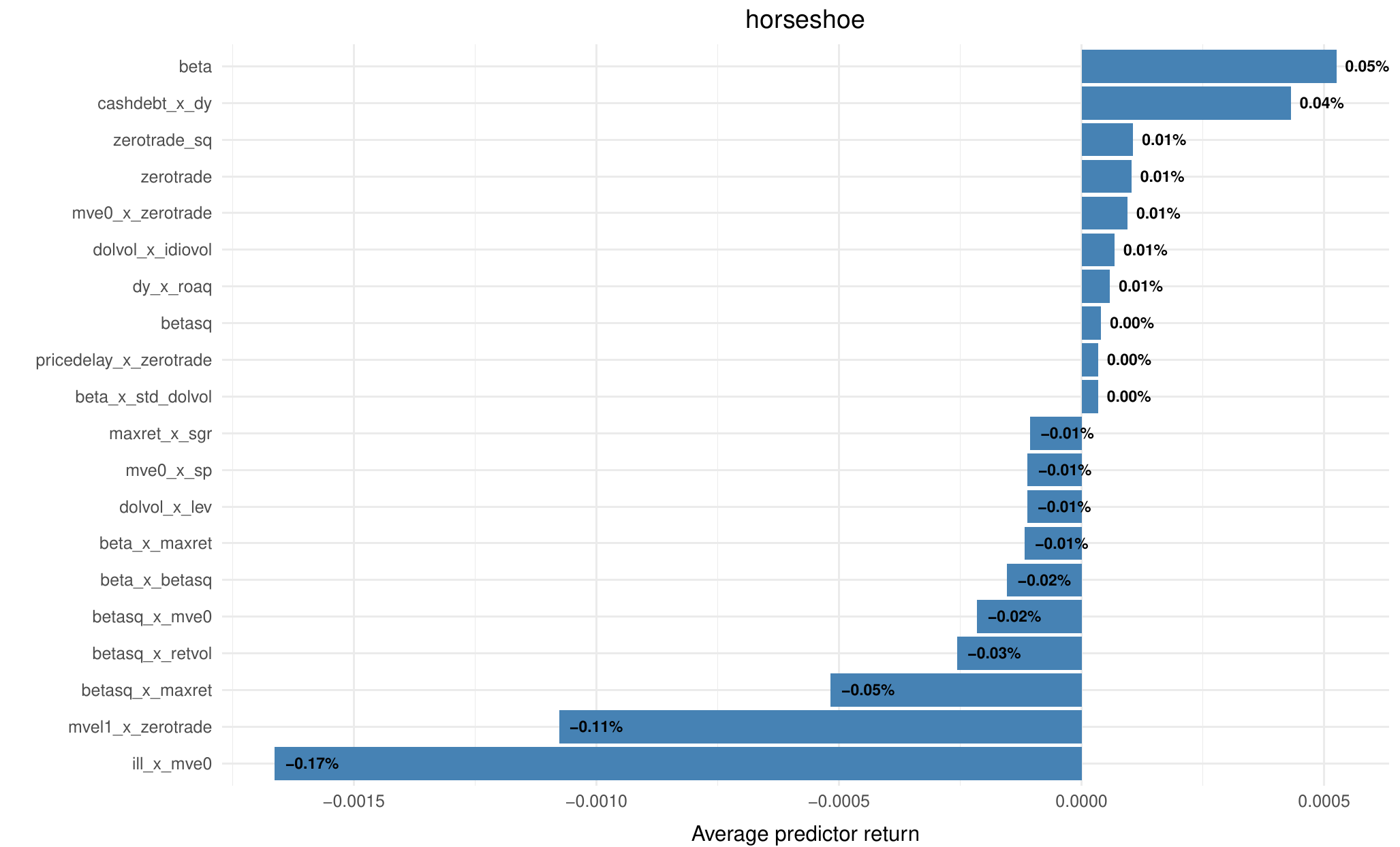}
	\includegraphics[width=3.3in]{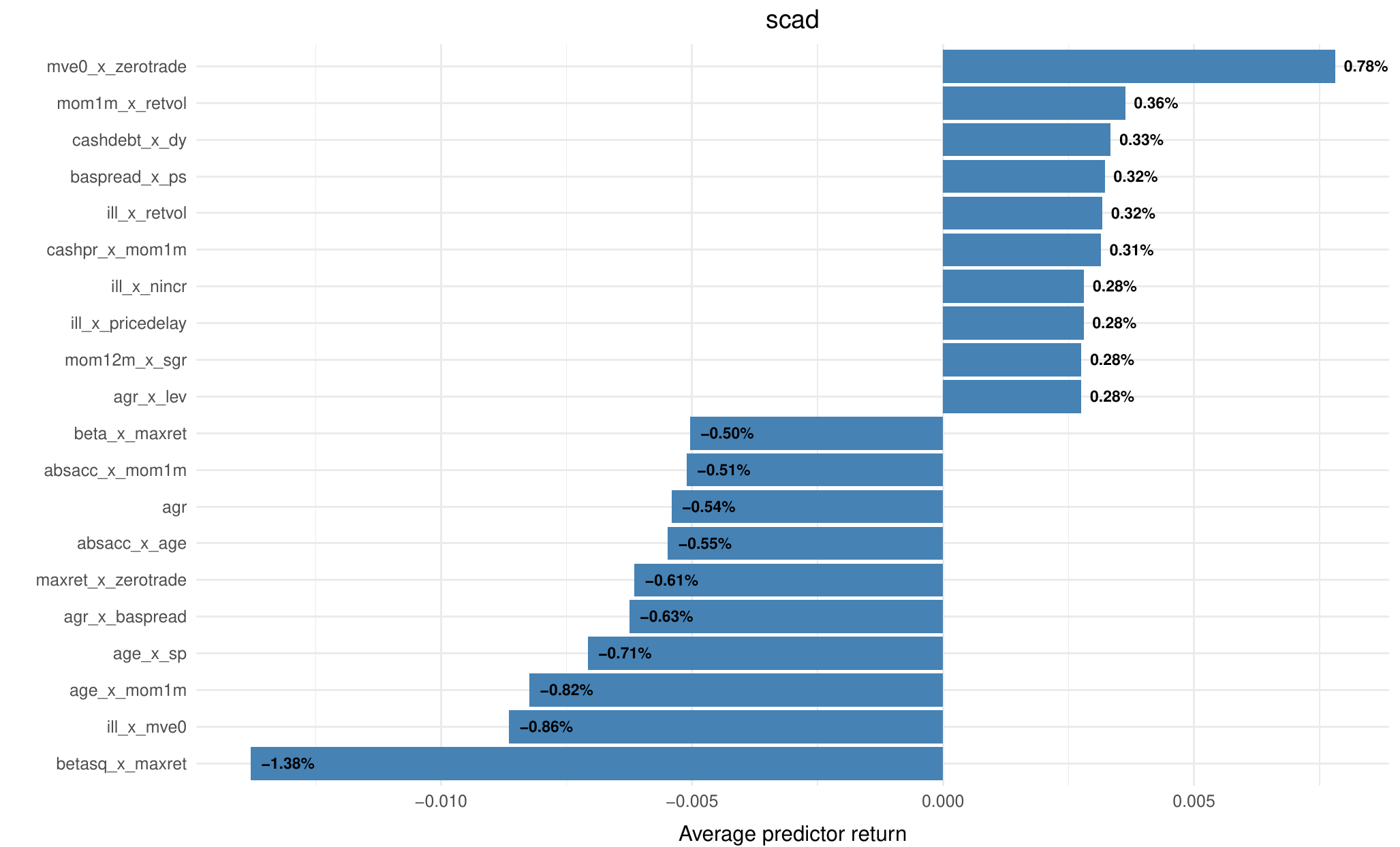}
	
\end{figure}

One important limitation of Figure \ref{figure:predictor_returns_all_methods} is the lack of information about which \textit{individual} predictor contributes the most to performance in terms of (aggregate) predictor returns. To complement the results in Figure \ref{figure:predictor_returns_all_methods}, we plot in Figure \ref{figure:individual_predictor_returns} the returns on the top-10 and bottom-10 individual predictors for each method. We observe that the interaction between size ($mve0$) and zero trading days ($zerotrade$) contributes the most to the positive risk-adjusted performance obtained with the $L2$-boosting method (0.22\% per month). For the other methods, the negative returns among the bottom-10 predictors are often larger in magnitude relative to those of the top-10 predictors, which helps explain not only the negative performance in terms of aggregate predictor return displayed in Figure \ref{figure:individual_predictor_returns}, but also the negative performance in terms of risk-adjusted returns reported in Table \ref{table:out_sample_performance}.

The fourth column of Table \ref{table:out_sample_performance} reports the average monthly portfolio turnover. We observe that portfolio turnover is significantly influenced by the choice of variable selection methods. Lasso, elastic net, and $L2$-boosting achieve portfolio turnovers around 1.5, whereas ridge-based portfolios have a much lower turnover of around 0.2.
All variable selection methods yield portfolios with much lower turnover relative to that obtained with OLS-based portfolios (5.14). Ad-hoc selected characteristic-based portfolios and traditional benchmark strategies, such as EW and VW portfolios, typically have lower turnover compared to ML methods. This is because these strategies are based on simpler allocation rules that do not change frequently with market conditions and, consequently, require less portfolio rebalancing.

We also report in Table \ref{table:portfolio_measure} the performance of the various portfolio strategies in terms of maximum drawdown and value-at-risk (VaR) based on the historical simulation method with 99\% confidence. The Table shows that adalasso and $L2$-boosting methods outperformed all other benchmarks in terms of both maximum drawdown and VaR. Out of all the methods, $L2$-boosting demonstrated the most superior performance, with a maximum drawdown of 38.6\% and a VaR of 7.4\%. The OLS method performed poorly compared to the other methods, with a maximum drawdown of 68.9\% and a VaR of 14.8\%. These results show that ML-based portfolios are also effective in minimizing tail risks.

In summary, the results reported in Tables \ref{table:out_sample_std_dev} to \ref{table:portfolio_measure} reveal that selecting predictors with variable selection methods leads to minimum-variance portfolios with superior performance relative to traditional OLS methods and ad-hoc characteristic-based portfolios, in terms of lower risk, smaller maximum drawdowns, and VaR. Moreover, the results obtained with the $L2$-boosting method showed that lower portfolio risk can be achieved without compromising risk-adjusted returns: the performance measured by the Sharpe ratio is substantially higher relative to all other competing strategies, even when transaction costs are taken into account.

\subsection{Predictor importance}

To further understand the aspects that contribute to the performance of minimum-variance portfolios, we study the profile of selected predictors as well as their distribution across predictor classes (main effects, second-order effects, and interaction effects.). The results reported in Table \ref{table:selection} reveal distinct patterns in predictor selection across different methods. Specifically, the average number of selected predictors varies from 16 (SCAD) to 75 ($L2$-boosting). The best performing method in delivering portfolios with lower risk (lasso) selects 33 predictors on average. This result suggests that the number of selected predictors vastly exceeds the number of variables used in popular factor models, suggesting that ad-hoc sparsity can be detrimental to portfolio performance.

The ``class-relative percentage'' panel in Table \ref{table:selection} reports the distribution of the selected predictors within each class (main effects, second-order effects, and interaction effects) for each method. Across all methods, interaction effects consistently account for the highest percentage of selected predictors. This finding is not surprising, given the composition of the predictor space: with a total of 4610 predictors, only 95 (2.1\%) are main effects, 88 (1.9\%) are second-order effects, and the remaining 4427 (96\%) are interaction terms. Thus, interaction effects naturally represent the largest pool of potential predictors and are expected to make up the majority of selected predictors simply due to their abundant quantity. Despite this expectation, the distribution across classes varies by method, reflecting each method's tendency to prioritize different classes of predictors. For example, SCAD places a much higher emphasis on main effects (16\%) compared to other methods, such as adalasso (0.8\%) and $L2$-boosting (1.4\%). In contrast, methods like adalasso and elastic net devote a higher percentage to interaction effects.

The ``difference from equal-importance'' panel in Table \ref{table:selection} shows how each method's selection deviates from a benchmark based on equal-importance weighting. If the classes were equally important, one would expect the selection of predictors to be proportional to the number of predictors available in each class. Therefore, deviations from this equal-importance benchmark reveal which classes are being over- or under-selected relative to what would be expected by chance. We observe that lasso, elastic net, horseshoe and SCAD \textit{underrepresent} interaction effects compared to an equal-importance distribution, as shown by the \textit{negative} deviations for this class, and \textit{overrepresent} main effects and second-order effects, as shown by the \textit{positive} deviations for these classes.



\begin{figure}[h]
	
	\caption{\textbf{Lasso predictor importance}}\label{figure:lasso}
	
	\scriptsize{The figure plots marginal contributions of the top-20 selected predictors for the lasso method. The figure breaks down the contributions into three components: the predictor's own variance, its covariance with other predictors, and its correlation with the benchmark portfolios (see Eq.~\eqref{eq:FOC}). Importance based on own-variance (left-hand plot) is ranked in ascending order: predictors whose own variance leads to smaller increases in the objective function (Eq.~\eqref{eq:RPP2}) are more important. Importance based on covariance with other predictors (center plot) and correlation with the benchmark portfolio (right-hand plot) is ranked in descending order: predictors with lower covariance with others and with the benchmark portfolio contribute more to decrease the objective function and are thus more important. Finally, predictors with a positive (negative) parametric-portfolio coefficient are in blue (red).}\\
	
	\centering
	\includegraphics[width=7in]{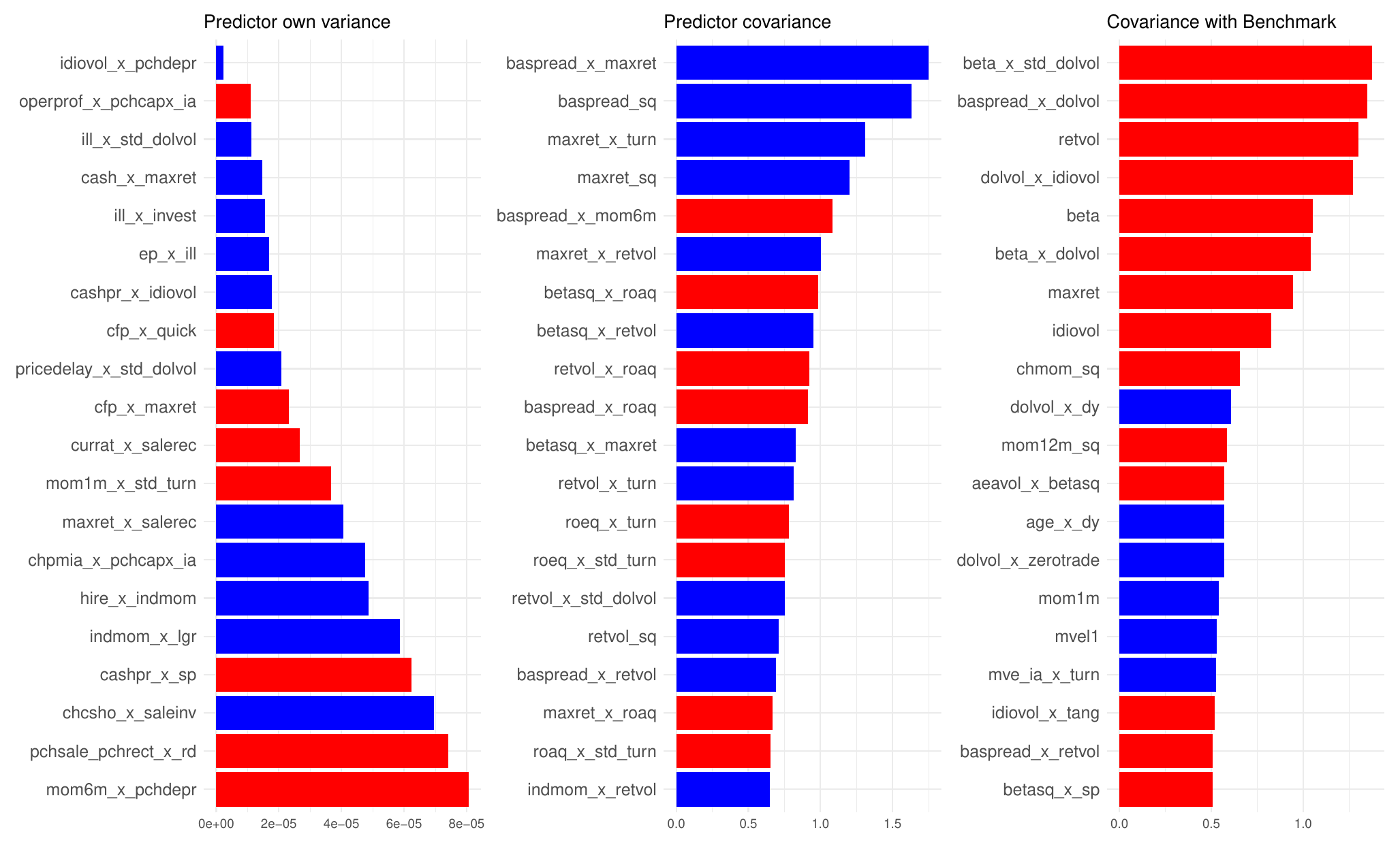}
\end{figure}

\begin{figure}[h]
	\caption{\textbf{Elastic net predictor importance}}\label{figure:enet}
	
	\scriptsize{The figure plots marginal contributions of the top-20 selected predictors for the elastic net method. The figure breaks down the contributions into three components: the predictor's own variance, its covariance with other predictors, and its correlation with the benchmark portfolios (see Eq.~\eqref{eq:FOC}). Importance based on own-variance (left-hand plot) is ranked in ascending order: predictors whose own variance leads to smaller increases in the objective function (Eq.~\eqref{eq:RPP2}) are more important. Importance based on covariance with other predictors (center plot) and correlation with the benchmark portfolio (right-hand plot) is ranked in descending order: predictors with lower covariance with others and with the benchmark portfolio contribute more to decrease the objective function and are thus more important. Finally, predictors with a positive (negative) parametric-portfolio coefficient are in blue (red).}\\
	
	\centering
	\includegraphics[width=7in]{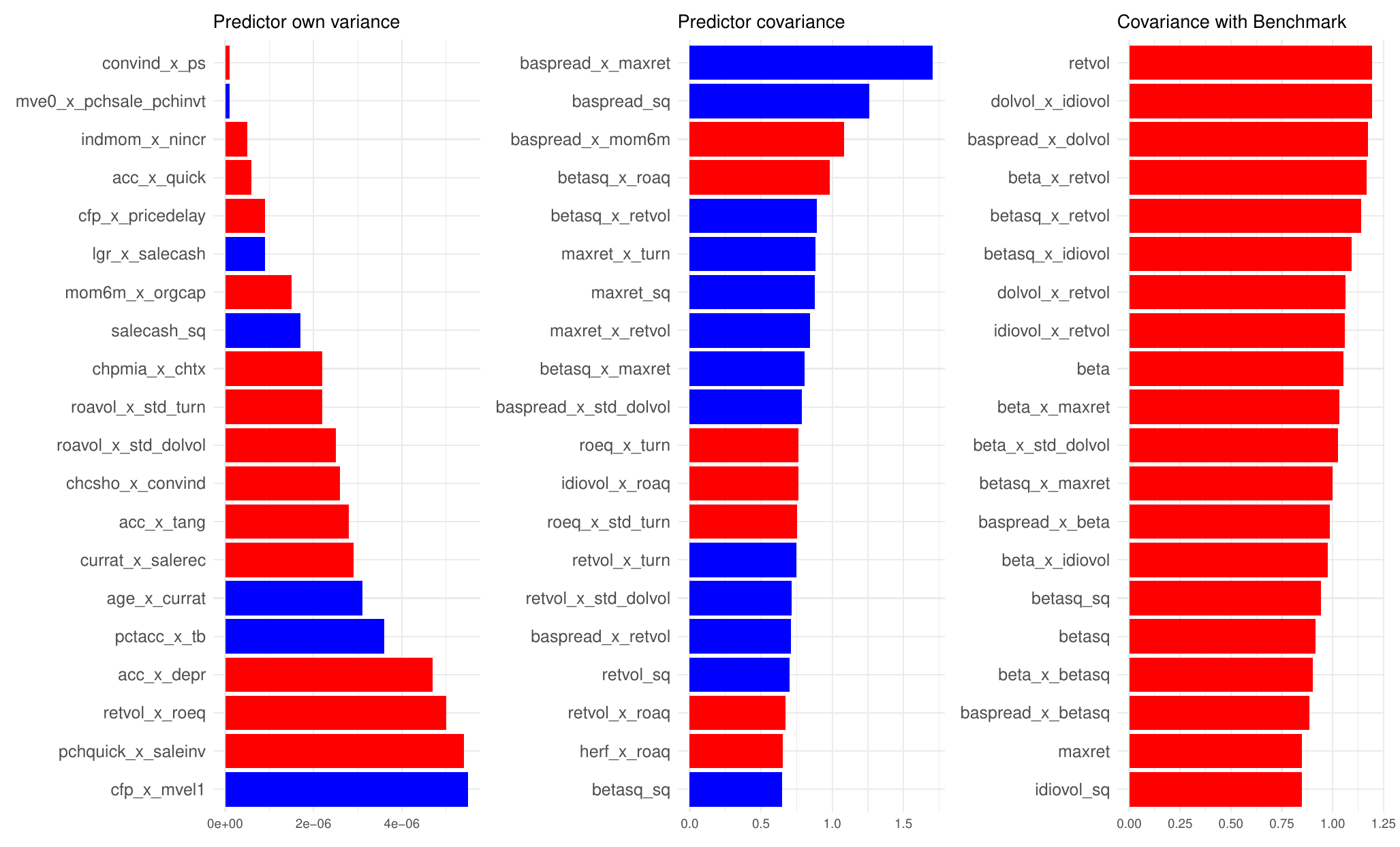}
\end{figure}

\begin{figure}[h]
	\caption{\textbf{$L2$-boosting predictor importance}}\label{figure:boosting}
	
	\scriptsize{The figure plots marginal contributions of the top-20 selected predictors for the $L2$-boosting method. The figure breaks down the contributions into three components: the predictor's own variance, its covariance with other predictors, and its correlation with the benchmark portfolios (see Eq.~\eqref{eq:FOC}). Importance based on own-variance (left-hand plot) is ranked in ascending order: predictors whose own variance leads to smaller increases in the objective function (Eq.~\eqref{eq:RPP2}) are more important. Importance based on covariance with other predictors (center plot) and correlation with the benchmark portfolio (right-hand plot) is ranked in descending order: predictors with lower covariance with others and with the benchmark portfolio contribute more to decrease the objective function and are thus more important. Finally, predictors with a positive (negative) parametric-portfolio coefficient are in blue (red).}\\
	
	\includegraphics[width=7in]{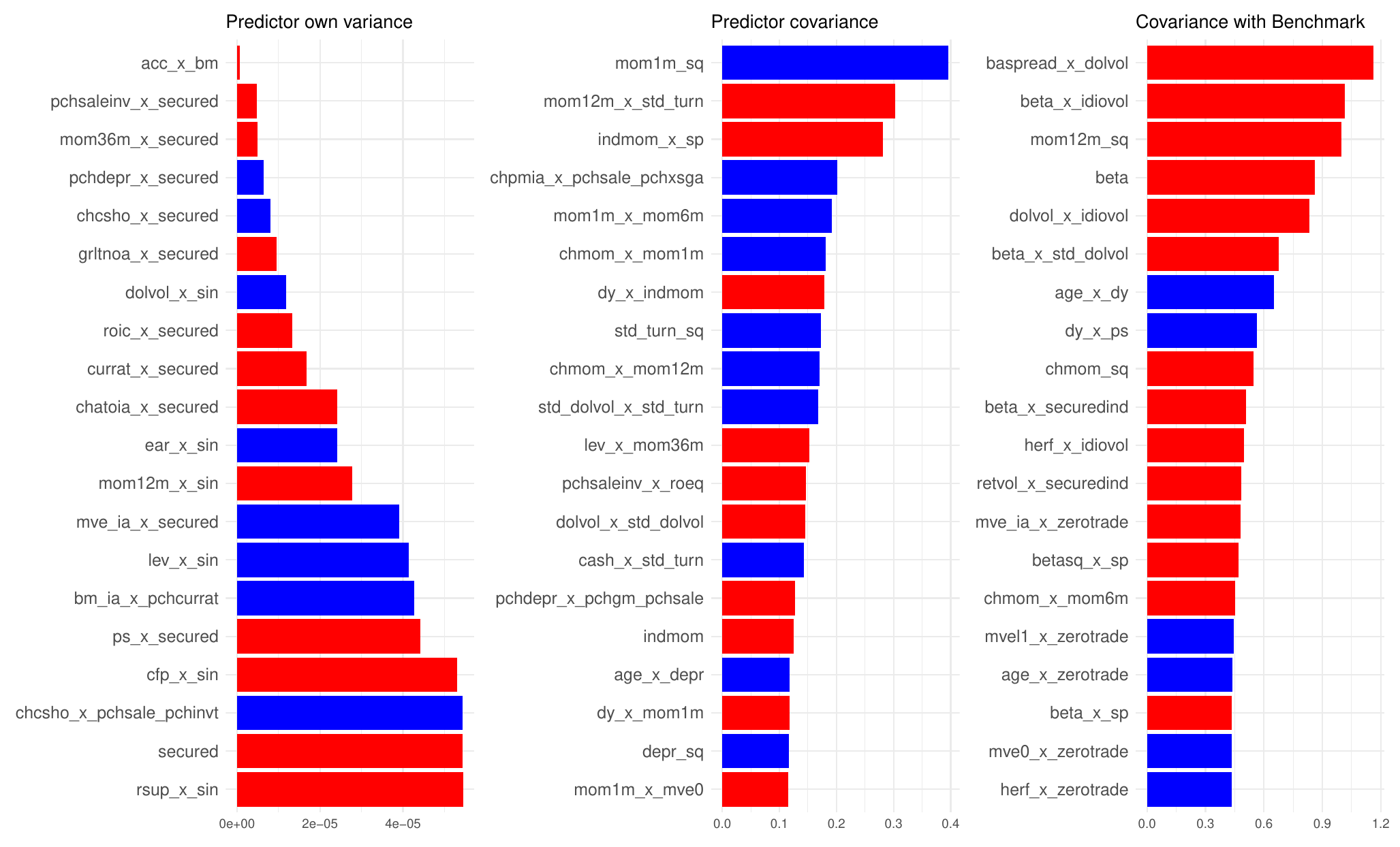}
\end{figure}

We also investigate the importance of individual predictors using the method described in Section~\ref{section:importance}. Figures~\ref{figure:lasso},~\ref{figure:enet}, and~\ref{figure:boosting} show the marginal contributions of the top-20 selected predictors for lasso, elastic net, and $L2$-boosting, respectively.\footnote{Marginal contributions are calculated across the out-of-sample period, and the reported values are averages across all estimation rounds.} Each Figure breaks down the contributions into three components: the predictor's own variance, its covariance with other predictors, and its covariance with the benchmark portfolios (see Eq.~\eqref{eq:FOC}). Predictor importance based on own-variance (left-hand plot) is ranked in ascending order: predictors whose own variance leads to smaller increases in the objective function (Eq.~\eqref{eq:RPP2}) are more important. Predictor importance based on the correlation with other predictors (center plot) and on the correlation with the benchmark portfolio (right-hand plot) is ranked in descending order: predictors with lower covariance with other predictors and with the benchmark portfolio contribute more to decrease the objective function and are thus more important. Finally, predictors with a positive (negative) parametric-portfolio coefficient are in blue (red).

Figures~\ref{figure:lasso}-\ref{figure:boosting} show that the importance of individual predictors varies across methods. Some predictors like market beta, return volatility (\textit{retvol}), idiosyncratic volatility (\textit{idiovol}), and maximum daily return (\textit{maxret}) are selected by lasso and elastic net mainly for their ability to improve investors' utility because of their covariance with the benchmark portfolio. These results align with those reported in \cite{scherer2011note}, who argue that risk-based characteristics are important drivers of the minimum-variance portfolio allocations. These four predictors have a negative parametric-portfolio coefficient, indicating that stocks with higher values on those predictors are assigned lower weights in the minimum-variance portfolio. The lasso method also prioritizes stocks with higher values of size (\textit{mvel1}) and short-term reversal (\textit{mom1m}). As for the $L2$-boosting method, square values of one-year momentum (\textit{mom12m\_sq}) are important because they covary with the benchmark portfolio, whereas square values of short-term reversals ($mom1m\_sq$) matter because they have low covariance with the other predictors. 

\begin{landscape}

	\begin{figure}
		\centering
		\caption{\textbf{Portfolio weights, liquidity and idiosyncratic volatility}}
			\scriptsize{The figure plots the average minimum-variance portfolio weights (vertical axis) across values of the standardized liquidity (\textit{dolvol}) and across quintiles of idiosyncratic volatility (\textit{idiovol})}.\\
		\label{fig:interaction_dolvol_idiovol}
		\begin{subfigure}[t]{0.6\textwidth} 
			\centering
			\includegraphics[width=1\linewidth]{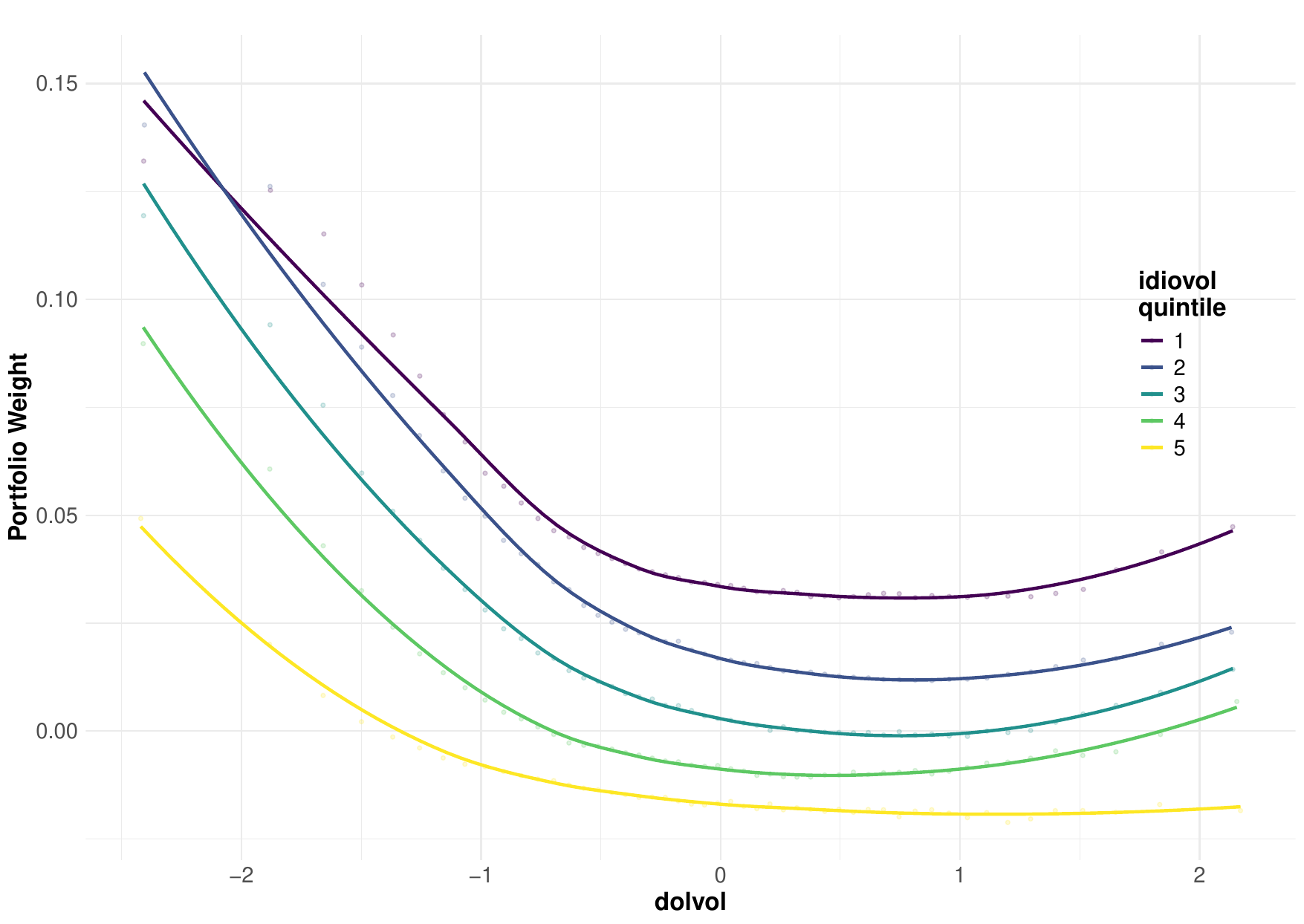}
			\caption{Lasso}
			\label{fig:sub1}
		\end{subfigure}
		\hspace{1cm}
		\begin{subfigure}[t]{0.6\textwidth}
			\centering
			\includegraphics[width=1\linewidth]{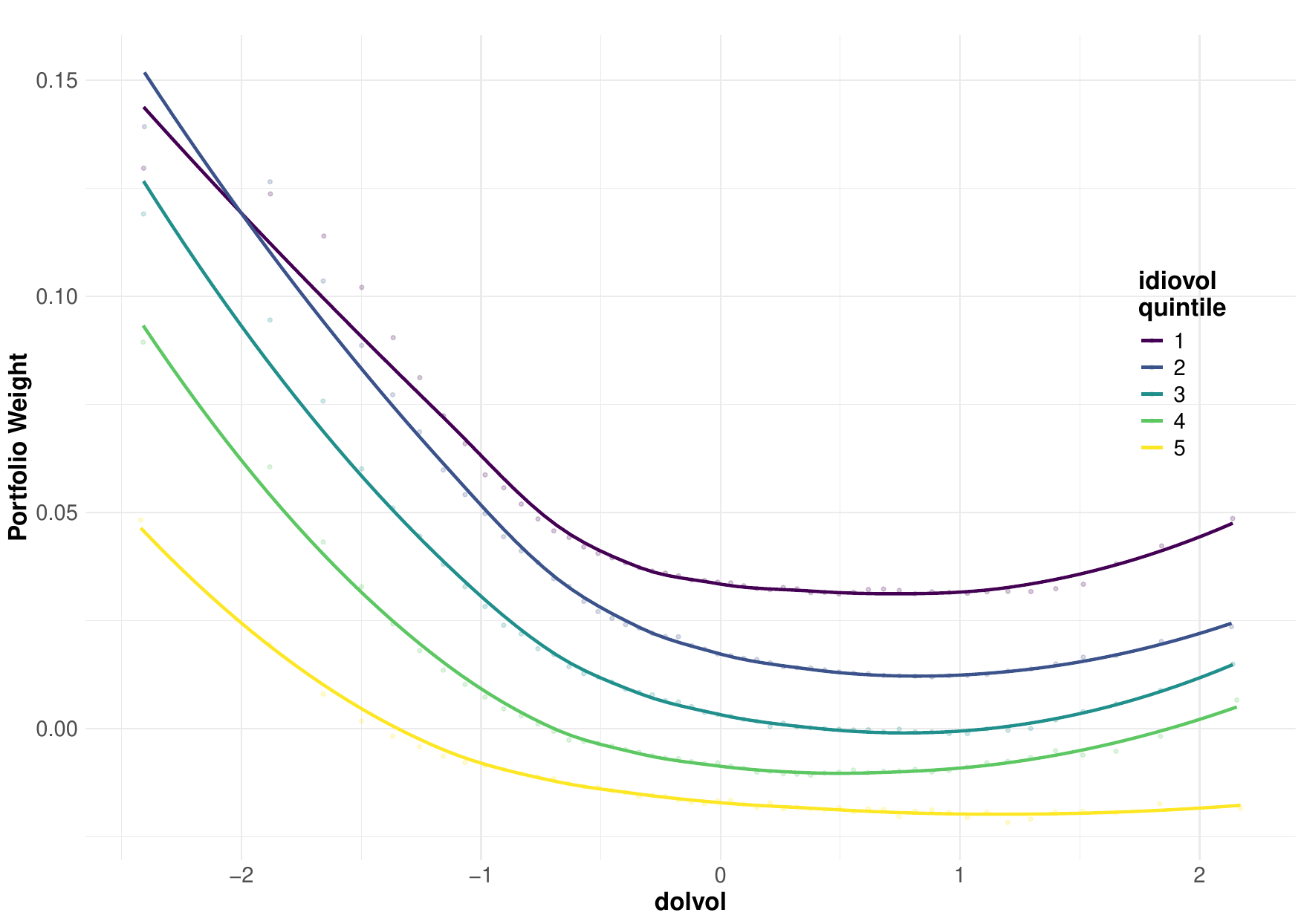}
			\caption{Elastic net}
			\label{fig:sub2}
		\end{subfigure}
		\hspace{1cm}
		\begin{subfigure}[t]{0.6\textwidth}
			\centering
			\includegraphics[width=1\linewidth]{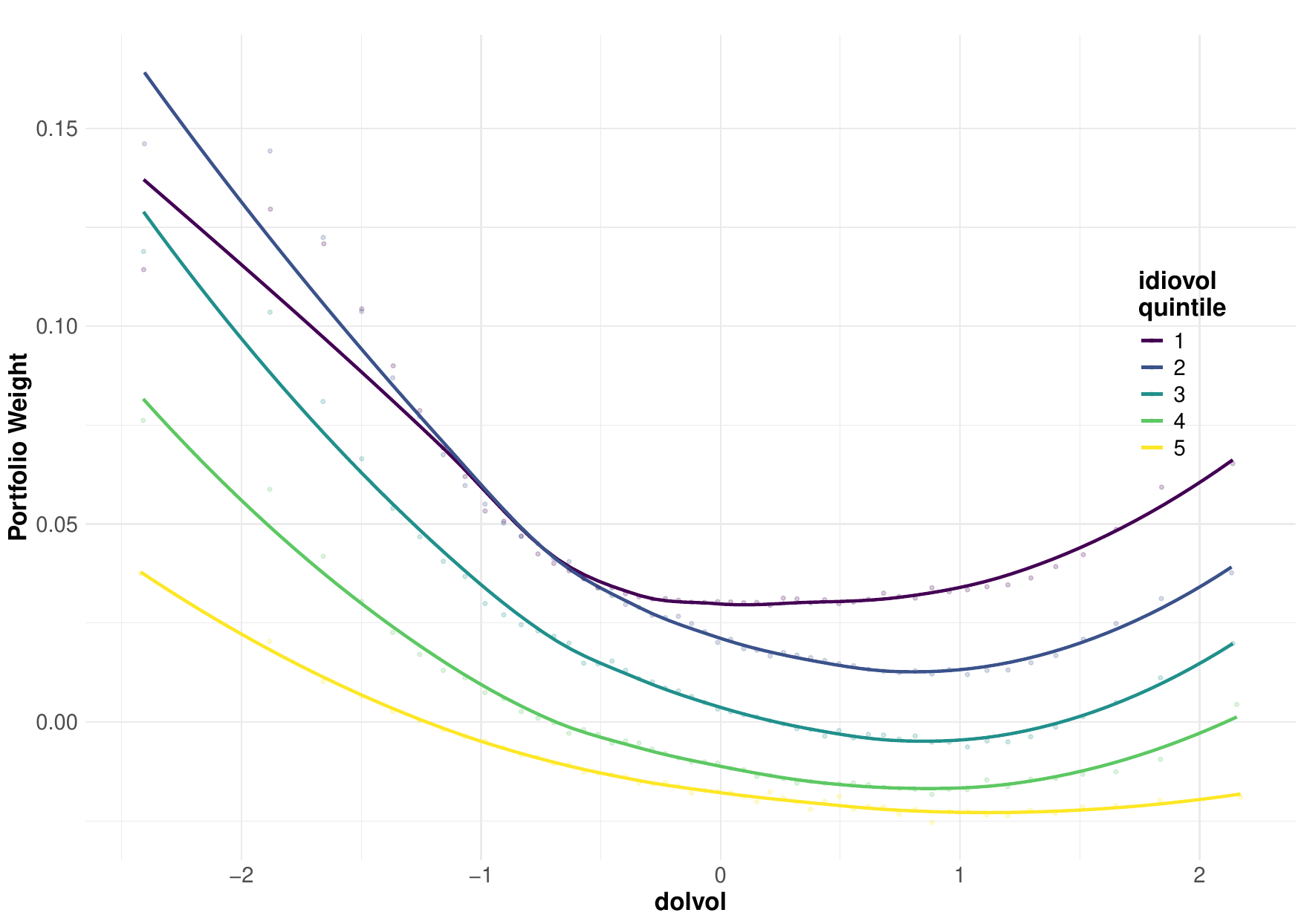}
			\caption{$L2$-boosting}
			\label{fig:sub3}
		\end{subfigure}
		
	\end{figure}
	
\end{landscape}

\begin{landscape}

	\begin{figure}
		\centering
		\caption{\textbf{Portfolio weights, market beta and standard deviation of liquidity}}
		\label{fig:interacion_beta_std_dolvol}
		\scriptsize{The figure plots the average minimum-variance portfolio weights (vertical axis) across values of the standardized market beta (\textit{beta}) and across quintiles of standard deviation of liquidity (\textit{std\_dolvol})}.\\
		\begin{subfigure}[t]{0.6\textwidth} 
			\centering
			\includegraphics[width=1\linewidth]{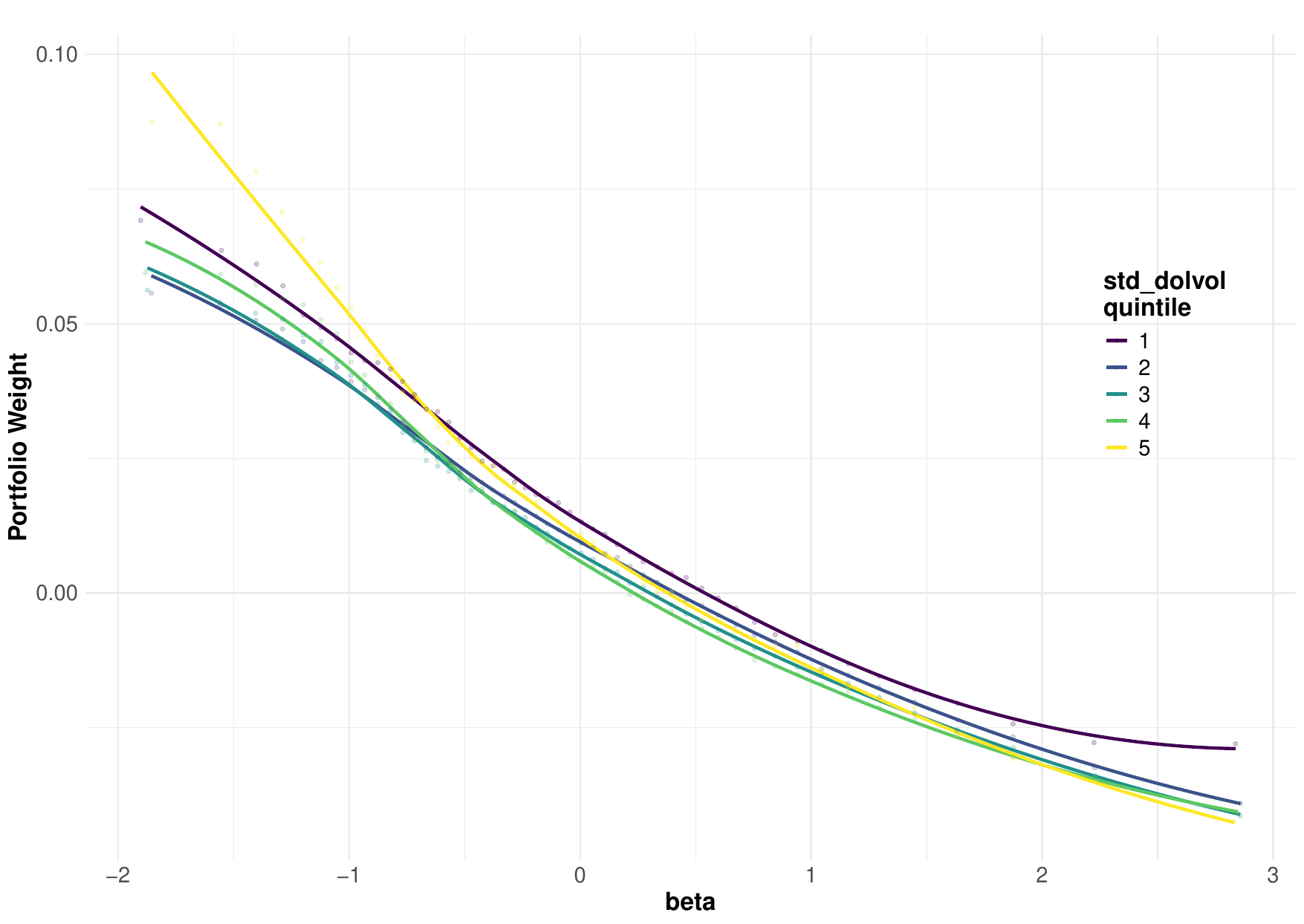}
			\caption{Lasso}
			\label{fig:sub1}
		\end{subfigure}
		\hspace{1cm}
		\begin{subfigure}[t]{0.6\textwidth}
			\centering
			\includegraphics[width=1\linewidth]{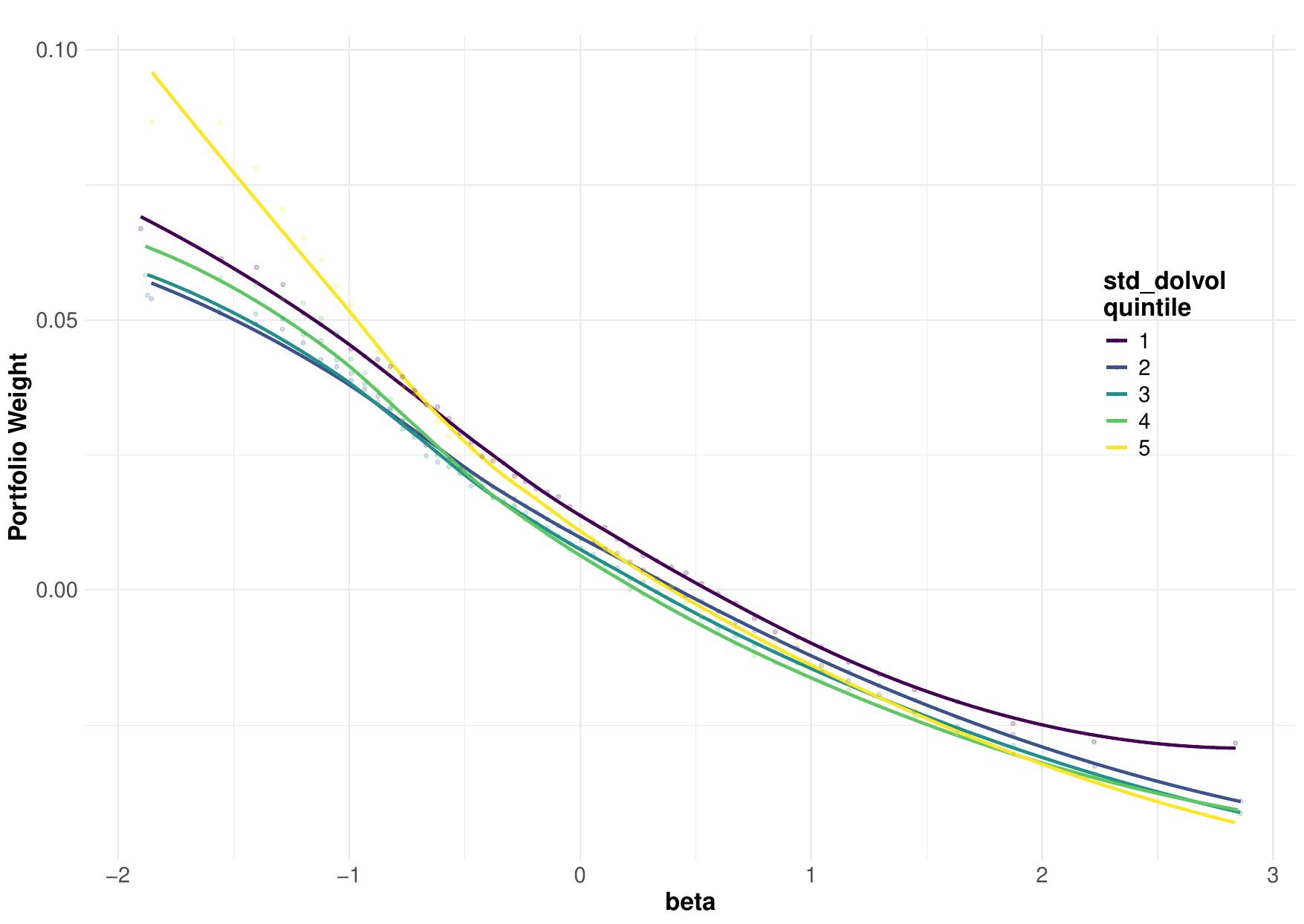}
			\caption{Elastic net}
			\label{fig:sub2}
		\end{subfigure}
		\hspace{1cm}
		\begin{subfigure}[t]{0.6\textwidth}
			\centering
			\includegraphics[width=1\linewidth]{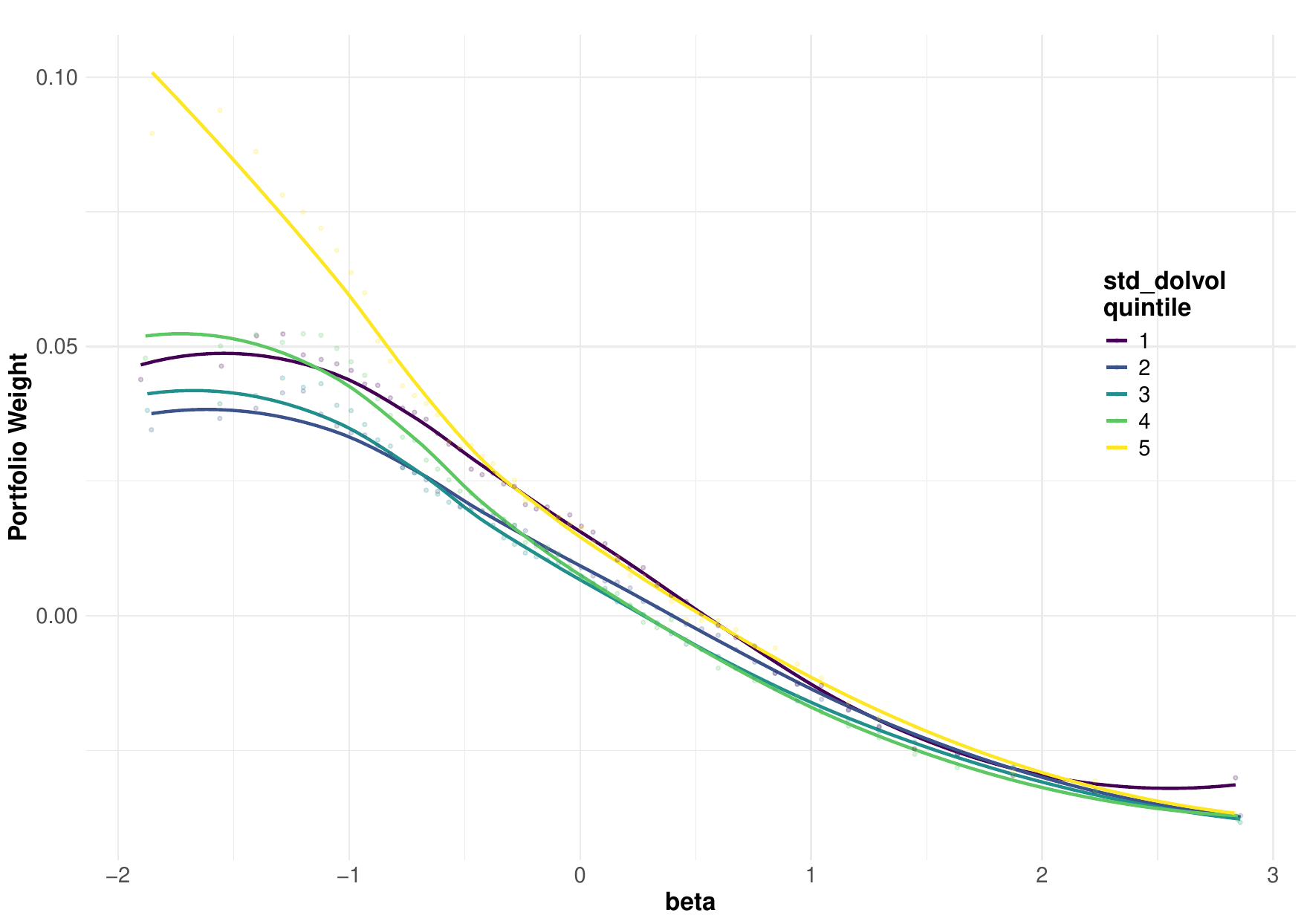}
			\caption{$L2$-boosting}
			\label{fig:sub3}
		\end{subfigure}
		
	\end{figure}
	
\end{landscape}

\begin{landscape}

	\begin{figure}
		\centering
		\caption{\textbf{Portfolio weights, bid-ask spread and liquidity}}
		\label{fig:interacion_baspread_dolvol}
		\scriptsize{The figure plots the average minimum-variance portfolio weights (vertical axis) across values of the standardized bid-ask spread (\textit{baspread}) and across quintiles of liquidity (\textit{dolvol})}.\\
		\begin{subfigure}[t]{0.6\textwidth} 
			\centering
			\includegraphics[width=1\linewidth]{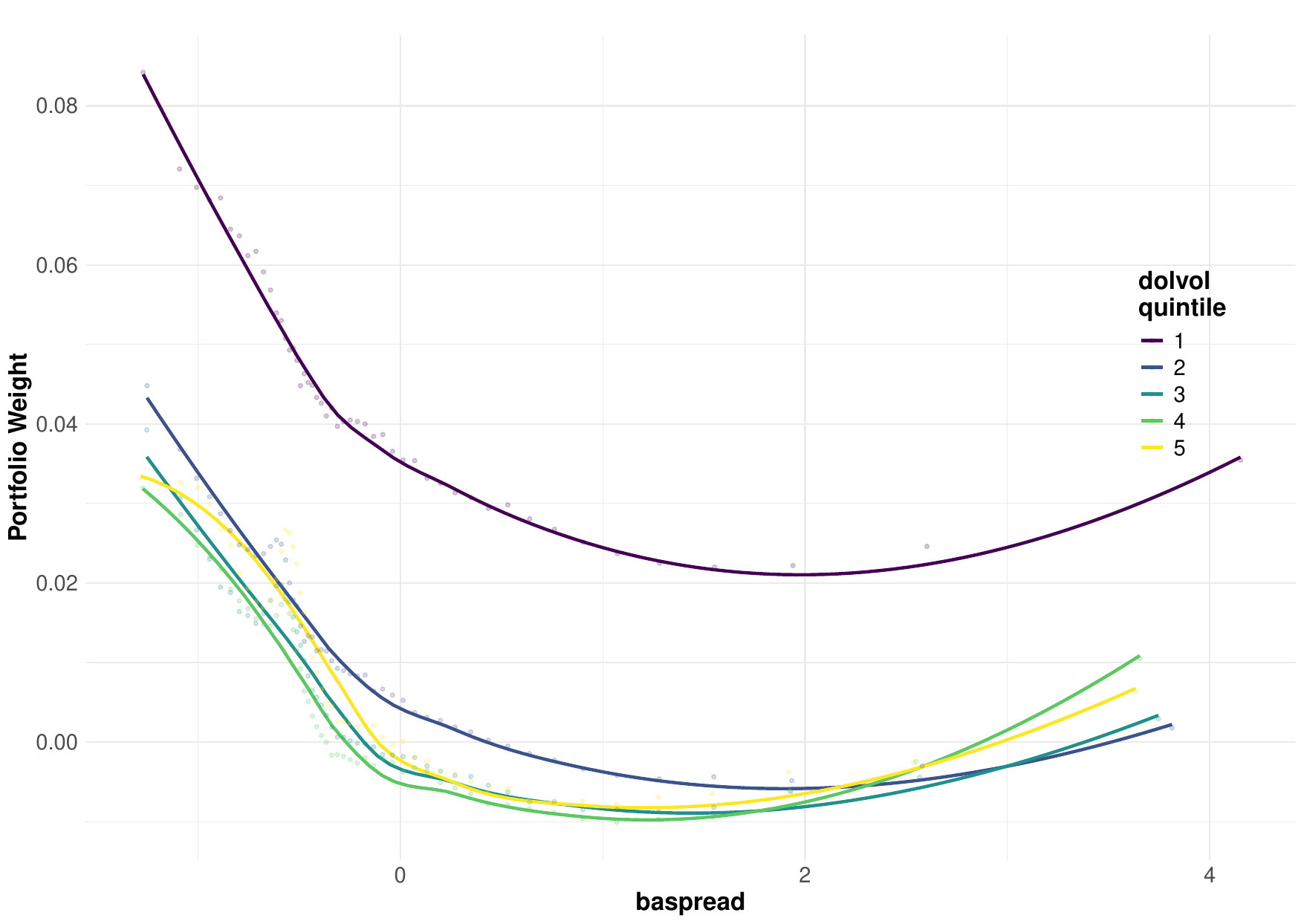}
			\caption{Lasso}
			\label{fig:sub1}
		\end{subfigure}
		\hspace{1cm}
		\begin{subfigure}[t]{0.6\textwidth}
			\centering
			\includegraphics[width=1\linewidth]{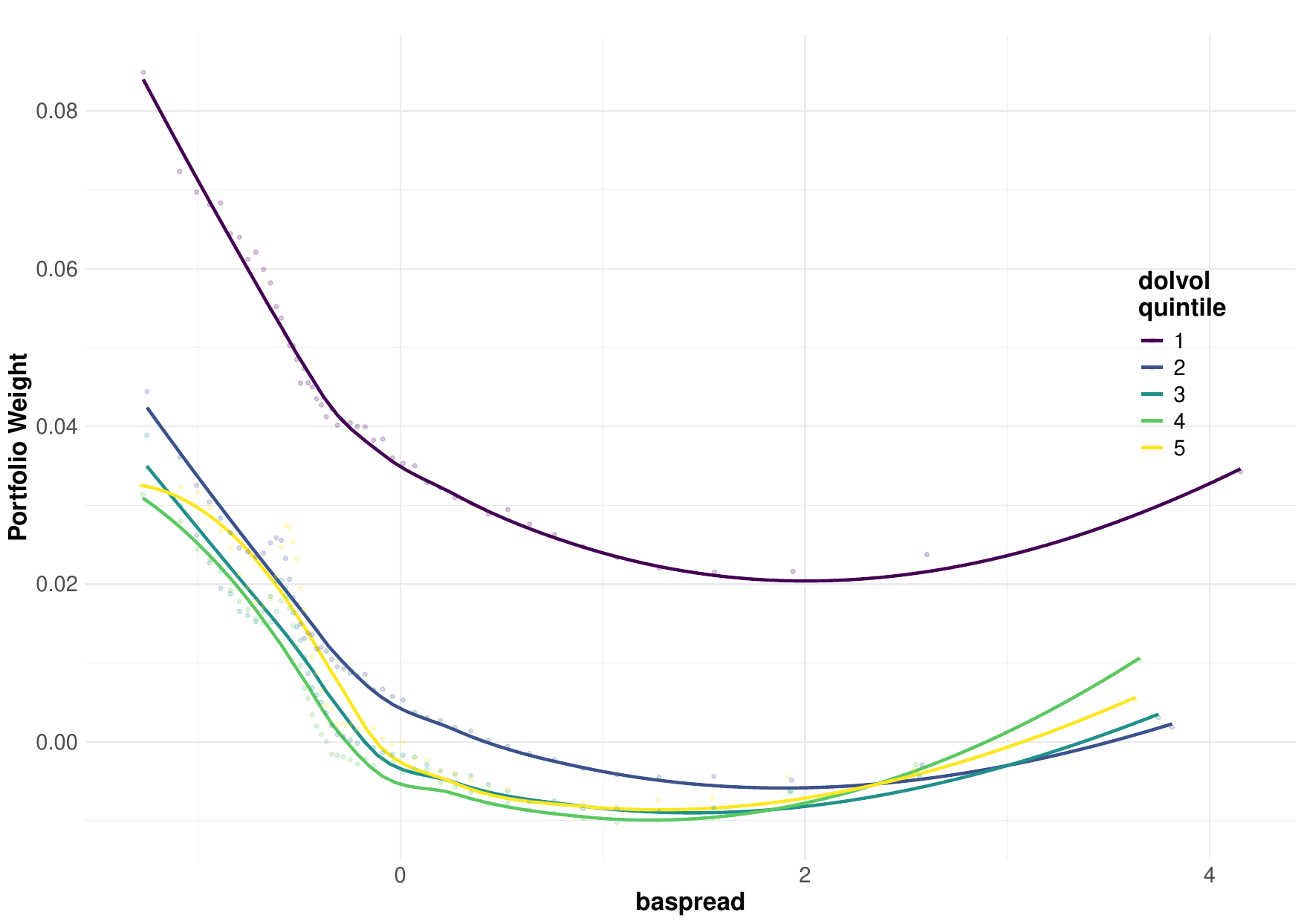}
			\caption{Elastic net}
			\label{fig:sub2}
		\end{subfigure}
		\hspace{1cm}
		\begin{subfigure}[t]{0.6\textwidth}
			\centering
			\includegraphics[width=1\linewidth]{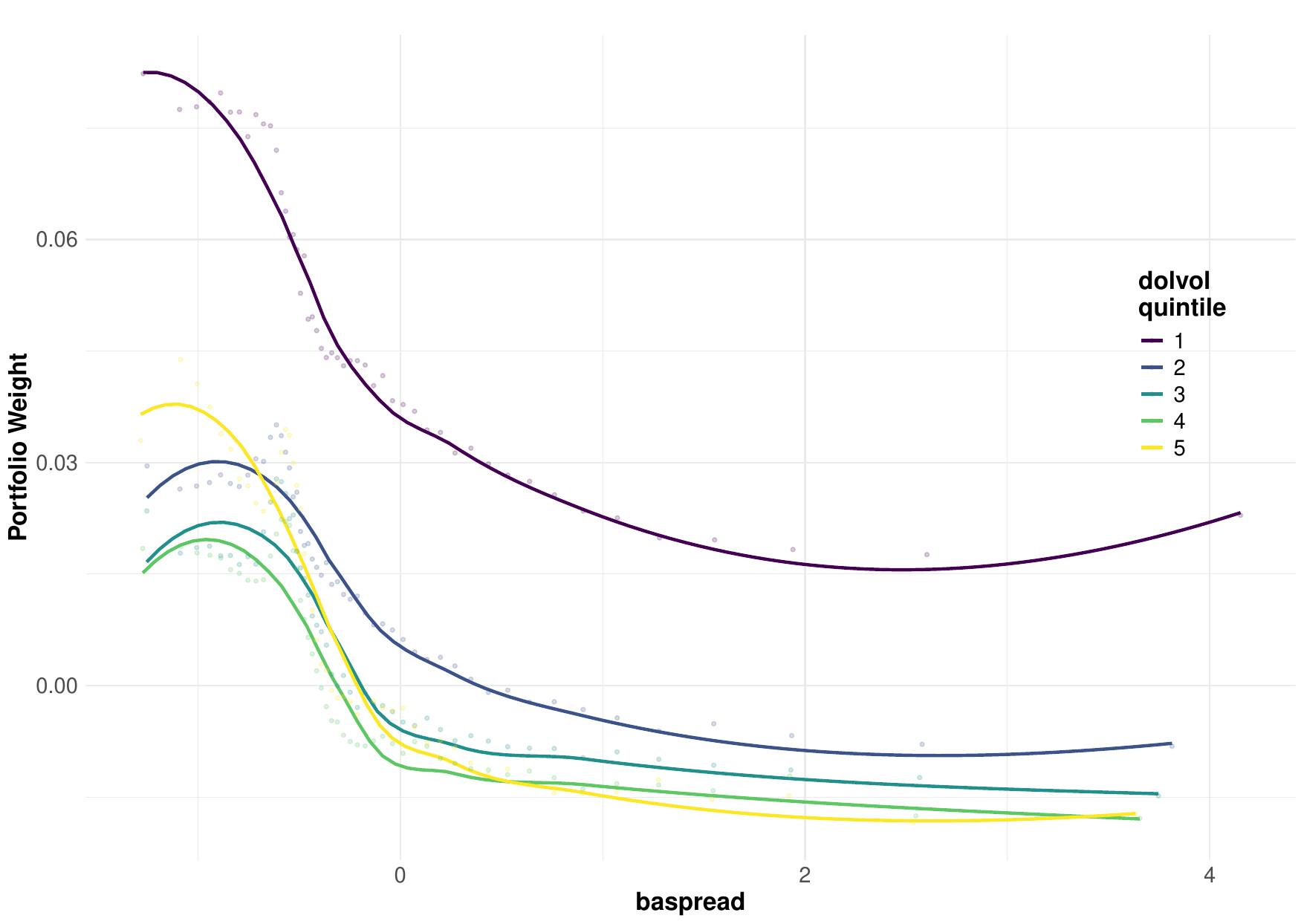}
			\caption{$L2$-boosting}
			\label{fig:sub3}
		\end{subfigure}
		
	\end{figure}
	
\end{landscape}

It is also remarkable that interactions between risk-based and liquidity measures are among the most important predictors for the three methods. For instance, interactions between market beta and standard deviation of liquidity ($beta \times std\_dolvol$), liquidity and idiosyncratic volatility ($dolvol\times idiovol$), and bid-ask spread and liquidity ($baspread\times dolvol$) are among the most important predictors for the three methods.  To understand how these three interactions connect to optimal portfolio allocations, we plot in Figures \ref{fig:interaction_dolvol_idiovol} to \ref{fig:interacion_baspread_dolvol} the average minimum-variance portfolio weight across standardized values of the first characteristic and across quintiles of the second characteristic. We find that stocks with lower liquidity get higher weights, and this result is stronger for stocks with lower idiosyncratic volatility. Stocks with low market betas get higher weights, and this result is stronger for stocks with higher standard deviation of liquidity. Finally, we observe a pronounced u-shaped non-linear relation between bid-ask spreads and portfolio weights, as well as a pronounced interaction with volatility-based predictors. 

\begin{landscape}
	\begin{figure}[h]
		\caption{\textbf{$L2$-boosting coefficients: second-order effects}}\label{figure:heatmap2_boosting}
		
		\scriptsize{The figure plots the estimated coefficients of the feature's second-order effects when using the $L2$-boosting method across all estimation rounds.}\\
		\centering
		\includegraphics[width=9in]{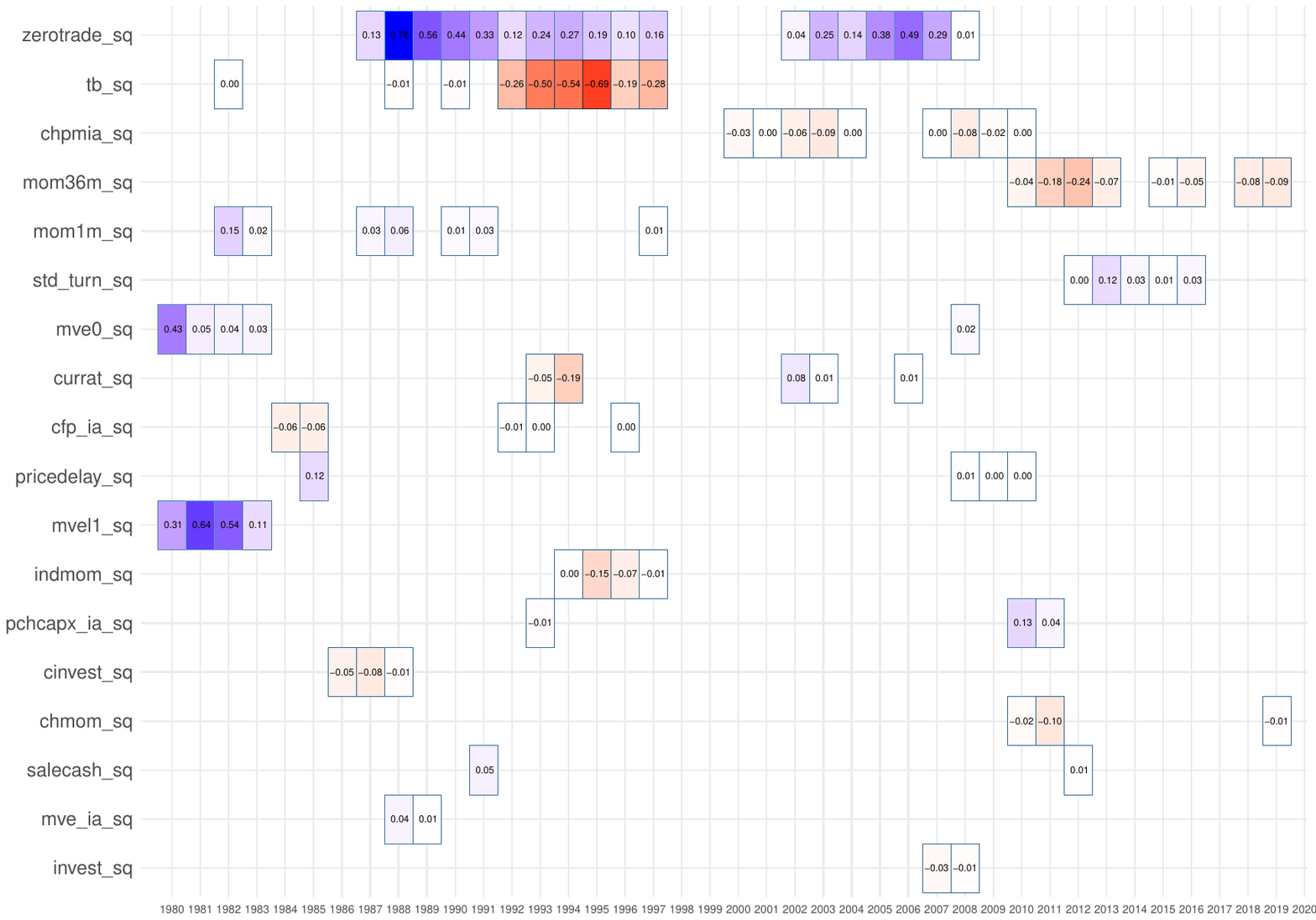}
	\end{figure}
\end{landscape}

The analysis of predictor importance shown in Figures \ref{figure:lasso}-\ref{figure:boosting} allows us to draw three main conclusions. First, we show that risk-based predictors are not the only class of important predictors for constructing minimum-variance portfolios. Our results reveal that size-, momentum-, and liquidity-based predictors are also among the most important predictors via their main effects as well as non-linear transformations (especially interactions). Second, we observe that some predictors that help decrease portfolio risk can also help increase portfolio returns in some situations. For instance, in the case of the $L2$-boosting method, Figure \ref{figure:heatmap2_boosting} shows that the interaction between size and zero trading days appears among the most important predictors that help decrease portfolio risk via its lower correlation with the benchmark portfolio. This predictor also contributes to \textit{increasing} portfolio \textit{returns}, as reported in Figure \ref{figure:individual_predictor_returns}.

We also study the economic significance of the most important risk-based and liquidity predictors as well as their interactions. Table \ref{table:state_variables} presents the regression results of selected parametric-portfolio predictor coefficients on three state variables: investor sentiment, market liquidity, and economic recession. Panel A shows the main effects of each predictor, while Panel B reports the interaction effects. For the predictor \textit{beta} the results indicate a significant relationship with investor sentiment across all methods, with consistently negative coefficients. Additionally, the variations in the \textit{beta} parametric-portfolio coefficient is also significantly associated with the recession indicator in the $L2$-boosting method, where it shows a negative coefficient. 

Moreover, some of the interaction effects show significant relationships with state variables even when the corresponding main effects are not significant. This suggests that interactions may capture more complex or nuanced dynamics that are not apparent in the main effects alone. For instance, The $dolvol\_x\_idiovol$ interaction also exhibits significant relationships with investor sentiment across several methods, all with positive coefficients. The liquidity variable has significant effects on the $beta\_x\_idiovol$ interaction the $L2$-boosting method, showing a negative coefficient.



\begin{landscape}
	\begin{figure}[h]
		\caption{\textbf{Lasso coefficients: first-order effects}}\label{figure:heatmap1_lasso}
		
		\scriptsize{The figure plots the estimated coefficients of the features' first-order effects when using the lasso method across all estimation rounds.}\\
		\centering
		\includegraphics[width=9in]{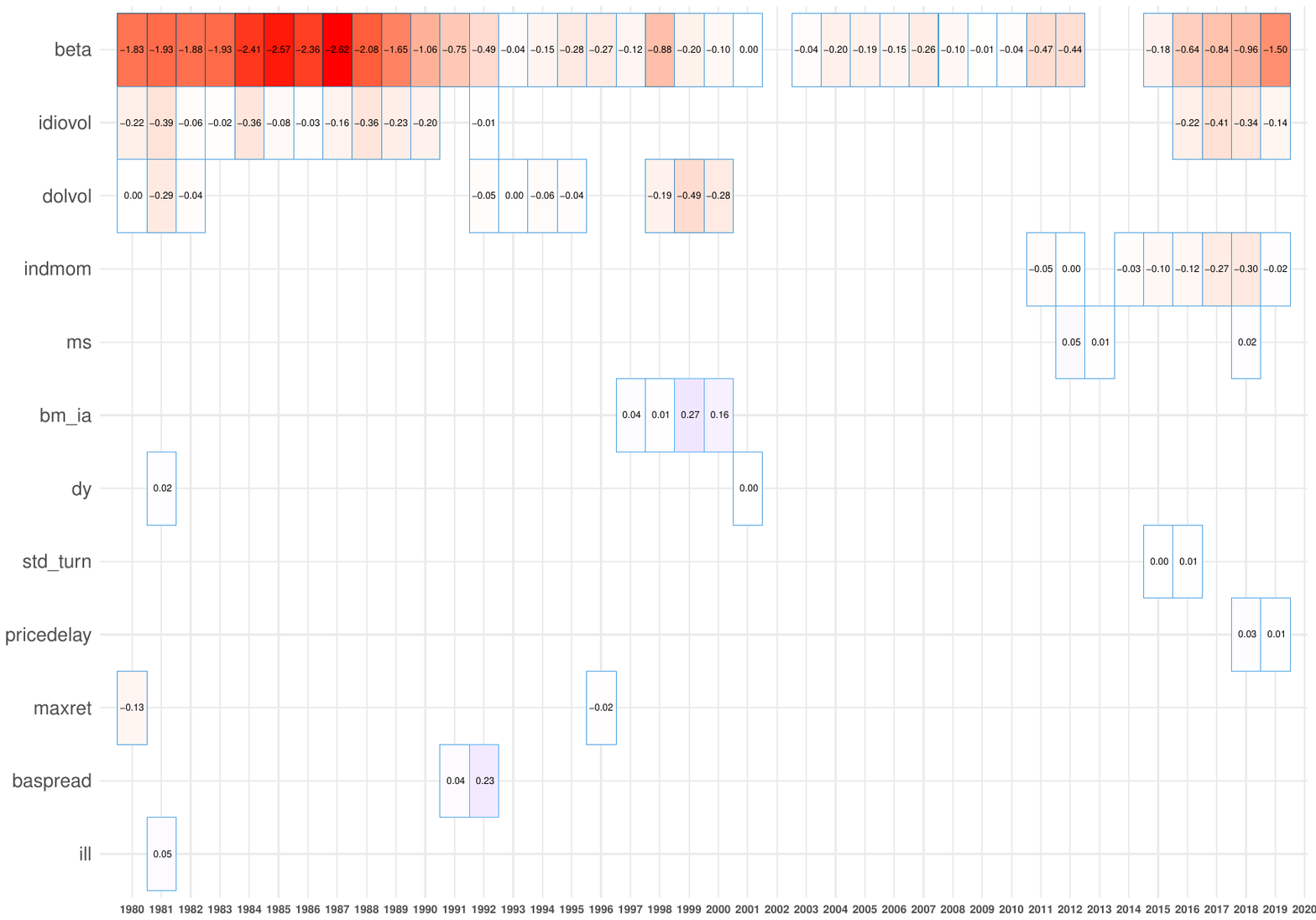}
	\end{figure}
\end{landscape}
\begin{landscape}
	\begin{figure}[h]
		\caption{\textbf{Lasso coefficients: second-order effects}}\label{figure:heatmap2_lasso}
		
		\scriptsize{The figure plots the estimated coefficients of the features' second-order effects when using the lasso method across all estimation rounds.}\\
		\centering
		\includegraphics[width=9in]{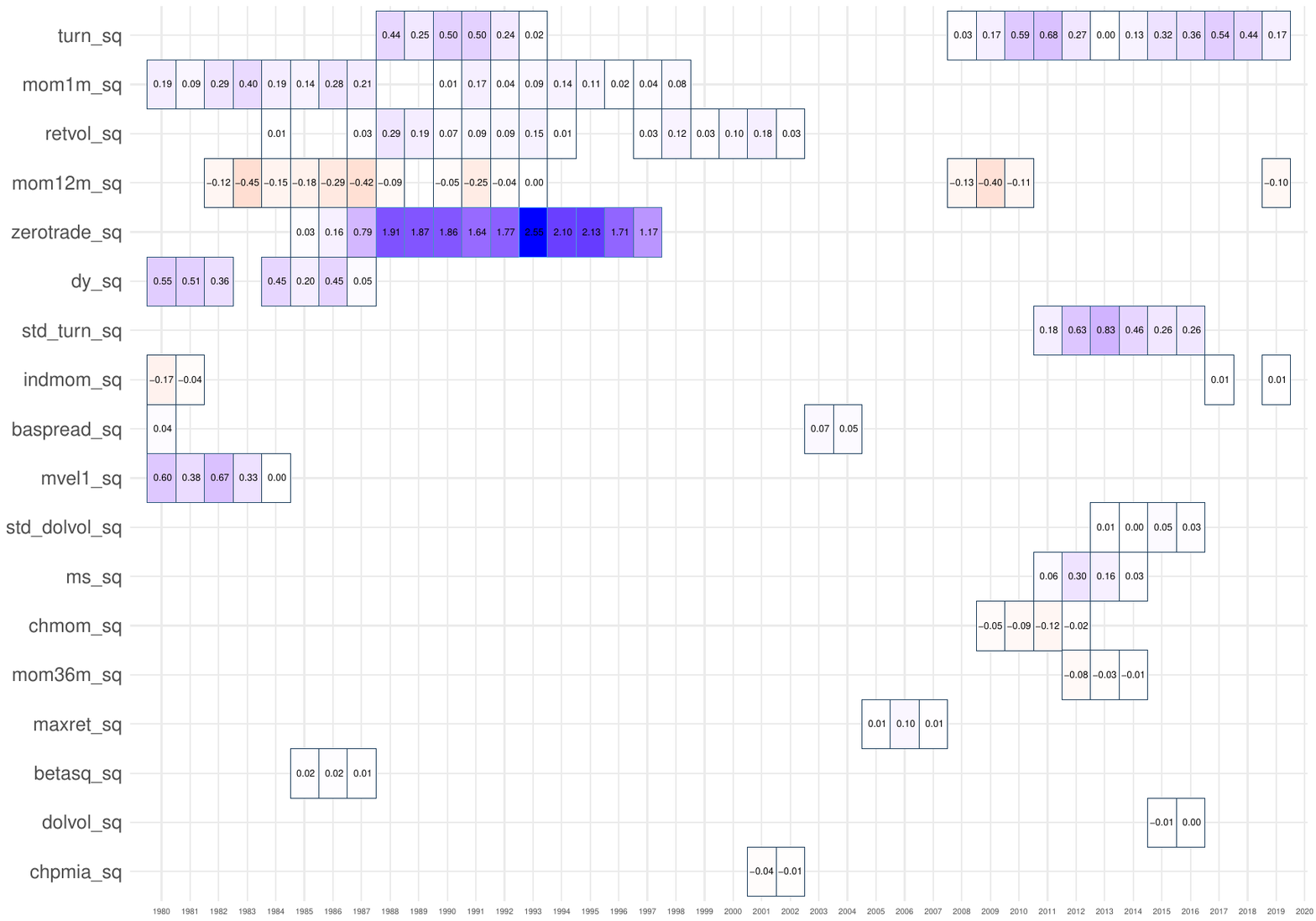}
	\end{figure}
\end{landscape}
\begin{landscape}
	\begin{figure}[h]
		\caption{\textbf{Lasso coefficients: interactions}}\label{figure:heatmap3_lasso}
		
		\scriptsize{The figure plots the estimated coefficients of the features' interaction effects when using the lasso method across all estimation rounds.}\\
		\centering
		\includegraphics[width=9in]{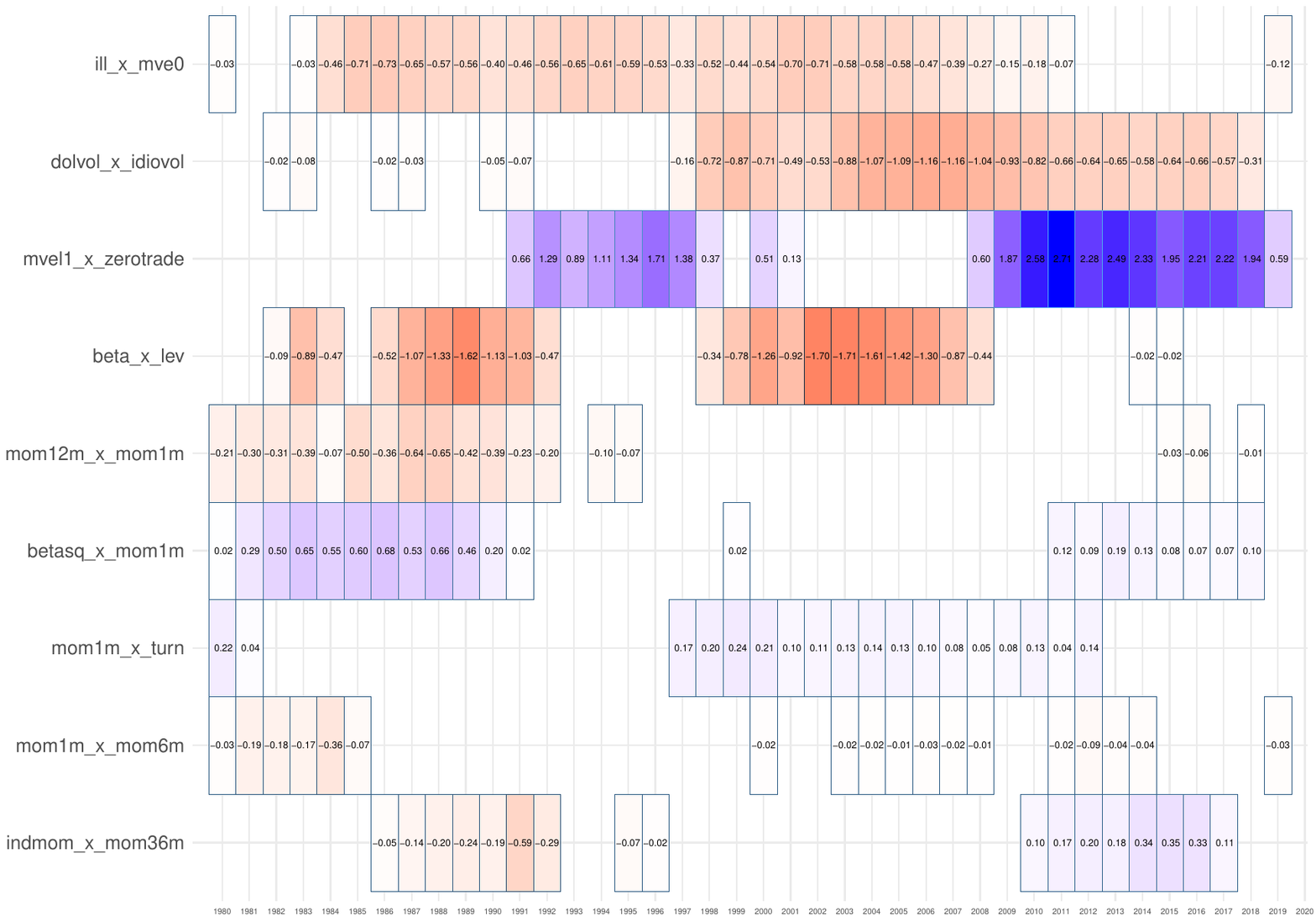}
	\end{figure}
\end{landscape}
\begin{landscape}
	\begin{figure}[h]
		\caption{\textbf{Elastic net coefficients: first-order effects}}\label{figure:heatmap1_enet}		
		\scriptsize{The figure plots the estimated coefficients of the feature's first-order effects when using the elastic net method across all estimation rounds.}\\
		\centering
		\includegraphics[width=9in]{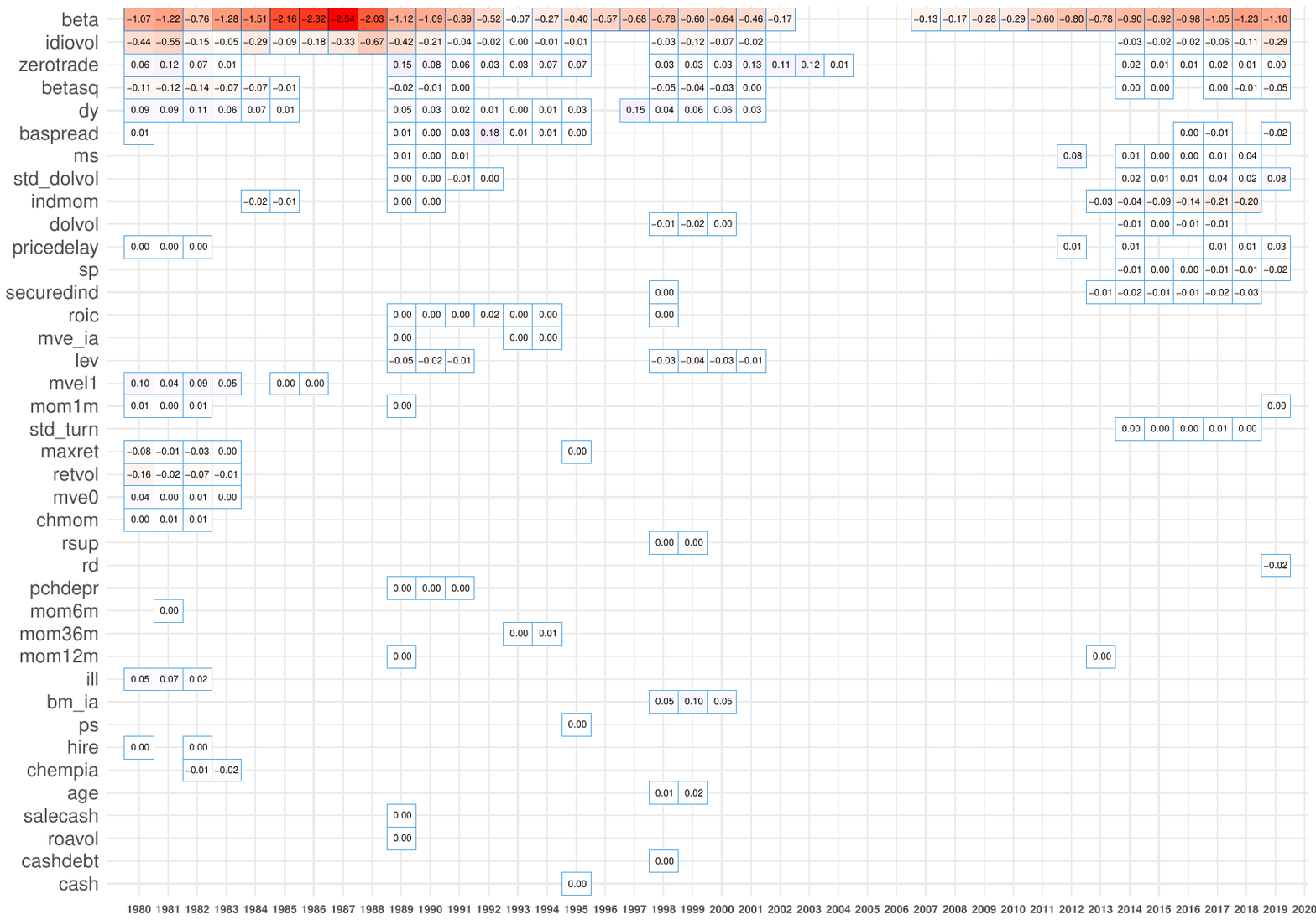}
	\end{figure}
\end{landscape}
\begin{landscape}
	\begin{figure}[h]
		\caption{\textbf{Elastic net coefficients: second-order effects}}\label{figure:heatmap2_enet}
		
		\scriptsize{The figure plots the estimated coefficients of the feature's second-order effects when using the elastic net method across all estimation rounds.}\\
		\centering
		\includegraphics[width=9in]{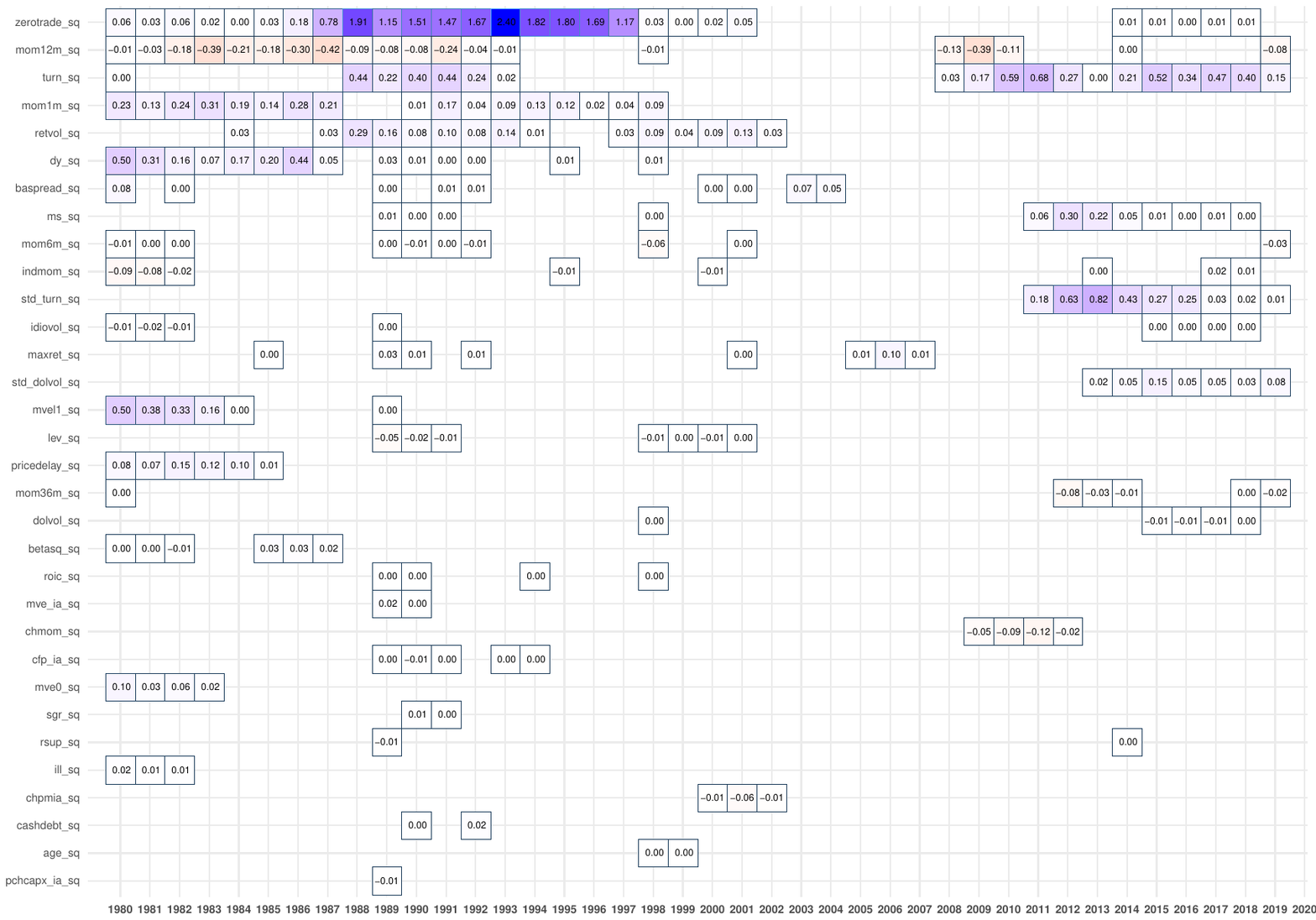}
	\end{figure}
\end{landscape}
\begin{landscape}
	\begin{figure}[h]
		\caption{\textbf{Elastic net coefficients: interactions}}\label{figure:heatmap3_enet}
		
		\scriptsize{The figure plots the estimated coefficients of the feature's interactions when using the elastic net method across all estimation rounds.}\\
		\centering
		\includegraphics[width=9in]{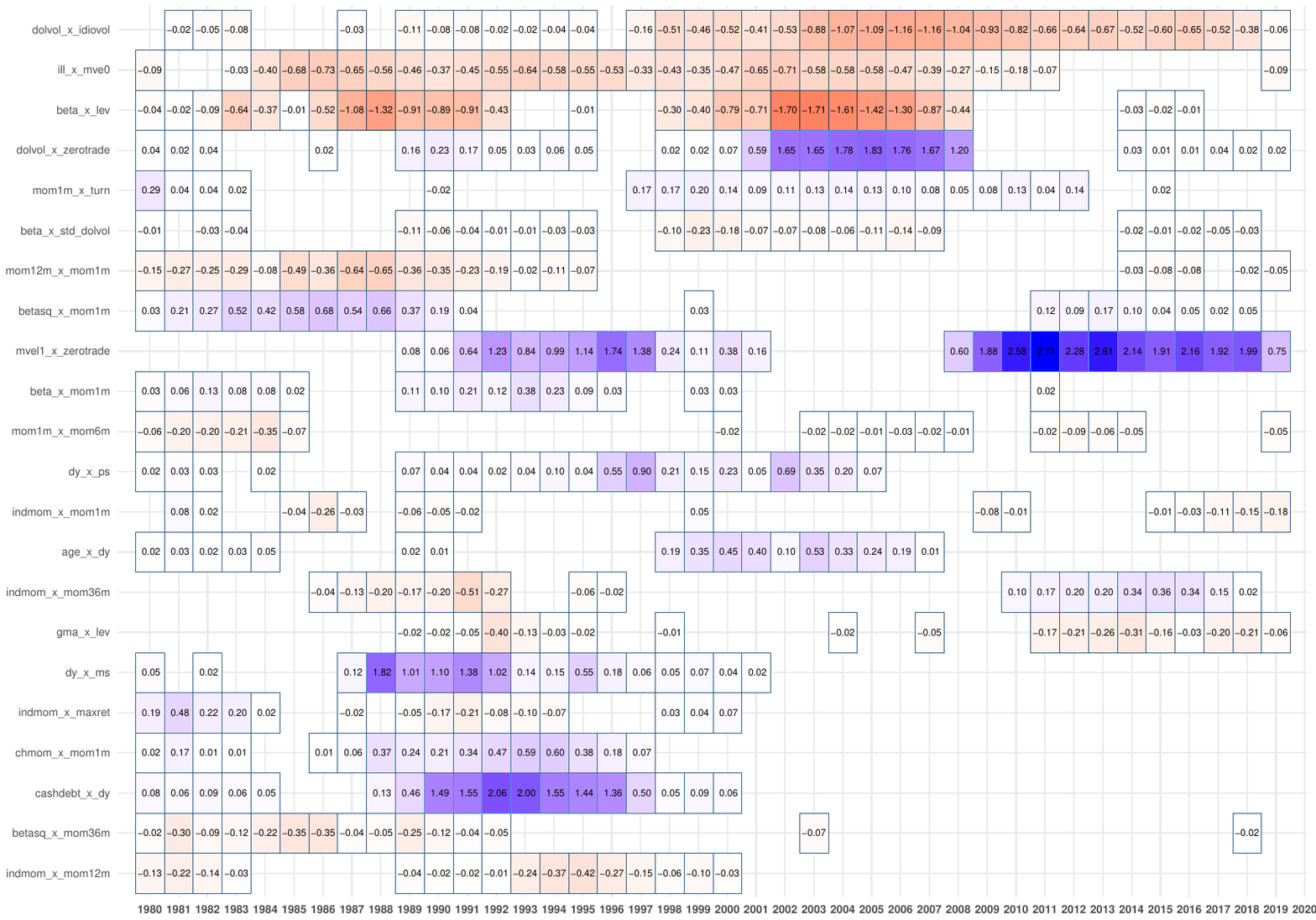}
	\end{figure}
\end{landscape}
\begin{landscape}
	\begin{figure}[h]
		\caption{\textbf{$L2$-boosting coefficients: first-order effects}}\label{figure:heatmap1_boosting}
		
		\scriptsize{The figure plots the estimated coefficients of the feature's first-order effects when using the $L2$-boosting method across all estimation rounds.}\\
		\centering
		\includegraphics[width=9in]{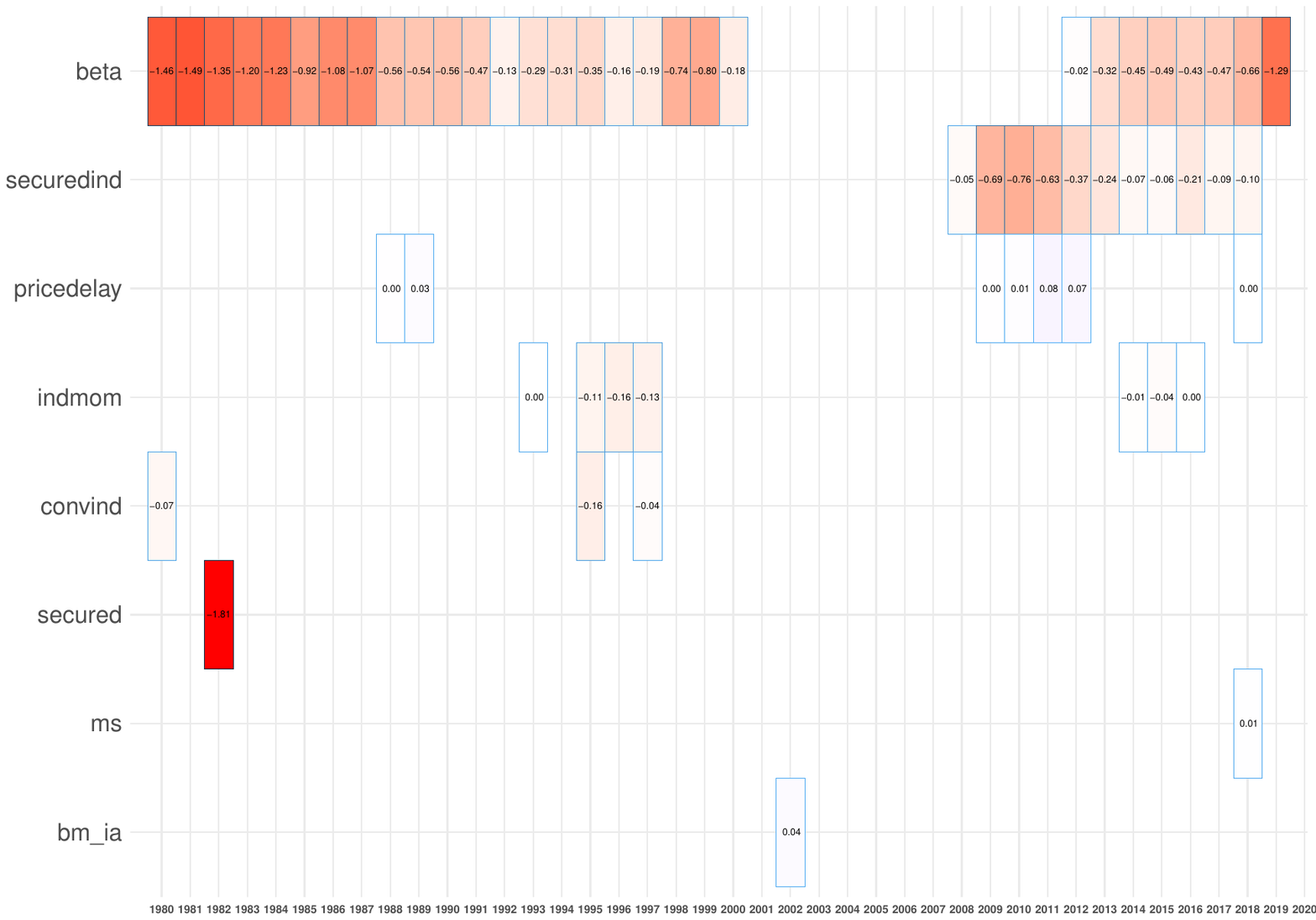}
	\end{figure}
\end{landscape}
\begin{landscape}
	\begin{figure}[h]
		\caption{\textbf{$L2$-boosting coefficients: interactions}}\label{figure:heatmap3_boosting}
		
		\scriptsize{The figure plots the estimated coefficients of the feature's interactions when using the $L2$-boosting method across all estimation rounds.}\\
		\centering
		\includegraphics[width=9in]{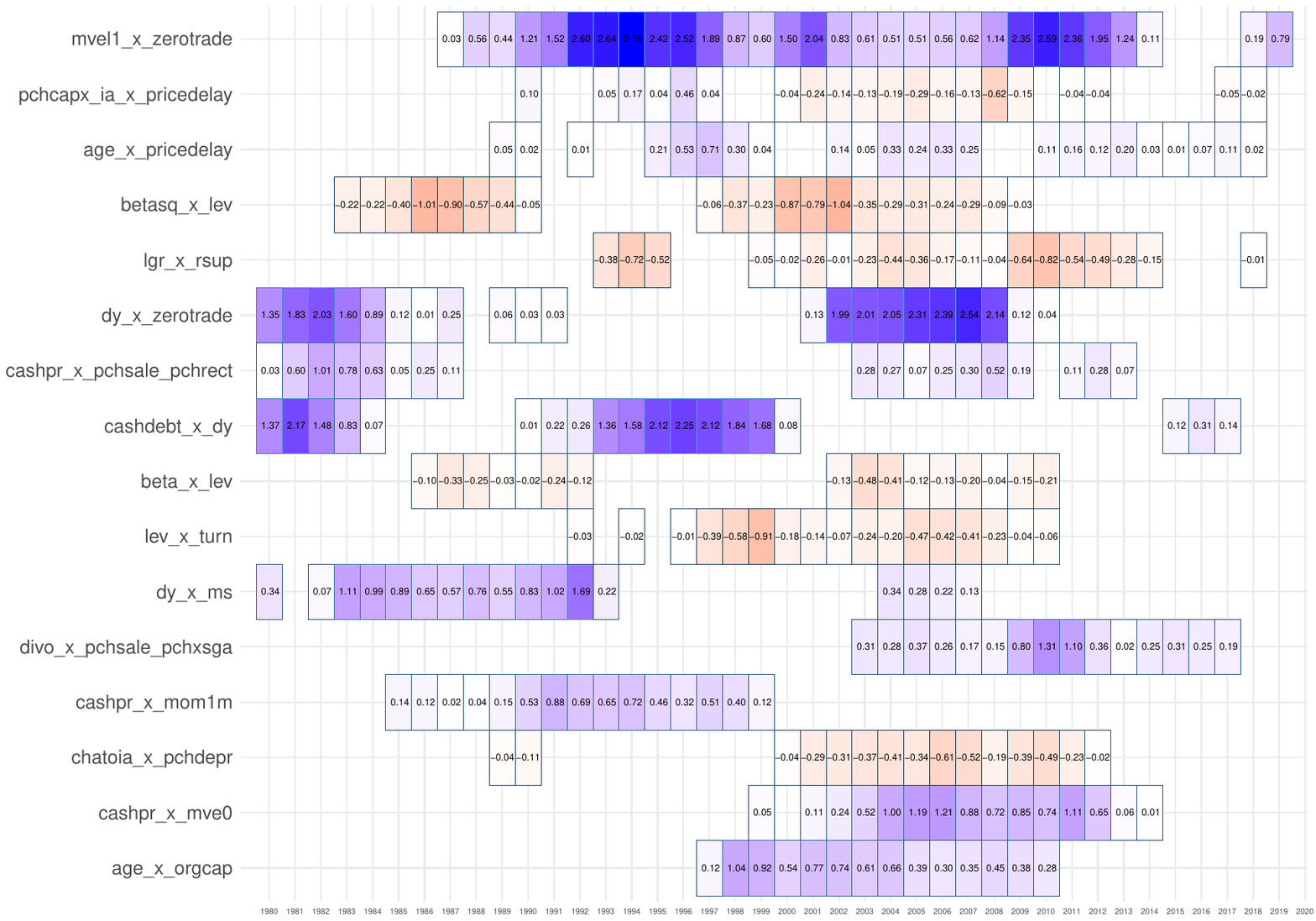}
	\end{figure}
\end{landscape}

Finally, we also examine the temporal dynamics of the estimated coefficients of the selected characteristics with the lasso, elastic net, and $L2$-boosting methods across all estimation rounds. For that purpose, we plot in Figures \ref{figure:heatmap1_lasso} to \ref{figure:heatmap3_boosting} a heatmap of the estimated coefficients for the features' main effect, second-order effect, and interactions, respectively. The Figures reveal that the selection of relevant characteristics and the strength of their effects exhibit substantial variations over time. Many characteristics appear and disappear from the chosen subset, suggesting that the importance of firm characteristics for portfolio choice is highly time-varying.

\subsection{Are the portfolios implementable?}

To assess the practicality of implementing the ML-based minimum-variance portfolios, Table \ref{table:portfolio_stats} reports descriptive statistics of the distribution of portfolio weights across all estimation rounds. The Table shows that the minimum and maximum weights of the ML-based portfolios are not extreme, ranging from -0.8\% to 0.8\%. These values are well within the acceptable range for traditional investment strategies. Moreover, the proportion of negative weights (short proportion) is also relatively low, hovering around 30\%. This suggests that the ML-based portfolios are long-biased and do not involve excessive short selling.

It is interesting to note that, according to \cite{green1992will}, the presence of a dominant first principal component (or factor) would result in extreme negative weights in minimum-variance portfolios. Our results seem to contradict this finding, as they indicate the absence of extreme weighting. One plausible explanation for our results is that our methodology involves constructing a minimum-variance portfolio using a \textit{multitude} of characteristics-based factors. In this situation, \cite{demiguel2017portfolio} and \cite*{demiguel2022can} show that exploiting multiple factors leads to trading diversification (i.e. netting of trades across factors), which helps reduce extreme positions in individual stocks.

\subsection{Screening predictors}

The results reported in Table \ref{table:out_sample_performance} show that selecting a few predictors ad hoc, as in the popular 3-factor and 5-factor models, leads to minimum-variance parametric portfolios with poor performance relative to those obtained with ML variable selection methods. To avoid selecting predictors ad hoc, we implement the marginal screening (MS) method of \cite{fan2008sure} which employs a data-driven approach to screen a vast pool of predictors based on their individual correlations with the target variable. MS is computationally efficient, making it well-suited for ultrahigh dimensions. Moreover, MS ensures the exact recovery of true non-zero coefficients under specific sparsity and signal strength conditions \citep*[see][]{genovese2012comparison}. Algorithm 2 describes the procedure for implementing the MS method. 

We implement the MS method to screen the augmented set of 4,610 features to select the best 3, 5, 10, and 50 predictors. We then use the selected predictors to construct minimum-variance portfolios. The results reported in Table \ref{table:screening} reveal that selecting only the top-3 predictors using MS leads to minimum-variance portfolio returns with an annualized standard deviation of 10.7\%, which is substantially \textit{lower} relative to those obtained when using ad-hoc selected characteristics, as reported in Table \ref{table:out_sample_performance}. The performance of the MS methods peaks when the top-10 predictors are selected (9.5\%). However, the performance of screening-based portfolios is worse than those obtained with the variable selection methods, as reported in Table \ref{table:out_sample_performance}. We conclude from this analysis that although MS is a much better alternative in comparison to selecting predictors ad-hoc, the use of more sophisticated variable selection methods can offer further improvements.

\section{Concluding remarks}\label{section:conclusions}

This paper provides evidence that machine learning (ML) can be a powerful tool for selecting variables relevant for optimal portfolio choice. We focus on directly parameterizing portfolio weights as a function of lagged firm-level predictors. By employing ML variable selection methods on a large pool of features and their second-order and cross-product transformations (4,610 predictors), we find that ML can identify a subset of important variables that outperform traditional low-dimensional factor models in terms of lower risk and higher risk-adjusted returns. We also find that the choice of variable selection method matters, with different methods selecting different subsets of features. The $L2$-boosting method emerged as the most comprehensive approach, as it was able to identify predictors that minimize risk, reduce covariance with the benchmark, and increase risk-adjusted portfolio returns.

Our findings have several implications for practitioners. First, they suggest that ML can be a valuable tool for improving portfolio performance. Second, they highlight the importance of accounting for the interaction between variables in the portfolio construction process. Third, they indicate that the number of selected features relevant to the portfolio construction problem is greater than the number of variables used in popular factor models. As a result, practitioners should consider using ML variable selection methods to identify a subset of important variables for their portfolios.


\section*{Acknowledgments}

This work was supported by Agencia Estatal de Investigación (Spain) under grant PID2022-138289NB-I00 and Comunidad Autonoma de Madrid Government through project 2022-T1/SOC-24167 (Convocatoria Talento).

\section*{Disclosure of interest}

There are no interests to declare.

\clearpage
\singlespacing
\begin{small}
\bibliography{References}
\end{small}	
\clearpage


\vspace{0.5cm}
\onehalfspacing
\begin{small}
	\begin{longtable}{lcccccc}
		\caption{\textbf{Descriptive statistics of predictor returns}\label{table:descriptive_stats} } \\[-0.5cm]
		\caption*{\scriptsize{The table reports descriptive statistics of  predictor returns. The following statistics are reported: average monthly return, the standard deviation of monthly returns, the 25th and 75th percentiles, and the correlation of the predictor return with the equally weighted and value-weighted portfolio returns ($\rho_{\text{ew}}$ and $\rho_{\text{vw}}$, respectively). The first four rows report aggregate values across i) all predictors, ii) predictors' main effects, iii) predictor's second-order effects, and iv) predictors' interaction effects. The remaining rows report individual statistics for the  top and bottom 10 predictors within each category sorted in terms of average monthly return.}}\\
		\hline
		& Mean & Std. & 25th (\%) & 75th (\%) & $\rho_{\text{ew}}$ & $\rho_{\text{vw}}$ \\
		& return (\%) & deviation (\%) & & & & \\
		\hline
		\endfirsthead
		
		\multicolumn{7}{r}%
		{{ \textit{Table \thetable\ continued from previous page}}} \\
		\addlinespace
		
		& Mean & Std. & 25th (\%) & 75th (\%) & $\rho_{\text{ew}}$ & $\rho_{\text{vw}}$ \\
		& return (\%) & deviation (\%) & & & & \\
		\hline
		\endhead
		
		\addlinespace
		\multicolumn{7}{r}{{\textit{Table \thetable\ continued on next page}}} \\ 
		\endfoot
		
		\hline
		\endlastfoot
		
		\addlinespace
		All predictors & -0.003 & 0.523 & -0.233 & 0.232 & 0.02 & 0.04 \\
		&       &       &       &       &       &  \\
		\multicolumn{1}{l}{Main effects} & \multicolumn{1}{c}{-0.009} & \multicolumn{1}{c}{0.797} & \multicolumn{1}{c}{-0.367} & \multicolumn{1}{c}{0.352} & \multicolumn{1}{c}{0.04} & \multicolumn{1}{c}{0.06} \\
		&       &       &       &       &       &  \\
		\multicolumn{1}{l}{Second-order effects} & \multicolumn{1}{c}{-0.024} & \multicolumn{1}{c}{0.706} & \multicolumn{1}{c}{-0.373} & \multicolumn{1}{c}{0.273} & \multicolumn{1}{c}{0.26} & \multicolumn{1}{c}{0.17} \\
		&       &       &       &       &       &  \\
		\multicolumn{1}{l}{Interaction effects} & \multicolumn{1}{c}{-0.002} & \multicolumn{1}{c}{0.513} & \multicolumn{1}{c}{-0.227} & \multicolumn{1}{c}{0.229} & \multicolumn{1}{c}{0.02} & \multicolumn{1}{c}{0.04} \\
		&       &       &       &       &       &  \\
		\multicolumn{1}{l}{Top-10 main effects} &       &       &       &       &       &  \\
		\multicolumn{1}{l}{indmom} & \multicolumn{1}{c}{0.309} & \multicolumn{1}{c}{1.545} & \multicolumn{1}{c}{-0.360} & \multicolumn{1}{c}{1.039} & \multicolumn{1}{c}{-0.12} & \multicolumn{1}{c}{-0.04} \\
		\multicolumn{1}{l}{agr} & \multicolumn{1}{c}{0.275} & \multicolumn{1}{c}{0.653} & \multicolumn{1}{c}{-0.063} & \multicolumn{1}{c}{0.541} & \multicolumn{1}{c}{-0.21} & \multicolumn{1}{c}{-0.34} \\
		\multicolumn{1}{l}{mom12m} & \multicolumn{1}{c}{0.270} & \multicolumn{1}{c}{1.863} & \multicolumn{1}{c}{-0.354} & \multicolumn{1}{c}{1.096} & \multicolumn{1}{c}{-0.33} & \multicolumn{1}{c}{-0.13} \\
		\multicolumn{1}{l}{ill} & \multicolumn{1}{c}{0.205} & \multicolumn{1}{c}{1.206} & \multicolumn{1}{c}{-0.424} & \multicolumn{1}{c}{0.555} & \multicolumn{1}{c}{0.25} & \multicolumn{1}{c}{-0.07} \\
		\multicolumn{1}{l}{mve0} & \multicolumn{1}{c}{0.186} & \multicolumn{1}{c}{0.917} & \multicolumn{1}{c}{-0.273} & \multicolumn{1}{c}{0.715} & \multicolumn{1}{c}{-0.57} & \multicolumn{1}{c}{-0.13} \\
		\multicolumn{1}{l}{sp} & \multicolumn{1}{c}{0.154} & \multicolumn{1}{c}{0.783} & \multicolumn{1}{c}{-0.301} & \multicolumn{1}{c}{0.506} & \multicolumn{1}{c}{0.20} & \multicolumn{1}{c}{-0.01} \\
		\multicolumn{1}{l}{mom6m} & \multicolumn{1}{c}{0.148} & \multicolumn{1}{c}{1.946} & \multicolumn{1}{c}{-0.399} & \multicolumn{1}{c}{0.941} & \multicolumn{1}{c}{-0.39} & \multicolumn{1}{c}{-0.21} \\
		\multicolumn{1}{l}{rd\_mve} & \multicolumn{1}{c}{0.139} & \multicolumn{1}{c}{0.658} & \multicolumn{1}{c}{-0.171} & \multicolumn{1}{c}{0.356} & \multicolumn{1}{c}{0.45} & \multicolumn{1}{c}{0.25} \\
		\multicolumn{1}{l}{orgcap} & \multicolumn{1}{c}{0.105} & \multicolumn{1}{c}{0.668} & \multicolumn{1}{c}{-0.227} & \multicolumn{1}{c}{0.364} & \multicolumn{1}{c}{0.34} & \multicolumn{1}{c}{0.13} \\
		\multicolumn{1}{l}{bm} & \multicolumn{1}{c}{0.102} & \multicolumn{1}{c}{0.639} & \multicolumn{1}{c}{-0.258} & \multicolumn{1}{c}{0.448} & \multicolumn{1}{c}{-0.21} & \multicolumn{1}{c}{-0.24} \\
		&       &       &       &       &       &  \\
		
		\multicolumn{1}{l}{Bottom-10 main effects} &       &       &       &       &       &  \\
		\multicolumn{1}{l}{mom1m} & \multicolumn{1}{c}{-0.561} & \multicolumn{1}{c}{1.859} & \multicolumn{1}{c}{-1.125} & \multicolumn{1}{c}{0.211} & \multicolumn{1}{c}{-0.40} & \multicolumn{1}{c}{-0.25} \\
		\multicolumn{1}{l}{maxret} & \multicolumn{1}{c}{-0.310} & \multicolumn{1}{c}{2.117} & \multicolumn{1}{c}{-1.353} & \multicolumn{1}{c}{0.479} & \multicolumn{1}{c}{0.69} & \multicolumn{1}{c}{0.43} \\
		\multicolumn{1}{l}{turn} & \multicolumn{1}{c}{-0.248} & \multicolumn{1}{c}{1.533} & \multicolumn{1}{c}{-1.002} & \multicolumn{1}{c}{0.475} & \multicolumn{1}{c}{0.59} & \multicolumn{1}{c}{0.60} \\
		\multicolumn{1}{l}{chmom} & \multicolumn{1}{c}{-0.248} & \multicolumn{1}{c}{1.289} & \multicolumn{1}{c}{-0.723} & \multicolumn{1}{c}{0.285} & \multicolumn{1}{c}{-0.33} & \multicolumn{1}{c}{-0.25} \\
		\multicolumn{1}{l}{invest} & \multicolumn{1}{c}{-0.212} & \multicolumn{1}{c}{0.547} & \multicolumn{1}{c}{-0.494} & \multicolumn{1}{c}{0.111} & \multicolumn{1}{c}{0.24} & \multicolumn{1}{c}{0.35} \\
		\multicolumn{1}{l}{retvol} & \multicolumn{1}{c}{-0.197} & \multicolumn{1}{c}{2.569} & \multicolumn{1}{c}{-1.472} & \multicolumn{1}{c}{0.708} & \multicolumn{1}{c}{0.73} & \multicolumn{1}{c}{0.47} \\
		\multicolumn{1}{l}{lgr} & \multicolumn{1}{c}{-0.192} & \multicolumn{1}{c}{0.454} & \multicolumn{1}{c}{-0.417} & \multicolumn{1}{c}{0.033} & \multicolumn{1}{c}{0.44} & \multicolumn{1}{c}{0.41} \\
		\multicolumn{1}{l}{chcsho} & \multicolumn{1}{c}{-0.179} & \multicolumn{1}{c}{0.506} & \multicolumn{1}{c}{-0.420} & \multicolumn{1}{c}{0.077} & \multicolumn{1}{c}{0.34} & \multicolumn{1}{c}{0.42} \\
		\multicolumn{1}{l}{sgr} & \multicolumn{1}{c}{-0.172} & \multicolumn{1}{c}{0.542} & \multicolumn{1}{c}{-0.433} & \multicolumn{1}{c}{0.127} & \multicolumn{1}{c}{0.41} & \multicolumn{1}{c}{0.41} \\
		\multicolumn{1}{l}{hire} & \multicolumn{1}{c}{-0.170} & \multicolumn{1}{c}{0.530} & \multicolumn{1}{c}{-0.413} & \multicolumn{1}{c}{0.118} & \multicolumn{1}{c}{0.27} & \multicolumn{1}{c}{0.36} \\
		
		&       &       &       &       &       &  \\
		
		\multicolumn{1}{l}{Top-10 second-order effects} &       &       &       &       &       &  \\
		\multicolumn{1}{l}{ill\_sq} & \multicolumn{1}{c}{0.208} & \multicolumn{1}{c}{0.985} & \multicolumn{1}{c}{-0.325} & \multicolumn{1}{c}{0.516} & \multicolumn{1}{c}{0.25} & \multicolumn{1}{c}{-0.04} \\
		\multicolumn{1}{l}{mom12m\_sq} & \multicolumn{1}{c}{0.170} & \multicolumn{1}{c}{1.506} & \multicolumn{1}{c}{-0.561} & \multicolumn{1}{c}{0.667} & \multicolumn{1}{c}{0.70} & \multicolumn{1}{c}{0.54} \\
		\multicolumn{1}{l}{indmom\_sq} & \multicolumn{1}{c}{0.132} & \multicolumn{1}{c}{1.303} & \multicolumn{1}{c}{-0.548} & \multicolumn{1}{c}{0.712} & \multicolumn{1}{c}{0.53} & \multicolumn{1}{c}{0.41} \\
		\multicolumn{1}{l}{rd\_mve\_sq} & \multicolumn{1}{c}{0.114} & \multicolumn{1}{c}{0.539} & \multicolumn{1}{c}{-0.161} & \multicolumn{1}{c}{0.290} & \multicolumn{1}{c}{0.42} & \multicolumn{1}{c}{0.21} \\
		\multicolumn{1}{l}{mom6m\_sq} & \multicolumn{1}{c}{0.106} & \multicolumn{1}{c}{1.817} & \multicolumn{1}{c}{-0.695} & \multicolumn{1}{c}{0.644} & \multicolumn{1}{c}{0.67} & \multicolumn{1}{c}{0.49} \\
		\multicolumn{1}{l}{sp\_sq} & \multicolumn{1}{c}{0.087} & \multicolumn{1}{c}{0.610} & \multicolumn{1}{c}{-0.265} & \multicolumn{1}{c}{0.385} & \multicolumn{1}{c}{0.26} & \multicolumn{1}{c}{0.02} \\
		\multicolumn{1}{l}{mve0\_sq} & \multicolumn{1}{c}{0.087} & \multicolumn{1}{c}{0.618} & \multicolumn{1}{c}{-0.231} & \multicolumn{1}{c}{0.460} & \multicolumn{1}{c}{-0.61} & \multicolumn{1}{c}{-0.20} \\
		\multicolumn{1}{l}{std\_dolvol\_sq} & \multicolumn{1}{c}{0.086} & \multicolumn{1}{c}{0.823} & \multicolumn{1}{c}{-0.307} & \multicolumn{1}{c}{0.561} & \multicolumn{1}{c}{-0.23} & \multicolumn{1}{c}{-0.44} \\
		\multicolumn{1}{l}{orgcap\_sq} & \multicolumn{1}{c}{0.072} & \multicolumn{1}{c}{0.551} & \multicolumn{1}{c}{-0.203} & \multicolumn{1}{c}{0.271} & \multicolumn{1}{c}{0.29} & \multicolumn{1}{c}{0.07} \\
		\multicolumn{1}{l}{ps\_sq} & \multicolumn{1}{c}{0.071} & \multicolumn{1}{c}{0.713} & \multicolumn{1}{c}{-0.193} & \multicolumn{1}{c}{0.416} & \multicolumn{1}{c}{-0.52} & \multicolumn{1}{c}{-0.26} \\
		&       &       &       &       &       &  \\
		\multicolumn{1}{l}{Bottom-10 second-order effects} &       &       &       &       &       &  \\
		\multicolumn{1}{l}{maxret\_sq} & \multicolumn{1}{c}{-0.322} & \multicolumn{1}{c}{1.749} & \multicolumn{1}{c}{-1.192} & \multicolumn{1}{c}{0.311} & \multicolumn{1}{c}{0.61} & \multicolumn{1}{c}{0.33} \\
		\multicolumn{1}{l}{turn\_sq} & \multicolumn{1}{c}{-0.257} & \multicolumn{1}{c}{1.175} & \multicolumn{1}{c}{-0.814} & \multicolumn{1}{c}{0.293} & \multicolumn{1}{c}{0.52} & \multicolumn{1}{c}{0.53} \\
		\multicolumn{1}{l}{retvol\_sq} & \multicolumn{1}{c}{-0.227} & \multicolumn{1}{c}{2.276} & \multicolumn{1}{c}{-1.366} & \multicolumn{1}{c}{0.539} & \multicolumn{1}{c}{0.66} & \multicolumn{1}{c}{0.39} \\
		\multicolumn{1}{l}{agr\_sq} & \multicolumn{1}{c}{-0.195} & \multicolumn{1}{c}{0.516} & \multicolumn{1}{c}{-0.457} & \multicolumn{1}{c}{0.072} & \multicolumn{1}{c}{0.61} & \multicolumn{1}{c}{0.52} \\
		\multicolumn{1}{l}{invest\_sq} & \multicolumn{1}{c}{-0.183} & \multicolumn{1}{c}{0.481} & \multicolumn{1}{c}{-0.437} & \multicolumn{1}{c}{0.062} & \multicolumn{1}{c}{0.47} & \multicolumn{1}{c}{0.40} \\
		\multicolumn{1}{l}{chcsho\_sq} & \multicolumn{1}{c}{-0.144} & \multicolumn{1}{c}{0.419} & \multicolumn{1}{c}{-0.339} & \multicolumn{1}{c}{0.081} & \multicolumn{1}{c}{0.29} & \multicolumn{1}{c}{0.37} \\
		\multicolumn{1}{l}{lgr\_sq} & \multicolumn{1}{c}{-0.139} & \multicolumn{1}{c}{0.415} & \multicolumn{1}{c}{-0.354} & \multicolumn{1}{c}{0.074} & \multicolumn{1}{c}{0.52} & \multicolumn{1}{c}{0.39} \\
		\multicolumn{1}{l}{grltnoa\_sq} & \multicolumn{1}{c}{-0.135} & \multicolumn{1}{c}{0.381} & \multicolumn{1}{c}{-0.323} & \multicolumn{1}{c}{0.053} & \multicolumn{1}{c}{0.39} & \multicolumn{1}{c}{0.35} \\
		\multicolumn{1}{l}{sgr\_sq} & \multicolumn{1}{c}{-0.130} & \multicolumn{1}{c}{0.538} & \multicolumn{1}{c}{-0.432} & \multicolumn{1}{c}{0.122} & \multicolumn{1}{c}{0.56} & \multicolumn{1}{c}{0.39} \\
		\multicolumn{1}{l}{hire\_sq} & \multicolumn{1}{c}{-0.127} & \multicolumn{1}{c}{0.465} & \multicolumn{1}{c}{-0.379} & \multicolumn{1}{c}{0.102} & \multicolumn{1}{c}{0.60} & \multicolumn{1}{c}{0.46} \\
		&       &       &       &       &       &  \\
		\multicolumn{1}{l}{Top-10 interaction effects} &       &       &       &       &       &  \\
		\multicolumn{1}{l}{ill $\times$ mve0} & \multicolumn{1}{c}{0.953} & \multicolumn{1}{c}{0.988} & \multicolumn{1}{c}{0.272} & \multicolumn{1}{c}{1.417} & \multicolumn{1}{c}{0.09} & \multicolumn{1}{c}{-0.15} \\
		\multicolumn{1}{l}{mom1m $\times$ mom6m} & \multicolumn{1}{c}{0.496} & \multicolumn{1}{c}{1.522} & \multicolumn{1}{c}{-0.150} & \multicolumn{1}{c}{0.888} & \multicolumn{1}{c}{0.31} & \multicolumn{1}{c}{0.19} \\
		\multicolumn{1}{l}{mom12m $\times$ mom1m} & \multicolumn{1}{c}{0.364} & \multicolumn{1}{c}{1.405} & \multicolumn{1}{c}{-0.223} & \multicolumn{1}{c}{0.754} & \multicolumn{1}{c}{0.26} & \multicolumn{1}{c}{0.13} \\
		\multicolumn{1}{l}{mve0 $\times$ std\_turn} & \multicolumn{1}{c}{0.333} & \multicolumn{1}{c}{0.934} & \multicolumn{1}{c}{-0.120} & \multicolumn{1}{c}{0.759} & \multicolumn{1}{c}{-0.17} & \multicolumn{1}{c}{0.19} \\
		\multicolumn{1}{l}{agr $\times$ maxret} & \multicolumn{1}{c}{0.326} & \multicolumn{1}{c}{0.764} & \multicolumn{1}{c}{-0.053} & \multicolumn{1}{c}{0.642} & \multicolumn{1}{c}{-0.31} & \multicolumn{1}{c}{-0.36} \\
		\multicolumn{1}{l}{agr $\times$ retvol} & \multicolumn{1}{c}{0.318} & \multicolumn{1}{c}{0.798} & \multicolumn{1}{c}{-0.058} & \multicolumn{1}{c}{0.660} & \multicolumn{1}{c}{-0.34} & \multicolumn{1}{c}{-0.38} \\
		\multicolumn{1}{l}{baspread $\times$ mve0} & \multicolumn{1}{c}{0.317} & \multicolumn{1}{c}{0.887} & \multicolumn{1}{c}{-0.124} & \multicolumn{1}{c}{0.750} & \multicolumn{1}{c}{-0.33} & \multicolumn{1}{c}{0.10} \\
		\multicolumn{1}{l}{agr $\times$ baspread} & \multicolumn{1}{c}{0.292} & \multicolumn{1}{c}{0.785} & \multicolumn{1}{c}{-0.077} & \multicolumn{1}{c}{0.611} & \multicolumn{1}{c}{-0.32} & \multicolumn{1}{c}{-0.33} \\
		\multicolumn{1}{l}{maxret $\times$ mom12m} & \multicolumn{1}{c}{0.291} & \multicolumn{1}{c}{1.869} & \multicolumn{1}{c}{-0.232} & \multicolumn{1}{c}{1.111} & \multicolumn{1}{c}{-0.30} & \multicolumn{1}{c}{-0.11} \\
		\multicolumn{1}{l}{mve0 $\times$ retvol} & \multicolumn{1}{c}{0.287} & \multicolumn{1}{c}{0.917} & \multicolumn{1}{c}{-0.124} & \multicolumn{1}{c}{0.753} & \multicolumn{1}{c}{-0.31} & \multicolumn{1}{c}{0.12} \\
		&       &       &       &       &       &  \\
		\multicolumn{1}{l}{Bottom-10 interaction effects} &       &       &       &       &       &  \\
		\multicolumn{1}{l}{mom1m $\times$ retvol} & \multicolumn{1}{c}{-0.562} & \multicolumn{1}{c}{1.890} & \multicolumn{1}{c}{-1.063} & \multicolumn{1}{c}{0.217} & \multicolumn{1}{c}{-0.35} & \multicolumn{1}{c}{-0.21} \\
		\multicolumn{1}{l}{maxret $\times$ mom1m} & \multicolumn{1}{c}{-0.491} & \multicolumn{1}{c}{1.705} & \multicolumn{1}{c}{-0.944} & \multicolumn{1}{c}{0.165} & \multicolumn{1}{c}{-0.30} & \multicolumn{1}{c}{-0.18} \\
		\multicolumn{1}{l}{baspread $\times$ mom1m} & \multicolumn{1}{c}{-0.478} & \multicolumn{1}{c}{1.907} & \multicolumn{1}{c}{-1.012} & \multicolumn{1}{c}{0.249} & \multicolumn{1}{c}{-0.33} & \multicolumn{1}{c}{-0.18} \\
		\multicolumn{1}{l}{idiovol $\times$ mom1m} & \multicolumn{1}{c}{-0.472} & \multicolumn{1}{c}{1.763} & \multicolumn{1}{c}{-1.012} & \multicolumn{1}{c}{0.262} & \multicolumn{1}{c}{-0.37} & \multicolumn{1}{c}{-0.23} \\
		\multicolumn{1}{l}{mom1m $\times$ tang} & \multicolumn{1}{c}{-0.443} & \multicolumn{1}{c}{1.396} & \multicolumn{1}{c}{-0.859} & \multicolumn{1}{c}{0.109} & \multicolumn{1}{c}{-0.37} & \multicolumn{1}{c}{-0.23} \\
		\multicolumn{1}{l}{beta $\times$ mom1m} & \multicolumn{1}{c}{-0.440} & \multicolumn{1}{c}{1.798} & \multicolumn{1}{c}{-1.040} & \multicolumn{1}{c}{0.289} & \multicolumn{1}{c}{-0.38} & \multicolumn{1}{c}{-0.26} \\
		\multicolumn{1}{l}{age $\times$ mom1m} & \multicolumn{1}{c}{-0.435} & \multicolumn{1}{c}{1.064} & \multicolumn{1}{c}{-0.810} & \multicolumn{1}{c}{0.061} & \multicolumn{1}{c}{-0.42} & \multicolumn{1}{c}{-0.27} \\
		\multicolumn{1}{l}{dolvol $\times$ mom1m} & \multicolumn{1}{c}{-0.410} & \multicolumn{1}{c}{1.730} & \multicolumn{1}{c}{-0.899} & \multicolumn{1}{c}{0.251} & \multicolumn{1}{c}{-0.39} & \multicolumn{1}{c}{-0.26} \\
		\multicolumn{1}{l}{mom1m $\times$ std\_dolvol} & \multicolumn{1}{c}{-0.409} & \multicolumn{1}{c}{1.484} & \multicolumn{1}{c}{-0.834} & \multicolumn{1}{c}{0.186} & \multicolumn{1}{c}{-0.34} & \multicolumn{1}{c}{-0.20} \\
		\multicolumn{1}{l}{mom1m $\times$ ps} & \multicolumn{1}{c}{-0.403} & \multicolumn{1}{c}{1.173} & \multicolumn{1}{c}{-0.779} & \multicolumn{1}{c}{0.105} & \multicolumn{1}{c}{-0.40} & \multicolumn{1}{c}{-0.27} \\
		\addlinespace
	\end{longtable}
\end{small}
\doublespacing
\clearpage

\onehalfspacing
\begin{table}[h]
	\caption{\textbf{Standard deviation of out-of-sample minimum-variance portfolio returns}}\label{table:out_sample_std_dev}
	\scriptsize{The table reports annualized out-of-sample (OOS) standard deviation of portfolio returns for the minimum-variance parametric portfolio strategy when the ML methods discussed in Section \ref{section:ML_methods} are used to perform variable selection and regularization. The table also reports the performance of the minimum-variance strategy when using only the market, size, value, profitability, and investment characteristics as well as the performance of the equally weighted and value-weighted strategies. The ML-based portfolios are implemented with two alternative configurations of the data set discussed in Section \ref{section:data}: using only the original 95 predictors used in \cite{gu2020empirical,gu2021autoencoder} and ii) all 4,610 predictors including the second-order and cross-products transformations. One, two, and three asterisks indicate that the differences in portfolio variance with respect to those of the lasso strategy with 4,610 predictors are significant at the 10\%, 5\% and 1\% level, respectively.}	
	\begin{center}
		\resizebox{11cm}{!}{
		
			\begin{tabular}{lc}
				\toprule
				& Std. dev. of OOS \\
				&  portfolio returns \\
				\cmidrule{2-2}market, size, value & 0.150*** \\
				market, size, value, profit., invest. & 0.146*** \\
				equally weighted & 0.183*** \\
				value-weighted & 0.158*** \\
				&  \\
				OLS   & 0.176*** \\
				&  \\
				Lasso &  \\
				(95 characteristics) & 0.091* \\
				(4,610 characteristics) & 0.083 \\
				&  \\
				Adalasso &  \\
				(95 characteristics) & 0.097** \\
				(4,610 characteristics) & 0.091** \\
				&  \\
				Elastic net  &  \\
				(95 characteristics) & 0.092 \\
				(4,610 characteristics) & 0.084 \\
				&  \\
				Ridge &  \\
				(95 characteristics) & 0.117*** \\
				(4,610 characteristics) & 0.158*** \\
				&  \\
				Horseshoe &  \\
				(95 characteristics) & 0.089* \\
				(4,610 characteristics) & 0.094** \\
				&  \\
				Scad  &  \\
				(95 characteristics) & 0.100*** \\
				(4,610 characteristics) & 0.091** \\
				&  \\
				$L2$-boosting &  \\
				(95 characteristics) & 0.090** \\
				(4,610 characteristics) & 0.087 \\
				\bottomrule
			\end{tabular}%
			
		}
	\end{center}
\end{table}
\doublespacing

\onehalfspacing
\begin{table}[h]
	\caption{\textbf{Out-of-sample performance statistics of various portfolio policies}}\label{table:out_sample_performance}
	\scriptsize{The table reports annualized out-of-sample risk-adjusted portfolio returns measured by the Sharpe ratio (SR). Minimum-variance  parametric portfolio are obtained when using the ML methods discussed in Section \ref{section:ML_methods} to perform variable selection and regularization. The table also reports the performance of the minimum-variance strategy when using only the market, size, value, profitability, and investment characteristics as well as the performance of the equally weighted and value-weighted strategies. The ML-based portfolios are implemented with two alternative configurations of the data set discussed in Section \ref{section:data}: using only the original 95 predictors used in \cite{gu2020empirical,gu2021autoencoder} and ii) all 4,610 predictors including the second-order and cross-products transformations. The table also reports the average monthly turnover of each strategy. Panels A and B report results assuming two alternative levels of transaction costs (T.C.): 0 basis points (bp) and 10 bp. One, two, and three asterisks indicate that the differences in portfolio Sharpe ratios with respect to those of the lasso strategy with 4,610 predictors are significant at the 10\%, 5\% and 1\% level, respectively.}	
	\begin{center}
		\resizebox{17cm}{!}{
			
			\begin{tabular}{lcccc}
				\toprule
				& SR  &       & Turnover & SR \\
				& before T.C. &       &       & after T.C. (10 bp) \\
				\midrule
				&       &       &       &  \\
				market, size, value &  0.690*** &       & 0.34  &  0.662*** \\
				market, size, value, profit., invest. &  0.991*** &       & 0.51  &  0.949*** \\
				equally weighted &  0.454 &       & 0.16  &  0.444** \\
				value-weighted &  0.509* &       & 0.11  &  0.500*** \\
				&       &       &       &  \\
				OLS   & -0.075 &       & 5.14  & -0.428* \\
				&       &       &       &  \\
				Lasso &       &       &       &  \\
				(95 characteristics) &  0.112* &       & 1.77  & -0.121 \\
				(4,610 characteristics) & -0.156 &       & 1.14  & -0.320 \\
				&       &       &       &  \\
				Adalasso &       &       &       &  \\
				(95 characteristics) &  0.292*** &       & 2.31  &  0.004* \\
				(4,610 characteristics) &  0.142** &       & 2.36  & -0.169 \\
				&       &       &       &  \\
				Elastic net  &       &       &       &  \\
				(95 characteristics) &  0.107* &       & 1.72  & -0.119 \\
				(4,610 characteristics) & -0.164 &       & 1.15  & -0.328 \\
				&       &       &       &  \\
				Ridge &       &       &       &  \\
				(95 characteristics) &  0.842*** &       & 0.27  &  0.814*** \\
				(4,610 characteristics) &  0.557** &       & 0.15  &  0.545*** \\
				&       &       &       &  \\
				Horseshoe &       &       &       &  \\
				(95 characteristics) &  0.378*** &       & 1.07  &  0.233*** \\
				(4,610 characteristics) &  0.194** &       & 3.48  & -0.251 \\
				&       &       &       &  \\
				Scad  &       &       &       &  \\
				(95 characteristics) &  0.087 &       & 2.21  & -0.178 \\
				(4,610 characteristics) & -0.339** &       & 2.46  & -0.664*** \\
				&       &       &       &  \\
				$L2$-boosting &       &       &       &  \\
				(95 characteristics) &  1.220*** &       & 1.04  &  1.076*** \\
				(4,610 characteristics) &  1.153*** &       & 1.67  &  0.925*** \\
				\bottomrule
			\end{tabular}%
			
		}
	\end{center}
\end{table}
\doublespacing

\onehalfspacing
\begin{table}[h]
	\caption{\textbf{Out-of-sample portfolio drawdown and value-at-risk}}\label{table:portfolio_measure}
	\scriptsize{The table reports the maximum drawdown and the value-at-risk (VaR) based on the historical simulation method with 99\% confidence for the alternative portfolios strategies implemented in the paper.}	
	\begin{center}
		\resizebox{14cm}{!}{
			\begin{tabular}{lcc}
				\hline
				& Maximun Drawdown & VaR (99\%) \bigstrut[t]\\
				&       &  \\
				market, size, value & 52.7\% & 11.3\% \\
				market, size, value, profit., invest. & 54.2\% & 12.2\% \\
				equally weighted & 56.8\% & 13.1\% \\
				value-weighted & 52.9\% & 11.4\% \\
				&       &  \\
				OLS (95 characteristics) & 68.9\% & 14.8\% \\
				&       &  \\
				Lasso &       &  \\
				(95 characteristics) & 53.7\% & 6.5\% \\
				(4,610 characteristics) & 55.9\% & 7.9\% \\
				&       &  \\
				Adalasso &       &  \\
				(95 characteristics) & 49.2\% & 8.3\% \\
				(4,610 characteristics) & 40.8\% & 7.7\% \\
				&       &  \\
				Elastic net  &       &  \\
				(95 characteristics) & 54.1\% & 7.1\% \\
				(4,610 characteristics) & 55.3\% & 7.9\% \\
				&       &  \\
				Ridge &       &  \\
				(95 characteristics) & 48.3\% & 9.8\% \\
				(4,610 characteristics) & 53.6\% & 11.4\% \\
				&       &  \\
				Horseshoe &       &  \\
				(95 characteristics) & 47.1\% & 6.7\% \\
				(4,610 characteristics) & 49.3\% & 7.4\% \\
				&       &  \\
				Scad  &       &  \\
				(95 characteristics) & 59.6\% & 8.2\% \\
				(4,610 characteristics) & 68.3\% & 7.1\% \\
				&       &  \\
				$L2$-boosting &       &  \\
				(95 characteristics) & 38.6\% & 7.4\% \\
				(4,610 characteristics) & 39.2\% & 6.0\% \bigstrut[b]\\
				\hline
			\end{tabular}%
		}
	\end{center}
\end{table}
\doublespacing

\begin{landscape}
\onehalfspacing
\begin{table}[h]
	\caption{\textbf{Predictor selection}}\label{table:selection}
	\scriptsize{The table presents the average total number of selected predictors for each method, alongside the percentage distribution across predictor classes—main effects, second-order effects, and interaction effects. The ``Class-relative percentage'' panel shows the proportion of selected predictors within each class. The ``Difference from equal-importance'' panel displays the deviation of each class's proportion from an equal-importance benchmark, indicating if certain predictor classes are over- or underrepresented relative to an equal distribution. All values are averaged across all estimation rounds.		
	}	
	\begin{center}
		\resizebox{18cm}{!}{
\begin{tabular}{rrrrrrrrrr}
	\hline
	&       &       &       &       &       &       & \multicolumn{3}{c}{} \\
	&       &       & \multicolumn{3}{c}{Class-relative percentage} &       & \multicolumn{3}{c}{Difference from equal-importance} \\
	\cmidrule{2-2}\cmidrule{4-6}\cmidrule{8-10}      & \multicolumn{1}{c}{Numer of selected } &       & \multicolumn{1}{c}{Main effects} & \multicolumn{1}{c}{Second-order} & \multicolumn{1}{c}{Interaction} &       & \multicolumn{1}{c}{Main effects} & \multicolumn{1}{c}{Second-order} & \multicolumn{1}{c}{Interaction} \\
	& \multicolumn{1}{c}{predictors} &       &       & \multicolumn{1}{c}{effects} & \multicolumn{1}{c}{effects} &       &       & \multicolumn{1}{c}{effects} & \multicolumn{1}{c}{effects} \\
	\cmidrule{2-2}\cmidrule{4-6}\cmidrule{8-10}      &       &       &       &       &       &       &       &       &  \\
	\multicolumn{1}{l}{Lasso} & \multicolumn{1}{c}{33} &       & \multicolumn{1}{c}{5.5\%} & \multicolumn{1}{c}{7.2\%} & \multicolumn{1}{c}{87.3\%} &       & \multicolumn{1}{c}{3.5\%} & \multicolumn{1}{c}{5.2\%} & \multicolumn{1}{c}{-8.7\%} \\
	&       &       &       &       &       &       &       &       &  \\
	\multicolumn{1}{l}{Adalasso} & \multicolumn{1}{c}{62} &       & \multicolumn{1}{c}{0.8\%} & \multicolumn{1}{c}{2.5\%} & \multicolumn{1}{c}{96.7\%} &       & \multicolumn{1}{c}{-1.3\%} & \multicolumn{1}{c}{0.6\%} & \multicolumn{1}{c}{0.7\%} \\
	&       &       &       &       &       &       &       &       &  \\
	\multicolumn{1}{l}{Elastic net } & \multicolumn{1}{c}{60} &       & \multicolumn{1}{c}{4.8\%} & \multicolumn{1}{c}{6.5\%} & \multicolumn{1}{c}{88.7\%} &       & \multicolumn{1}{c}{2.7\%} & \multicolumn{1}{c}{4.6\%} & \multicolumn{1}{c}{-7.3\%} \\
	&       &       &       &       &       &       &       &       &  \\
	\multicolumn{1}{l}{Horseshoe} & \multicolumn{1}{c}{47} &       & \multicolumn{1}{c}{3.5\%} & \multicolumn{1}{c}{2.8\%} & \multicolumn{1}{c}{93.7\%} &       & \multicolumn{1}{c}{1.5\%} & \multicolumn{1}{c}{0.9\%} & \multicolumn{1}{c}{-2.3\%} \\
	&       &       &       &       &       &       &       &       &  \\
	\multicolumn{1}{l}{Scad} & \multicolumn{1}{c}{16} &       & \multicolumn{1}{c}{16.0\%} & \multicolumn{1}{c}{4.4\%} & \multicolumn{1}{c}{79.5\%} &       & \multicolumn{1}{c}{14.0\%} & \multicolumn{1}{c}{2.5\%} & \multicolumn{1}{c}{-16.5\%} \\
	&       &       &       &       &       &       &       &       &  \\
	\multicolumn{1}{l}{L2-boosting} & \multicolumn{1}{c}{75} &       & \multicolumn{1}{c}{1.4\%} & \multicolumn{1}{c}{1.8\%} & \multicolumn{1}{c}{96.8\%} &       & \multicolumn{1}{c}{-0.7\%} & \multicolumn{1}{c}{-0.2\%} & \multicolumn{1}{c}{0.8\%} \\
	&       &       &       &       &       &       &       &       &  \\
	\bottomrule
\end{tabular}%

		}
\end{center}
\end{table}
\end{landscape}

\onehalfspacing
\begin{table}[h]
	\caption{\textbf{Economic significance of the most important predictors}}\label{table:state_variables}
	\scriptsize{This table reports the regression results of selected predictor coefficients on state variables, including the investor sentiment index from \cite{baker2006investor}, the liquidity index from \cite{pastor2003liquidity}, and the NBER recession indicator. The results are presented for six different estimation methods: Lasso, Adalasso, Elastic-Net, Scad, L2-Boosting, and Horseshoe. Panel A reports the results for the main effects of each predictor, while Panel B shows the interaction effects. One, two, and three asterisks indicate that the estimated coefficient is significant at the 10\%, 5\% and 1\% level, respectively. A dash (``-'') indicates that the predictor was not selected by the method or that the estimation coefficient is equal to zero.
	}	
	\begin{center}
		\resizebox{16cm}{!}{
			\begin{tabular}{rcccccc}
				\toprule
				& Lasso & Adalasso & Elastic-net & Scad  & L2-Boosting & Horseshoe \\
				\cmidrule{2-7}\multicolumn{1}{l}{\textbf{Panel A: Main effects}} &       &       &       &       &       &  \\
				&       &       &       &       &       &  \\
				\multicolumn{1}{l}{\textit{\textbf{beta}}} &       &       &       &       &       &  \\
				\multicolumn{1}{l}{sentiment} & -0.208*** & -0.097*** & -0.152*** & -0.475*** & -0.135*** & -0.312** \\
				&       &       &       &       &       &  \\
				\multicolumn{1}{l}{recession} & -0.087 & 0.025 & 0.155 & 0.077 & -0.173*** & 0.205 \\
				&       &       &       &       &       &  \\
				\multicolumn{1}{l}{liquidity} & -0.127 & -0.058 & -0.060 & 0.222 & -0.199 & 0.166 \\
				&       &       &       &       &       &  \\
				\multicolumn{1}{l}{\textit{\textbf{dolvol}}} &       &       &       &       &       &  \\
				\multicolumn{1}{l}{sentiment} & -     & -     & 0.001 & 0.031 & -     & -0.004 \\
				&       &       &       &       &       &  \\
				\multicolumn{1}{l}{recession} & -     & -     & 0.002 & 0.064 & -     & 0.003 \\
				&       &       &       &       &       &  \\
				\multicolumn{1}{l}{liquidity} & -0.001 & -     & 0.001 & -0.016 & -     & -0.074 \\
				&       &       &       &       &       &  \\
				\multicolumn{1}{l}{\textit{\textbf{idiovol}}} &       &       &       &       &       &  \\
				\multicolumn{1}{l}{sentiment} & 0.022 & -     & 0.003 & -0.067 & -     & 0.068* \\
				&       &       &       &       &       &  \\
				\multicolumn{1}{l}{recession} & -0.040 & -     & -0.045* & 0.047 & -     & 0.051 \\
				&       &       &       &       &       &  \\
				\multicolumn{1}{l}{liquidity} & 0.008 & -     & 0.004 & -0.264 & -     & 0.205 \\
				&       &       &       &       &       &  \\
				\multicolumn{1}{l}{\textbf{Panel B: Interactions}} &       &       &       &       &       &  \\
				&       &       &       &       &       &  \\
				\multicolumn{1}{l}{\textit{\textbf{beta\_x\_dolvol}}} &       &       &       &       &       &  \\
				\multicolumn{1}{l}{sentiment} & -0.012*** & 0.020*** & -0.014* & -0.038*** & -     & 0.055 \\
				&       &       &       &       &       &  \\
				\multicolumn{1}{l}{recession} & -0.017*** & 0.019 & -0.012 & -0.059** & -     & 0.086 \\
				&       &       &       &       &       &  \\
				\multicolumn{1}{l}{liquidity} & -0.022 & -0.083 & -0.03 & -0.170 & -     & -0.082 \\
				&       &       &       &       &       &  \\
				\multicolumn{1}{l}{\textit{\textbf{dolvol\_x\_idiovol}}} &       &       &       &       &       &  \\
				\multicolumn{1}{l}{sentiment} & 0.184*** & 0.001 & 0.198*** & 0.219*** & 0.046*** & 0.027 \\
				&       &       &       &       &       &  \\
				\multicolumn{1}{l}{recession} & 0.005 & 0.001 & -0.016 & 0.016 & 0.023 & 0.076 \\
				&       &       &       &       &       &  \\
				\multicolumn{1}{l}{liquidity} & 0.257 & -0.004 & 0.154 & 0.072 & -0.115 & -0.477 \\
				&       &       &       &       &       &  \\
				\multicolumn{1}{l}{\textit{\textbf{beta\_x\_idiovol}}} &       &       &       &       &       &  \\
				\multicolumn{1}{l}{sentiment} & -     & -     & -0.007 & -0.013 & 0.050*** & 0.024 \\
				&       &       &       &       &       &  \\
				\multicolumn{1}{l}{recession} & -     & -     & -0.048*** & -0.011 & 0.052*** & 0.008 \\
				&       &       &       &       &       &  \\
				\multicolumn{1}{l}{liquidity} & -     & -     & -0.040 & -0.134 & -0.243** & -0.050 \\
				&       &       &       &       &       &  \\
				\bottomrule
			\end{tabular}%
  	}
	\end{center}
\end{table}

\doublespacing

\onehalfspacing
\begin{table}[h]
	\caption{\textbf{Portfolio weight statistics}}\label{table:portfolio_stats}
	\scriptsize{The table reports descriptive statistics of the distribution of portfolio weights of the various strategies considered. The columns reports the time series averages of the minimum weight, maximum weight, and fraction of negative weights (short proportion).}	
	\begin{center}
		\resizebox{14cm}{!}{
			\begin{tabular}{lccc}
				\hline
				& Minimum  & Maximum & Short \bigstrut[t]\\
				& weight (\%) & weight (\%) & proportion (\%) \bigstrut[b]\\
				\hline
				market, size, value & -0.143 & 0.578 & 0.426 \bigstrut[t]\\
				market, size, value, profit., invest. & -0.352 & 0.543 & 0.379 \\
				equally weighted & 0.000 & 0.067 & 0.000 \\
				value-weighted & 0.000 & 0.432 & 0.000 \\
				&       &       &  \\
				OLS (95 characteristics) & -1.760 & 1.753 & 0.442 \\
				&       &       &  \\
				Lasso &       &       &  \\
				(95 characteristics) & -0.422 & 0.477 & 0.379 \\
				(4,610 characteristics) & -0.318 & 0.395 & 0.329 \\
				&       &       &  \\
				Adalasso &       &       &  \\
				(95 characteristics) & -0.663 & 0.710 & 0.386 \\
				(4,610 characteristics) & -0.834 & 0.821 & 0.333 \\
				&       &       &  \\
				Elastic net  &       &       &  \\
				(95 characteristics) & -0.401 & 0.443 & 0.371 \\
				(4,610 characteristics) & -0.315 & 0.382 & 0.324 \\
				&       &       &  \\
				Ridge &       &       &  \\
				(95 characteristics) & -0.105 & 0.075 & 0.152 \\
				(4,610 characteristics) & -0.029 & 0.051 & 0.022 \\
				&       &       &  \\
				Horseshoe &       &       &  \\
				(95 characteristics) & -0.297 & 0.532 & 0.384 \\
				(4,610 characteristics) & -0.494 & 0.825 & 0.431 \\
				&       &       &  \\
				Scad  &       &       &  \\
				(95 characteristics) & -0.431 & 0.566 & 0.397 \\
				(4,610 characteristics) & -0.446 & 0.534 & 0.371 \\
				&       &       &  \\
				$L2$-boosting &       &       &  \\
				(95 characteristics) & -0.406 & 0.454 & 0.325 \\
				(4,610 characteristics) & -0.696 & 0.657 & 0.328 \bigstrut[b]\\
				\hline
			\end{tabular}%
			
		}
	\end{center}
\end{table}
\doublespacing
\clearpage

\onehalfspacing
\begin{table}[h]
	\caption{\textbf{Performance of minimum-variance parametric portfolios with marginal screening}}\label{table:screening}
	\scriptsize{The table reports annualized out-of-sample (OOS) standard deviation of the minimum-variance parametric portfolio returns when the marginal screening (MS) method of \cite{fan2008sure} is used to screen the pool of 4,610 predictors to select the best 5, 10, and 50 predictors. The table also reports the average monthly turnover of each strategy. Panels A and B report results assuming two alternative levels of transaction costs: 0 basis points (b.p.) and 10 b.p. One, two, and three asterisks indicate that the differences in portfolio variance and in Sharpe ratios with respect to those of the lasso strategy with 4,610 predictors are significant at the 10\%, 5\% and 1\% level, respectively.}	
	\begin{center}
		\resizebox{8cm}{!}{

\begin{tabular}{rr}
	\toprule
	& \multicolumn{1}{c}{Std. dev. of OOS} \\
	& \multicolumn{1}{c}{ portfolio returns} \\
	\cmidrule{2-2}      &  \\
	\multicolumn{1}{l}{Top-3 predictors} & \multicolumn{1}{c}{0.107***} \\
	&  \\
	\multicolumn{1}{l}{Top-5 predictors} & \multicolumn{1}{c}{0.105***} \\
	&  \\
	\multicolumn{1}{l}{Top-10 predictors} & \multicolumn{1}{c}{0.095**} \\
	&  \\
	\multicolumn{1}{l}{Top-50 predictors} & \multicolumn{1}{c}{0.103***} \\
	&  \\
	\bottomrule
\end{tabular}%

		}
	\end{center}
\end{table}%
\doublespacing


\vspace{0.2cm}

%


\section*{Algorithms}

The $L2$-boosting algorithm commences by initializing the 
parameter vector, \( \hat \theta^{[0]} \), to zero and setting the initial portfolio weights, 
\( w_t(\hat \theta_{0}) \), equal to the benchmark weights, \( w_{b_t} \). The algorithm then 
enters a loop that continues until a predefined stopping iteration \( m^{*} \) is reached. In each iteration \( m \), the algorithm systematically goes through each of the \( K \) 
characteristics. After evaluating all \( K \) characteristics, the algorithm selects the characteristic, \( k^{*} \), that minimizes the empirical objective function in eq.~\eqref{eq:emp.obj.fun}. It then updates the parameter vector \( \hat \theta^{[m]} \) to be zero for all elements except for the \( k^{*} \)-th element, which is set to \( \hat \theta_{k^{*}}^{[m]} \). The algorithm also updates the parameter vector \( \hat \theta_{m} \) 
using a step size \( \nu \in (0, 1) \), and accordingly adjusts the portfolio weights, \( w_t(\hat \theta_{m}) \), 
based on these updated parameters. The iteration counter \( m \) is incremented, and the process 
repeats until the termination criterion is met.

\begin{algorithm}[h]	
	\caption{$L2$-boosting for minimum-variance parametric portfolios}\label{algo_L2}
	\begin{algorithmic}[1]  
		\State Initialize $\hat \theta^{[0]} = 0,$ $w_t(\hat \theta_{0}) = w_{b_t}$ and
		$r_{b_{0}} = w_t(\hat \theta_{0})^{\intercal} r_{t+1}$
		\State Set $m = 1$
		\While{$m \leq m^{*}$}
		\For{$k = 1$ to $K$}
		\State Solve the empirical investor problem for the $k$-th characteristic:
		\State Minimize the objective function 
		\[
		\frac{1}{2}
		\frac{1}{(T - 1)} \sum_{t = 1}^{T - 1} 
		\left(\dot{r}_{b_m} + \theta_{k}^{[m]} \dot{r}_{c_k, t + 1}\right)^2
		\] 
		\State using the estimator: 
		\[
		\hat \theta_{k}^{[m]} = - \frac{\hat \sigma_{b_{m-1} c_{k}}}{\hat \sigma_{c_k}^2}
		\]
		\EndFor
		\State Choose $k^{*}$ that minimizes the objective function for all $k$
		\State Set $\hat \theta^{[m]}$ to zero except for the $k^{*}$-th element, which is $\hat \theta_{k^{*}}^{[m]}$
		\State Update $\hat \theta_{m}$ using a step size $\nu$:
		\[
		\hat \theta_{m} = \hat \theta_{m-1} + \nu \hat \theta^{[m]}, m \ge 1
		\]
		\State Update $w_t(\hat \theta_{m})$:
		\[
		w_t(\hat \theta_{m}) =
		w_{b_t} + \hat \theta_{m}^{\intercal} X_t / N_t, \ m \ge 1
		\]
		\State Update $r_{b_m}:$
		\[
			r_{b_m} = w_t(\hat \theta_{m})^{\intercal} r_{t+1}
		\]	
		\State $m = m + 1$
		\EndWhile
	\end{algorithmic}
\end{algorithm}

\begin{algorithm}
	\caption{Marginal Screening}
	\label{alg:marginal_screening}
	\begin{algorithmic}[1]
		\Require Design matrix $X \in \mathbb{R}^{n \times p}$, response vector $y \in \mathbb{R}^n$, model size $k$
		\Ensure Estimated coefficients $\hat{\beta}_{\hat{S}}$ for the selected variables
		
		\State Compute the product $X^T y$ to obtain a vector in $\mathbb{R}^p$
		\State Compute the absolute values $\left|X^T y\right|$ to get the marginal correlations
		\State Identify $\hat{S}$, the index set of the $k$ largest entries in $\left|X^T y\right|$
		\State Extract the submatrix $X_{\hat{S}}$ from $X$ using the indices in $\hat{S}$
		\State Compute $\hat{\beta}_{\hat{S}} = \left(X_{\hat{S}}^T X_{\hat{S}}\right)^{-1} X_{\hat{S}}^T y$ to estimate the coefficients for the selected variables. \Return $\hat{\beta}_{\hat{S}}$
	\end{algorithmic}
\end{algorithm}
	
\end{document}